\documentclass[11pt, notitlepage]{article} % {article}
\usepackage{a4}
\usepackage{epstopdf}
\usepackage{xcolor}

\definecolor{TUMblue}{RGB}{0, 101, 189}
\definecolor{TUMlightblue}{RGB}{100,160,200}
\definecolor{TUMgreen}{RGB}{162,173,0}
\definecolor{TUMorange}{RGB}{227,114,034}
\definecolor{TUMivory}{RGB}{218,215,203}
\definecolor{TUMivory}{RGB}{218,215,203}
\definecolor{firebrick2}{RGB}{238,44,44}
\definecolor{deepskyblue4}{RGB}{0,104,139}
\definecolor{brown3}{RGB}{251,51,51}
\definecolor{darkorchid3}{RGB}{154,50,205}
\definecolor{fuchsia}{rgb}{0.57, 0.36, 0.51}

\usepackage{natbib}

\usepackage{hyperref}
\hypersetup{
	colorlinks=true,
	linkcolor=TUMblue,
	citecolor=TUMblue,
	filecolor=TUMblue,
	urlcolor=TUMblue
}

\usepackage{etoolbox}
\makeatletter

\pretocmd{\NAT@citex}{%
	\let\NAT@hyper@\NAT@hyper@citex
	\def\NAT@postnote{#2}%
	\setcounter{NAT@total@cites}{0}%
	\setcounter{NAT@count@cites}{0}%
	\forcsvlist{\stepcounter{NAT@total@cites}\@gobble}{#3}}{}{}
\newcounter{NAT@total@cites}
\newcounter{NAT@count@cites}
\def\NAT@postnote{}

\def\NAT@hyper@citex#1{%
	\stepcounter{NAT@count@cites}%
	\hyper@natlinkstart{\@citeb\@extra@b@citeb}#1%
	\ifnumequal{\value{NAT@count@cites}}{\value{NAT@total@cites}}
	{\ifNAT@swa\else\if*\NAT@postnote*\else%
		\NAT@cmt\NAT@postnote\global\def\NAT@postnote{}\fi\fi}{}%
	\ifNAT@swa\else\if\relax\NAT@date\relax
	\else\NAT@@close\global\let\NAT@nm\@empty\fi\fi% avoid compact citations
	\hyper@natlinkend}
\renewcommand\hyper@natlinkbreak[2]{#1}

\patchcmd{\NAT@citex}
{\ifNAT@swa\else\if*#2*\else\NAT@cmt#2\fi
	\if\relax\NAT@date\relax\else\NAT@@close\fi\fi}{}{}{}
\patchcmd{\NAT@citex}
{\if\relax\NAT@date\relax\NAT@def@citea\else\NAT@def@citea@close\fi}
{\if\relax\NAT@date\relax\NAT@def@citea\else\NAT@def@citea@space\fi}{}{}

\makeatother

\usepackage{courier}
\usepackage{amssymb, graphicx}
\usepackage{subfigure}
\usepackage{amsmath}
\usepackage{amsthm}
\usepackage{calligra}
\usepackage{verbatim}
\usepackage{dsfont}
\usepackage{geometry}
\usepackage{pdflscape}
\usepackage{multirow}
\usepackage{aliascnt}
\usepackage{verbatim}

\usepackage{graphicx}
\usepackage{float}
\usepackage{tikz}

\usetikzlibrary{calc,shapes,arrows,decorations.pathmorphing,graphs,positioning,backgrounds, chains,arrows.meta}
\usepackage{latexsym}
\usepackage{mathtools}
\usepackage{mathrsfs}
\usepackage{bbm}
\usepackage{shadethm}
\usepackage{enumerate}
\usepackage{colortbl}
\usepackage{framed}
\colorlet{shadecolor}{gray!25}
\usepackage{booktabs}
\usepackage{longtable}
\usepackage{multirow}
\usepackage{rotating}
   \usepackage{pgfplots}
\pgfplotsset{compat=newest}
\usepackage{pgfplotstable}
\usepackage{chngpage}
\usepackage[colorinlistoftodos]{todonotes}

\newcommand{\mynewtheorem}[2]{
	\newaliascnt{#1}{dummy}
	\newtheorem{#1}[#1]{#2}
	\aliascntresetthe{#1}
	\expandafter\def\csname #1autorefname\endcsname{#2}
}

\bibpunct[\textcolor{TUMblue}{, }]{\textcolor{TUMblue}{(}}{\textcolor{TUMblue}{)}}{\textcolor{TUMblue}{;}}{\textcolor{TUMblue}{a}}{\textcolor{TUMblue}{}}{\textcolor{TUMblue}{,}}
\makeatletter
\renewcommand\eqref[1]{%
	\textup{\color{TUMblue}\tagform@{\ref{#1}}}%
}

\theoremstyle{definition}
\mynewtheorem{thm}{Theorem}
\mynewtheorem{defi}{Definition}%[section]
\mynewtheorem{lem}{Lemma}%[section]
\mynewtheorem{cor}{Corollary}%[section]
\mynewtheorem{prop}{Proposition}%[section]
\mynewtheorem{exa}{Example}%[section]
\mynewtheorem{alg}{Algorithm}%[section]
\mynewtheorem{rem}{Remark}%[section]
\mynewtheorem{bsp}{Example}

\newcommand{\aref}[1]{\hyperref[#1]{Section~\ref{#1}}}

\usepackage{sectsty}
\allsectionsfont{\sffamily}

\usepackage{titlesec}
\titleformat{\chapter}[display]
{\normalfont\sffamily\LARGE\bfseries\centering}
{\chaptertitlename\ \thechapter}{20pt}{\LARGE}

\usepackage{caption}
\usepackage{footnote}

\newcommand{\sC}{\mathbbm{C}} % sC for survival copula
\newcommand{\scd}{\mathbbm{c}} % sc for survival copula density

\begin{document}

{	\renewcommand*{\thefootnote}{\fnsymbol{footnote}}
	\title{\textbf{\sffamily A partial correlation vine based approach for modeling and forecasting multivariate volatility time-series}}

	\date{\small \today}
\newcounter{savecntr1}% Save footnote counter
\newcounter{restorecntr1}% Restore footnote counter
\newcounter{savecntr2}% Save footnote counter
\newcounter{restorecntr2}% Restore footnote counter	

%	\author{
%		NICOLE BARTHEL$^*$
%		\\%\vspace*{-.25cm}
%		\textit{Department of Mathematics, Technische Universit{\"a}t M{\"u}nchen, Garching, Germany}
%		\\%\vspace*{-.25cm}
%		{Boltzmannstra{\ss}e 3, 85748 Garching bei M{\"u}nchen, email: nicole.barthel@tum.de}
%		\\[4pt]
%		CLAUDIA CZADO
%		\\%\vspace*{-.25cm}
%		\textit{Department of Mathematics, Technische Universit{\"a}t M{\"u}nchen, Garching, Germany}
%		\\[4pt]
%		YAREMA OKHRIN
%		\\%\vspace*{-.25cm}
%		\textit{Department of Statistics, Universit{\"a}t Augsburg, Germany}
%		\\[4pt]
%	}
	\author{Nicole Barthel\setcounter{savecntr1}{\value{footnote}}\thanks{Department of Mathematics, Technische Universit{\"a}t M{\"u}nchen, Germany (email: \href{mailto:nicole.barthel@tum.de}{nicole.barthel@tum.de} (corresponding author), \href{mailto:cczado@ma.tum.de}{cczado@ma.tum.de})}, \  Claudia Czado\setcounter{restorecntr1}{\value{footnote}}%
		\setcounter{footnote}{\value{savecntr1}}\footnotemark% Print footnotemark
		\setcounter{footnote}{\value{restorecntr1}} \ and Yarema Okhrin\setcounter{savecntr2}{\value{footnote}}\thanks{Department of Statistics, Faculty of Business and Economics, Universit{\"a}t Augsburg, Germany (email: \href{mailto:yarema.okhrin@wiwi.uni-augsburg.de}{yarema.okhrin@wiwi.uni-augsburg.de})}\setcounter{restorecntr2}{\value{footnote}}%
		\setcounter{footnote}{\value{savecntr2}}\footnotemark% Print footnotemark
		\setcounter{footnote}{\value{restorecntr2}}}
%	\date{}
	\maketitle
%\vspace*{-.5cm}

}

%\vspace*{-.25cm}
%\paragraph{{\small Conflict of interest}} There is no conflict of interest to declare.
%\vspace*{-.25cm}	
%\paragraph{{\small Abstract}}	
\begin{abstract} 
	A novel approach for dynamic modeling and forecasting of realized covariance matrices is proposed. Realized variances and realized correlation matrices are jointly estimated. The one-to-one relationship between a positive definite correlation matrix and its associated set of partial correlations corresponding to any vine specification is used for data transformation. The model components therefore are realized variances as well as realized standard and partial correlations corresponding to a daily log-return series. As such, they have a clear practical interpretation. A method to select a regular vine structure, which allows for parsimonious time-series and dependence modeling of the model components, is introduced. Being algebraically independent the latter do not underlie any algebraic constraint. The proposed model approach is outlined in detail and motivated along with a real data example on six highly liquid stocks. The forecasting performance is evaluated both with respect to statistical precision and in the context of portfolio optimization. Comparisons with Cholesky decomposition based benchmark models support the excellent prediction ability of the proposed model approach.\bigskip\newline
	\bigskip
\end{abstract}

  % \newpage
  % \toccontents
  %\listoftodos
  %  \newpage
		
	\section{Introduction}

The increasing availability of high-frequency data makes volatility modeling and forecasting to one of the most vividly discussed topics in financial econometrics. Also, the strongly increasing interaction and interconnectedness between financial markets have stimulated the need for reliable modeling and forecasting techniques to capture the cross-sectional and temporal dependencies of financial asset returns. Especially during negative economic phases and periods of financial turmoil, assets become more dependent and linkages between asset market volatility tighten \citep{Cappiello2006}. This affects fields such as asset pricing, portfolio allocation and evaluation of risk. 

High-frequency data allow to consistently estimate ex-post realized volatility and realized covariances using the sum of squared intra-day returns \citep{doleans1970integrales, jacod1994limit}. By making naturally latent variables, namely volatilities and covariances, observable and measurable, standard time-series approaches can be applied to model their realized counterparts. Building upon the aforementioned classical estimator first used in the context of high-frequency data by \cite{Barndorff2004ApproxRCOV}, many refinements were investigated to improve its overall quality and precision \citep{Zhang2011SubsamplingRCOV}, to reduce market microstructure noise \citep{Daracogna2001PreviousTick, zhang2005tale} and to take into account jumps \citep{christensen2010pre} and asynchronicity \citep{hayashi2005covariance}. 

The main modeling challenge when developing prediction tools for realized covariance matrices are the algebraic restrictions of symmetry and positive definiteness the forecasts need to satisfy. Direct modeling of the components using univariate time-series models does not meet this constraint \citep{andersen2006volatility} and neglects e.g.\ dynamic volatility spillovers among the series of variances and covariances \citep{voev2008dynamic}. \textcolor{black}{Several multivariate approaches such as the Wishart Autoregressive (WAR) model \citep{gourieroux2009wishart} and its dynamic counterpart the Conditional Autoregressive Wishart (CAW) model \citep{golosnoy2012conditional} have been developed. \cite{andersen2006volatility} propose a multivariate generalization of the realized GARCH model \citep{Hansen2012} by modifying the Dynamic Conditional Correlation (DCC) model of \cite{Engle2002}. The basic idea of the latter model is to split up the estimation problem into the two simpler tasks of modeling the conditional volatilities and the correlation dynamics. \cite{halbleib2014forecasting} adopt this strategy using high-frequency data in the volatility part and daily data in the correlation part at the expense of less flexible correlation specifications.} As an alternative, data transformation is one of the most frequently used approaches. \cite{Bauer2011} apply the matrix logarithm function and a factor model approach to the individual components, which, however, leads to a computationally demanding model. First proposed by \cite{andersen2003modeling} and having evolved to one of the standard ways to proceed, the Cholesky decomposition is a proven tool to guarantee symmetry and positive definiteness of the forecasts. For example, \cite{ChiriacVoev2011ForecastingMultVol} decompose the series of realized covariance matrices via the Cholesky factorization and model the so-obtained series of Cholesky elements with a vector autoregressive fractionally integrated moving average (VARFIMA) process. \cite{Brechmann2015Cholesky} build upon this model approach, but pay special attention to the specific dependencies among the Cholesky series induced by the nonlinear data transformation. While Cholesky decomposition based models are straightforward and easy to implement, they also come with drawbacks. There is no clear interpretation of the \textcolor{black}{model components obtained after data transformation} and the latter induces an additive bias in the forecasts of the original data due to its nonlinear nature. Also, the Cholesky decomposition depends on the ordering of the data within the realized covariance matrices with no obvious way to fix the order in advance. Complete enumeration leads to a computationally expensive estimation problem. \textcolor{black}{On the other hand, fixing the order upfront ignores a possible changing behavior of the data over time.}

Irrespective of the considered data transformation, multivariate approaches for time-series modeling often suffer from lacking flexibility in the parameters. Further, they barely allow for convenient modeling of non-Gaussianity and conditional heteroscedasticity, which, however, are typical features of volatility data. In comparison, univariate time-series models allow for various extensions and refinements to tackle these problems. Besides ARFIMA processes \citep{andersen2006volatility}, heterogeneous autoregressive (HAR) processes are most commonly applied to (log-transformed) realized volatility time-series capturing their long-memory behavior. They include volatility measured over different time horizons and account for multifractal scaling \citep{corsi2009simple}. Both ARFIMA and HAR models can be extended by e.g.\ GARCH augmentations to account for non-Gaussianity and volatility clustering \citep{corsi2008volatility}. By considering skewed error distributions for the residuals, typically observed high skewness and kurtosis can be additionally captured \citep{bai2003kurtosis,fernandez1998bayesian}. 

In the light of the above discussion, \textcolor{black}{a tool to transform the realized covariance matrices, which allows for reasonable computational effort, interpretability of the model components obtained after data transformation and to exploit the beneficial features of univariate time-series modeling, is desirable.} A promising candidate which meets these requirements are partial correlation vines. The latter assign partial correlations to the edges of an R-vine tree structure. The latter is a graph theoretical object first proposed by \cite{bedford2002vines}, which consists of a set of linked trees specifying bivariate conditional constraints. The set of standard and partial correlations specified through an R-vine structure has attractive properties. \cite{bedford2002vines} proof that there is a bijection between the specified (partial) correlations and the set of symmetric and positive definite correlation matrices. Further, \cite{kurowicka2003parameterization} find that any partial correlation vine specifies algebraically independent (partial) correlations, i.e.\ the latter can take arbitrary values in $(-1,1)$ while still guaranteeing positive definiteness of the corresponding correlation matrix. This result advocates partial correlation vines to be a useful tool in several applications. \cite{kurowicka2006completion} use them to solve the completion problem for positive definite matrices, whereas \cite{lewandowski2009generating} introduce a method to uniformly generate random correlation matrices from the space of positive definite correlation matrices. \cite{brechmann2014parsimonious} base a parsimonious parameterization of correlation matrices on partial correlation vines in combination with factor analysis and \cite{brechmann2015truncationFitIndices} use these findings to capture the dependence structure in multivariate data. Considering financial data, \cite{poignard2017new} introduces a vine-GARCH approach as flexible multivariate GARCH-type model, which parametrizes the latent correlations appearing in the DCC model of \cite{Engle2002} in terms of a partial correlation vine. Based on the specific nature of an R-vine tree structure, their estimation technique proceeds iteratively by evoking only bivariate GARCH models in each tree level and thus allows for dimension reduction as compared to computationally highly demanding classical multivariate GARCH models.

\textcolor{black}{To our knowledge, data transformation using partial correlation vines has not yet been investigated to model and forecast multivariate realized volatility time-series. We propose a joint estimation and prediction model of the realized variance times-series and a subset of realized standard and partial correlation time-series. The latter are obtained after transforming the series of realized correlation matrices based on an R-vine structure as first step of the model approach. To select among the large number of possible R-vine structures the one used for data transformation, we propose a selection method, which exclusively relies on historical information of the modeled time-series and thus dynamically adapts to changing data behavior over time. We will show that data transformation based on this R-vine structure further allows for parsimony in the resulting multivariate time-series models, which are to be estimated as second step of the model approach. We opt for a copula based time-series model to exploit the beneficial features of elaborate univariate time-series models. By considering flexible copulas for the dependence between the model components possible asymmetry and nonlinearity can be captured. Combining in a third step the predicted realized variances and the predicted realized correlation matrix obtained after back-transformation of the underlying realized partial correlation vine guarantees a symmetric and positive definite realized covariance matrix forecast.}	

The paper is structured as follows. In \autoref{Sec:PartialCorrelationVines}, we introduce partial correlation vines combining the notion of partial correlations and an R-vine structure. \textcolor{black}{The transformation of a correlation matrix to a partial correlation vine based on a given R-vine structure and vice versa is explained in detail. In \autoref{Sec:DataSettingBenchmark}, we introduce the general data setting and motivate the choice of Cholesky decomposition based models as our main benchmarks.} In \autoref{Sec:PCVapproach}, we outline in detail the three main steps of the proposed partial correlation vine data transformation approach including R-vine structure selection in \autoref{Sec:R-VineStructureSelection} and multivariate time-series modeling in \autoref{Subsec:Step2TimeSeries}. Supported by the analysis of high-frequency data for six stocks listed on the NYSE, AMEX and NASDAQ beneficial properties of the proposed modeling strategy will be explored. In \autoref{Sec:EmpiricalStudy}, detailed investigation of the real data example will be continued. \textcolor{black}{\autoref{Sec:EmpStepS3} shows the excellent forecasting performance of the partial correlation vine data transformation approach both with respect to statistical precision and mean-variance balance in portfolio optimization.} This paper comes with extensive supplementary material.

\section{Partial correlation vines}\label{Sec:PartialCorrelationVines}
First, we provide necessary background on the two main ingredients of the proposed model approach -- partial correlations and regular vines.   

\subsection{Partial correlations}\label{subsec: PCors}

We consider a random vector $\boldsymbol{X}_{\mathcal{I}} \coloneqq \left(X_1,\ldots,X_d\right)$, $d \geq 2$, with zero mean, where $\mathcal{I}$ is the index set  $\lbrace 1, \ldots, d \rbrace$. We denote the $d \times d$ covariance matrix by $\boldsymbol{Y}$ and obtain the corresponding $d \times d$ correlation matrix $\boldsymbol{R}$ as
$\boldsymbol{R} = \boldsymbol{D}^{-1/2}\boldsymbol{Y}\boldsymbol{D}^{-1/2}$,
where $\boldsymbol{D} = \text{diag}\left(y_{1,1}, \ldots, y_{d,d}\right)$ is the diagonal matrix of variances. Further, we define a subset 
$L \subseteq \mathcal{I}$ having at least cardinality 2, i.e.\ $|L| \geq 2$. For a pair $i,j \in L, i \neq j,$ we denote $L$ with the subset $\lbrace i,j \rbrace$ removed by 
$D_{\{i,j\}} \coloneqq L_{-\{i,j\}} = L \backslash \lbrace i, j \rbrace$    
and the corresponding random vector by
$\boldsymbol{X}_{D_{\{i,j\}}} \coloneqq \lbrace X_k, k \in D_{\{i,j\}} \rbrace$.
The partial regression coefficients
$ b_{i,j;D_{\{i,j\}}}$
are defined as the quantities that minimize

\begin{align*}
\mathbb{E}\big[(X_i - \sum_{j \in L_{-\lbrace i\rbrace}}b_{i,j;D_{\{i,j\}}}X_j)^2\big]. 
\end{align*} 
The corresponding partial correlation coefficients $\rho_{i,j;D_{\{i,j\}}}$ quantify the dependence between $X_i$ and $X_j$ without the linear effect of $\boldsymbol{X}_{D_{\{i,j\}}}$ and are defined by \citep[p.\ 47]{kurowicka2011dependence}

\begin{align*}
\rho_{i,j;D_{\{i,j\}}} \coloneqq \textup{sgn}(b_{i,j;D_{\{i,j\}}})\left(b_{i,j;D_{\{i,j\}}}b_{j,i;D_{\{i,j\}}}\right)^{1/2}.
\end{align*}  
We refer to the cardinality of $D_{\{i,j\}}$ as order of the partial correlation coefficient. For order zero, i.e.\ $|L| = |\lbrace i,j \rbrace| = 2$ and thus $D_{\{i,j\}} = \emptyset$, we obtain pairwise standard correlations between $X_i$ and $X_j$, $i,j \in \mathcal{I}, i \neq j$. We write $\rho_{i,j;\emptyset} = \rho_{i,j}$. Now, consider for a subset $L \subseteq \mathcal{I}$ of at least cardinality 3, a set of distinct indices $\lbrace i,j,k\rbrace \subseteq L$, $i \neq j \neq k$. We define $\tilde{D} \coloneqq L_{-\{i,j,k\}}$ such that $D_{\{i,j\}} = \tilde{D} \cup k$. \cite{anderson1958introduction} derives a formula to recursively calculate the partial correlations of any order $|D_{\{i,j\}}|$ with $|D_{\{i,j\}}| \geq 1$ in terms of (partial) correlations of lower order. With $\rho_{i,k;\tilde{D}}^2 < 1$ and $\rho_{j,k;\tilde{D}}^2 < 1$ it holds that 

\begin{align}\label{eq:recursionCor2Pcor}
\rho_{i,j;D_{\{i,j\}}} = \frac{\rho_{i,j;\tilde{D}}-\rho_{i,k;\tilde{D}}\rho_{j,k;\tilde{D}}}{\sqrt{1-\rho_{i,k;\tilde{D}}^2}\sqrt{1-\rho_{j,k;\tilde{D}}^2}}.
\end{align}  
%Thus, for the pair $\left(i,j\right)$, $i,j \in \mathcal{I}$, $i \neq j$, partial correlations of any order %$p, 1 \leq p \leq d-2$, can be calculated from standard correlations through recursive application of \eqref{eq:recursionCor2Pcor}. 

\textcolor{black}{Since the evaluation of higher order partial correlations gets too involved when exclusively relying on this recursion formula, in practice typically a more efficient calculation procedure is used (see e.g.\ \cite{whittaker2009graphical}). Let $\boldsymbol{\Omega}$ be the submatrix of standard correlations with indices $L \subseteq \mathcal{I}$, i.e.\ $\boldsymbol{\Omega} = \left(\omega_{k,\ell}\right)_{k,\ell = 1,\ldots,|L|} = \left(\rho_{l_k l_\ell}\right)_{k,\ell=1,\ldots,|L|}$, where $l_k$ is the $k$-th element in $L$. Let $\boldsymbol{P}$ be its inverse, i.e.\ $\boldsymbol{P} = \boldsymbol{\Omega}^{-1} =\left(p_{k,\ell}\right)_{k,\ell = 1,\ldots,|L|}$. Then, it holds}

\textcolor{black}{\begin{align}\label{eq:PCorsMatrixInversion}
	\rho_{l_k,l_\ell;D_{\{l_k,l_\ell\}}} = - \frac{p_{k,\ell}}{\sqrt{p_{k,k}p_{\ell,\ell}}}.
	\end{align}
	Thus, through inversion of $\boldsymbol{\Omega}$ all partial correlations between $X_i$ and $X_j$ ($i,j \in L$, $i \neq j$) given all other variables $\boldsymbol{X}_{D_{\{i,j\}}}$ are simultaneously calculated. If interest is in a single partial correlation $\rho_{i,j;D_{\{i,j\}}}$ for  $i,j \in L$ with $i \neq j$ fixed, computing complexity can be reduced by assorting $\boldsymbol{\Omega}$ blockwise with indices $(i, j)$ and $D_{\{i,j\}}$, i.e.\ } 

\textcolor{black}{\begin{align*}
	\boldsymbol{\Omega}^{-1} = \left(\begin{matrix} \boldsymbol{\Omega}_{1,1} & \boldsymbol{\Omega}_{1,2} \\ 
	\boldsymbol{\Omega}_{2,1} & \boldsymbol{\Omega}_{2,2}\\
	\end{matrix} \right)^{-1} = \boldsymbol{P} = \left(\begin{matrix} \boldsymbol{P}_{1,1} & \boldsymbol{P}_{1,2} \\ 
	\boldsymbol{P}_{2,1} & \boldsymbol{P}_{2,2}\\
	\end{matrix} \right),
	\end{align*} 
	where $\boldsymbol{\Omega}_{1,1}$ is a $2 \times 2$ matrix with elements $\omega_{1,1} = \omega_{2,2} = 1$, $\omega_{1,2} = \omega_{2,1} = \rho_{i,j}$ and $\boldsymbol{P}_{1,1}$ its counterpart with elements $p_{1,1}$, $p_{1,2} = p_{2,1}$, $p_{2,2}$. Using standard results for block matrix inversion (see e.g.\ \cite{bernstein2005matrix}) we have $\boldsymbol{P}_{1,1}^{-1} = \boldsymbol{\Omega}_{1,1} - \boldsymbol{\Omega}_{1,2}\boldsymbol{\Omega}_{2,2}^{-1}\boldsymbol{\Omega}_{2,1}$ with elements $\tilde{p}_{1,1}$, $\tilde{p}_{1,2} = \tilde{p}_{2,1}$, $\tilde{p}_{2,2}$. We conclude that}

\textcolor{black}{\begin{align}\label{eq:PCorsSINGLEMatrixInversion}
	\rho_{i,j;D_{\{i,j\}}} \stackrel{\eqref{eq:PCorsMatrixInversion}}{=} - \frac{p_{1,2}}{\sqrt{p_{1,1}p_{2,2}}} = - \frac{- \frac{1}{\text{det}\boldsymbol{P}_{1,1}}\tilde{p}_{1,2}}{\sqrt{\frac{1}{\text{det}\boldsymbol{P}_{1,1}}\tilde{p}_{1,1} \frac{1}{\text{det}\boldsymbol{P}_{1,1}}\tilde{p}_{2,2}}} =
	\frac{\tilde{p}_{1,2}}{\sqrt{\tilde{p}_{1,1}\tilde{p}_{2,2}}} .
	\end{align}}

From now on, we refer to $\mathcal{C}_d$ as the set of all standard correlations and to $\mathcal{C}_d^{\text{p}}$ as the set of all pairwise standard and partial correlations. \textcolor{black}{The $1 \times \binom{d}{2}$ vector $\boldsymbol{P}_{\mathcal{C}_d}$ and the  $1 \times \binom{d}{2}2^{d-2}$ vector $\boldsymbol{P}_{\mathcal{C}_d^\textup{p}}$ record all standard correlations and all standard and partial correlations, respectively, of the random vector $\boldsymbol{X}_{\mathcal{I}}$ in lexicographical order with increasing subset $L \subseteq \mathcal{I}$, i.e.\ } 

\begin{align*}
\boldsymbol{P}_{\mathcal{C}_d} \coloneqq &\left(\rho_{1,2}, \ldots, \rho_{1,d},\rho_{2,3},\ldots,\rho_{2,d},\ldots, \rho_{(d-1),d}\right)\\   
\text{and} \hspace*{.5cm}& \\
\boldsymbol{P}_{\mathcal{C}_d^\textup{p}} \coloneqq & \left(\boldsymbol{P}_{\mathcal{C}_d},\right.\\
&\ \ \rho_{1,2;3},\ldots,\rho_{1,2;d}, \rho_{1,3;2}, \ldots,\rho_{1,d;(d-1)}, \rho_{2,3;1},\ldots,\rho_{(d-1),d;(d-2)},\\
%&\ \ \rho_{1,2;3,4},\ldots, \rho_{1,2;3,d}, \rho_{1,2;4,5},\ldots, \rho_{1,2;4,d}, \ldots, \rho_{1,d;(d-2),(d-1)},\rho_{2,3;1,4},\ldots,\rho_{(d-1),d;(d-3),(d-2)},\\
& \ \ ...,\\ 
&\ \left. \rho_{1,2;3,\ldots,d}, \ldots, \rho_{1,d;2,\ldots,(d-1)},\ldots,\rho_{(d-1),d;1,\ldots,(d-2)}\right).
\end{align*} 
\noindent
\textcolor{black}{To conclude and as illustrated in \autoref{fig:FlowchartData1}, from a $d \times d$ covariance matrix $\boldsymbol{Y}$ the $1 \times d$ vector of variances $\boldsymbol{y}$ and the $d \times d$ correlation matrix $\boldsymbol{R}$ can be obtained. The latter fully determines the vector $\boldsymbol{P}_{\mathcal{C}_d^\text{p}}$, which takes values in $\left(-1,1\right)$ and collects all $\binom{d}{2}2^{d-2}$ standard and partial correlations. In the following, we will show that the other way round the correlation matrix $\boldsymbol{R}$ can be uniquely determined from only a few elements of $\boldsymbol{P}_{\mathcal{C}_d^\text{p}}$, which are defined through a regular vine.}

\tikzstyle{ClassicalNode} = [rectangle, fill = lightgray!43, draw = black, text = black, align = center]
\tikzstyle{TreeLabels} = [draw = none, fill = none, text = black, font = \bf]
\tikzstyle{DummyNode}  = [draw = none, fill = none, text = white]
\tikzset{
	block/.style = {rectangle, draw,  text width=9.0cm,
		line width=.5pt,minimum height=1.75cm},
	block_half/.style = {block, text width=5cm},
}
\newcommand{\yshift}{-1cm}
\newcommand{\yshiftlabel}{-0.1cm}
\newcommand{\labelsize}{\small}
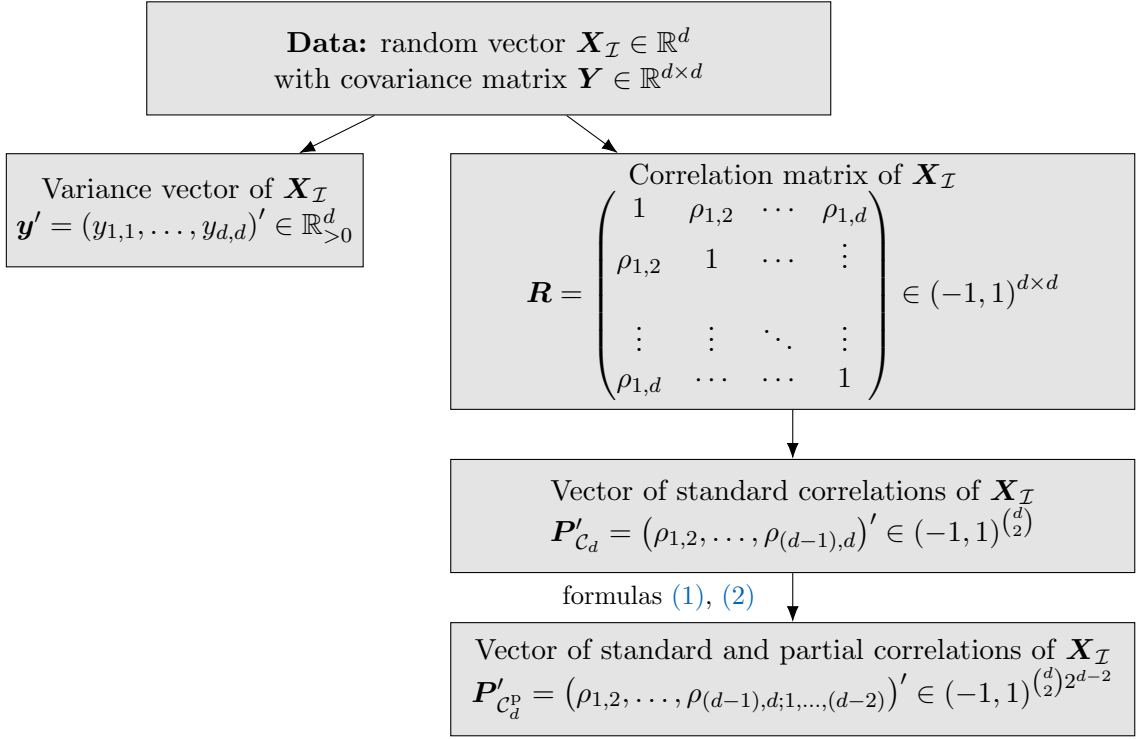
\begin{figure}[h!]
	\centering
	\begin{tikzpicture}	[every node/.style = ClassicalNode, node distance = .5cm, font = \normalsize, minimum height=1.5cm, minimum width=9cm]
	\node (Data) {
		\textbf{Data:} random vector $\boldsymbol{X}_{\mathcal{I}} \in \mathbb{R}^{d}$\\ with
		covariance matrix $\boldsymbol{Y}\in\mathbb{R}^{d \times d}$
	};
	\node [below = of Data, xshift = -4cm, minimum width=0cm] (vectorVars) {
		Variance vector of  $\boldsymbol{X}_{\mathcal{I}}$\\ $\boldsymbol{y}'  = \left(y_{1,1}, \ldots, y_{d,d}\right)' \in \mathbb{R}_{>0}^{d}$};
	\node [below = of Data, xshift = 4cm] (CorMatrix) {Correlation matrix of  $\boldsymbol{X}_{\mathcal{I}}$\\ $\boldsymbol{R}  = \left(\begin{matrix} 1 & \rho_{1,2} & \cdots & \rho_{1,d} \\\bigskip  
		\rho_{1,2} & 1 & \cdots & \vdots \\
		\vdots & \vdots & \ddots & \vdots \\
		\rho_{1,d} & \cdots & \cdots & 1 \\
		\end{matrix} \right) \in \left(-1,1\right)^{d \times d}$};		 
	\node[below =  of CorMatrix, yshift = -.15cm] (vecStandardCors)  {Vector of standard correlations of $\boldsymbol{X}_\mathcal{I}$\\
		$\boldsymbol{P}'_{\mathcal{C}_d} = \left(\rho_{1,2},\ldots,\rho_{\left(d-1\right),d}\right)' \in \left(-1,1\right)^{\binom{d}{2}}$};	
	\node[below = of vecStandardCors, yshift = -.15cm] (vecCorsPCors)  {Vector of standard and partial correlations of $\boldsymbol{X}_\mathcal{I}$\\
		$\boldsymbol{P}'_{\mathcal{C}_d^\text{p}}  = \left(\rho_{1,2},\ldots,\rho_{\left(d-1\right),d;1,\ldots,\left(d-2\right)}\right)' \in \left(-1,1\right)^{\binom{d}{2}2^{d-2}}$};
	
	\draw[-{Latex[scale=1.25]}] (Data) to node[draw=none, fill = none] {} (vectorVars);
	\draw[-{Latex[scale=1.25]}] (Data) to node[draw=none, fill = none] {} (CorMatrix);
	\draw[-{Latex[scale=1.25]}] (CorMatrix) to node[draw=none, fill = none] {} (vecStandardCors);
	\draw[-{Latex[scale=1.25]}] (vecStandardCors) to node[draw=none, fill = none, font = \labelsize, left, xshift = +2.75cm] {formulas \eqref{eq:recursionCor2Pcor}, \eqref{eq:PCorsMatrixInversion}} (vecCorsPCors);
	\end{tikzpicture}
	\caption{Data prespecified by a given covariance matrix $\boldsymbol{Y}$.}
	\label{fig:FlowchartData1}
	%\vspace*{-1cm}
\end{figure}

\subsection{Regular vines} 
According to \cite{bedford2002vines}, a regular vine (R-vine) on $d$ elements is a set of $d-1$ linked trees, i.e.\ undirected and acyclic graphs, $\mathcal{V}_d\coloneqq \left(\mathcal{T}_1,\dots,\mathcal{T}_{d-1}\right)$ with the set of edges $E\left(\mathcal{V}_d\right) \coloneqq E_1 \cup \cdots \cup E_{d-1}$ and the set of nodes $N\left(\mathcal{V}_d\right) \coloneqq N_1 \cup \cdots \cup N_{d-1}$ such that \begin{enumerate}
	\item[(i)] $\mathcal{T}_1$ is a tree with nodes $N_1=\lbrace 1,\dots,d \rbrace$ and edges $E_1$,
	\item[(ii)] for $\ell=2,\dots,d-1$, $\mathcal{T}_\ell$ is a tree with nodes $N_\ell=E_{\ell-1}$ and edges $E_\ell$,
	\item[(iii)] the \textit{proximity condition} holds: For $\ell=2,\dots,d-1$, whenever two nodes of $\mathcal{T}_{\ell}$ are connected by an edge, the corresponding edges of $\mathcal{T}_{\ell-1}$ share a node.
\end{enumerate}	   
According to property (ii), the $d-(\ell-1)$ edges $E_{\ell-1}$ in $\mathcal{T}_{\ell-1}$ become nodes in $\mathcal{T}_\ell$. Based on this linkage,  each sequence of trees of an R-vine -- from now on referred to as R-vine structure -- allows to identify a set of $\binom{d}{2}$ (conditional) bivariate constraints. We refer to \cite{kurowicka2003parameterization} and consider an arbitrary edge $e=\lbrace a,b \rbrace \in E_\ell$ of $\mathcal{V}_d$, $2 \leq \ell \leq d-1$, with $a,b \in N_\ell$. Its \textit{complete union} $U_e^{*}$ is the subset of nodes in $\mathcal{T}_1$, i.e.\ the subset of $\lbrace 1,\dots,d \rbrace$, reachable from $e$ by the membership relation, i.e.\

\begin{align*}
U_e^{*} \coloneqq \lbrace n \in N_1: \exists e_1 \in E_1,\dots,e_{\ell-1} \in E_{\ell-1}: n \in e_1 \in \dots \in e_{\ell-1} \in e  \rbrace.
\end{align*} 
The \textit{conditioning set $D_e$} corresponding to $e = \lbrace a, b \rbrace$ is the intersection of the complete unions $U_a^{*}$ and $U_b^{*}$ corresponding to the edges $a, b \in E_{\ell-1}$, i.e.\

\begin{align*}
D_e \coloneqq U_a^{*} \cap U_b^{*}.
\end{align*} 
The corresponding symmetric difference is referred to as \textit{conditioned set} 

\begin{align*}
\lbrace C_{e,a}, C_{e,b} \rbrace \coloneqq \lbrace U_a^{*}\backslash D_e, U_b^{*}\backslash D_e \rbrace.
\end{align*} 
By definition, each conditioned set in $\mathcal{V}_d$ consists of two single elements and forms a unique pair of variables $\{i,j\}$ with $i,j \in \{1,\ldots,d\}$, $i \neq j$. \textcolor{black}{Thus, each pair is modeled by $\mathcal{V}_d$ exactly once either unconditioned, if it appears in the first tree level, or via conditioning if it appears in tree levels $\ell= 2,\ldots,d-1$. If $\mathcal{V}_d$ specifies in each tree level one central node being attached to all edges, we speak of a C-vine. The latter reflects an ordering by importance.}

\begin{bsp}\label{Ex:RVineStructure} 
	\autoref{Fig:6D_RVine_Example} shows an R-vine structure on six elements labeled with the conditioned set and the conditioning set corresponding to each edge. The latter is indicated by a leading ``$\vert$''. The bold tree segment in $\mathcal{T}_2$ corresponds to the edge $e = \lbrace \lbrace 1, 2 \rbrace, \lbrace 2, 6 \rbrace \rbrace$. Reachable from edge $\{1,2\}\in \mathcal{T}_1$ and $\{2,6\}\in \mathcal{T}_1$ are the nodes $1, 2 \in N_1$ and $2,6 \in N_1$, respectively. Thus,
	%\begin{align*}
	$D_{e} = U_{\lbrace 1,2\rbrace}^{*} \cap U_{\lbrace 2,6\rbrace}^{*} = \lbrace 1, 2 \rbrace \cap \lbrace 2, 6 \rbrace = \lbrace 2 \rbrace$
	%\end{align*}
	is the conditioning set corresponding to $e$ and the conditioned set is
	%	\begin{align*}
	$\lbrace C_{e,\lbrace 1, 2 \rbrace}, C_{e,\lbrace 2, 6 \rbrace} \rbrace = \lbrace \lbrace 1, 2\rbrace \backslash \lbrace 2 \rbrace, \lbrace 2, 6\rbrace \backslash \lbrace 2 \rbrace \rbrace = \lbrace 1, 6 \rbrace$.
	%	\end{align*}
	%	The highlighted edge in $\mathcal{T}_2$ therefore identifies the conditional bivariate constraint ``$1,6\vert 2$''. 
	
\end{bsp} 
\tikzstyle{ClassicalVineNode} = [rectangle, fill = lightgray!43, draw = black, text = black, align = center, minimum height = 0.75cm, minimum width = 0.75cm]
\tikzstyle{TreeLabels} = [draw = none, fill = none, text = black, font = \bfseries]
\tikzstyle{DummyNode}  = [draw = none, fill = none, text = white]
\newcommand{\yshiftTreeLabel}{.8cm}
\newcommand{\labelsizeRVine}{\footnotesize}
\newcommand{\shiftlabel}{.075cm}  
\newcommand{\dummyxShift}{.45cm}
\begin{figure}[h!]
	%\vspace*{1.5cm}
	\centering	
	%\begin{minipage}{.5\linewidth}						
	\begin{tikzpicture}	[every node/.style = ClassicalVineNode, node distance = 1.55cm, font = \small]
	
	%%%%% Tree 1 %%%%%
	\node[very thick] (6){\textbf{6}}
	node[very thick]             (2)         [below of = 6] {\textbf{2}}			
	node             (3)         [below of = 2] {3}
	node             (5)         [below of = 3] {5}
	node[very thick]             (1)         [above right of = 2] {\textbf{1}}
	node             (4)         [below right of = 2] {4}
	node[TreeLabels] (T1)        [above  of = 6, yshift = -\yshiftTreeLabel] {$\mathcal{T}_1$}
	;
	
	\draw[very thick] (1) to node[draw=none, text = black, fill = none, font = \labelsizeRVine, right, xshift = -\shiftlabel, yshift = -\shiftlabel] {\textbf{1,2}} (2);
	\draw (2) to node[draw=none, text = black, fill = none, font = \labelsizeRVine, left, xshift = \shiftlabel] {2,3} (3);
	\draw (2) to node[draw=none, text = black, fill = none, font = \labelsizeRVine, right, xshift = -\shiftlabel] {\hspace{.05cm}2,4} (4);
	\draw[very thick] (2) to node[draw=none, text = black, fill = none, font = \labelsizeRVine, left, xshift = \shiftlabel] {\textbf{2,6}} (6);
	\draw (3) to node[draw=none, text = black, fill = none, font = \labelsizeRVine, left, xshift = \shiftlabel] {3,5} (5); 
	
	%%%%%% Tree 2 %%%%%
	\node[DummyNode](Dummy1)     [right of = T1, xshift = -\dummyxShift] {}
	node[TreeLabels] (T2)        [right  of = Dummy1] {$\mathcal{T}_2$}
	node[very thick]           (26)         [below of = T2, yshift = \yshiftTreeLabel]{\textbf{2,6}}         
	node[very thick]           (12)         [below of = 26] {\textbf{1,2}}
	node           (23)         [below of = 12] {2,3}
	node             (35)       [below of = 23] {3,5}
	node           (24)         [below right of = 23] {2,4}
	;
	
	\draw[very thick] (26) to node[draw=none, text = black, fill = none, font = \labelsizeRVine, left, xshift = 1.25*\shiftlabel] {$\boldsymbol{1,6|2}$} (12);   
	\draw (12) to node[draw=none, text = black, fill = none, font = \labelsizeRVine, left, xshift = 1.25*\shiftlabel] {$1,3|2$} (23) ;    
	\draw(23) to node[draw=none, text = black, fill = none, font = \labelsizeRVine, right, xshift = -\shiftlabel] {$3,4|2$} (24) ;   
	\draw (23) to node[draw=none, text = black, fill = none, font = \labelsizeRVine, left, xshift = 1.25*\shiftlabel] {$2,5|3$} (35) ;  
	
	%%%%%% Tree 3 %%%%%
	\node[DummyNode](Dummy2)     [right of = T2, xshift = -\dummyxShift] {}
	node[TreeLabels] (T3)        [right  of = Dummy2] {$\mathcal{T}_3$}
	node            (162)         [below of = T3, yshift = \yshiftTreeLabel] {$1,6|2$}         
	node             (132)        [below of = 162] {$1,3|2$}
	node             (342)        [below of = 132] {$3,4|2$}
	node             (253)        [below of = 342] {$2,5|3$}
	;
	
	\draw (162) to node[draw=none, text = black, fill = none, font = \labelsizeRVine, left, xshift = 1.5*\shiftlabel] {$3,6|1,2$} (132);   
	\draw (132) to node[draw=none, text = black, fill = none, font = \labelsizeRVine, left, xshift = 1.5*\shiftlabel]  {$1,4|2,3$} (342);    
	\draw (342) to node[draw=none, text = black, fill = none, font = \labelsizeRVine, left, xshift = 1.5*\shiftlabel] {$4,5|2,3$} (253);    																	    							
	%%%%%% Tree 4 %%%%%
	\node[TreeLabels] (T4)        [right  of = T3, xshift = \dummyxShift] {$\mathcal{T}_4$}
	node              (3612)        [below of = T4, yshift = \yshiftTreeLabel] {$3,6|1,2$}         
	node              (1423)        [below of = 3612] {$1,4|2,3$}
	node              (4523)        [below of = 1423] {$4,5|2,3$}
	;
	
	\draw (3612) to node[draw=none, text = black, fill = none, font = \labelsizeRVine, left, xshift = 1.75*\shiftlabel] {$4,6|1,2,3$} (1423);   
	\draw (1423) to node[draw=none, text = black, fill = none, font = \labelsizeRVine, left, xshift = 1.75*\shiftlabel]  {$1,5|2,3,4$} (4523);    								
	
	%%%%%% Tree 5 %%%%%
	\node[TreeLabels] (T5)        [right  of =T4, xshift = 1.75*\dummyxShift] {$\mathcal{T}_5$}
	node           (46123)        [below of = T5, yshift = \yshiftTreeLabel] {$4,6|1,2,3$}         
	node            (15234)       [below of = 46123] {$1,5|2,3,4$}
	;
	
	\draw (46123) to node[draw=none, text = black, fill = none, font = \labelsizeRVine, left, xshift = 2*\shiftlabel] {$5,6|1,2,3,4$} (15234);  																    					 
	\end{tikzpicture}
	%\end{minipage}
	%\hfill
	%\begin{minipage}{.20\linewidth}
	%	\begin{align*}	
	%	\left(\begin{matrix} 
	%		6 &  & & & & \\
	%		5 & 1 & & & & \\
	%		4 & 5 & 4 & & & \\
	%		3 & 4 & 5 & 2 & & \\
	%		1 & 3 & 3 & 5 & 3  & \\
	%		2 & 2 & 2 & 3 & 5 & 5 \\
	%	\end{matrix} \right)
	%\end{align*}
	%\end{minipage}
	\caption{Example of a 6-dimensional R-vine structure with conditioning and conditioned sets corresponding to each edge.}% on the left and R-vine structure matrix $M$ on the right.}
	\label{Fig:6D_RVine_Example}						
\end{figure}
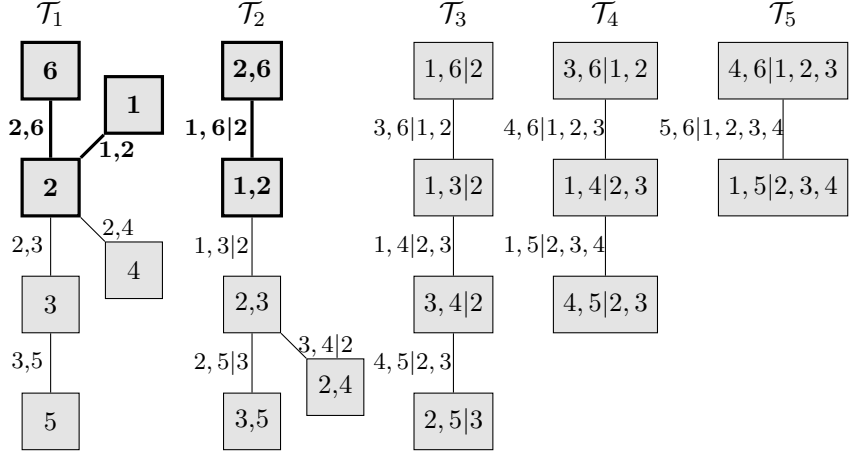

\subsection{Partial correlation vines}
Now, building the bridge between partial correlations determined by a $d \times d$ correlation matrix and an R-vine structure $\mathcal{V}_{d}$ is straightforward: in a partial correlation vine with R-vine structure $\mathcal{V}_{d}$ each edge $e=\lbrace a, b \rbrace \in E\left(\mathcal{V}_{d}\right)$ is identified with the partial correlation coefficient $\rho_{C_{e,a},C_{e,b};D_e}$ that coincides with the conditioned and conditioning set specified by $e$. Thus, to each edge in $\mathcal{V}_{d}$ a value in $(-1,1)$ is assigned. We define the set of the $\binom{d}{2}$ standard and partial correlations specified by $\mathcal{V}_{d}$ as $\mathcal{C}\left(\mathcal{V}_{d}\right)$ and denote by  $\boldsymbol{P}_{\mathcal{C}\left(\mathcal{V}_{d}\right)}$ the $1 \times \binom{d}{2}$ vector that collects the corresponding values specified by the correlation matrix $\boldsymbol{R}$ in lexicographical order.

\cite{bedford2002vines} provide the fundamental result that for any R-vine structure $\mathcal{V}_{d}$ there is a one-to-one relationship between the set of $d \times d$ positive definite correlation matrices and its set $\mathcal{C}\left(\mathcal{V}_{d}\right)$, i.e.\ for each R-vine structure $\mathcal{V}_{d}$ there exists a bijection

\begin{align}\label{eq:bijectionRVinePCor2Cor}
F_{\text{Cor2PCor}}: \left(-1, 1\right)^{\binom{d}{2}} \to \left(-1, 1\right)^{\binom{d}{2}}, \  F_{\text{Cor2PCor}}\left(\boldsymbol{P}_{\mathcal{C}_d}\right) = \boldsymbol{P}_{\mathcal{C}\left(\mathcal{V}_{d}\right)}.
\end{align}
\textcolor{black}{In particular, according to \cite{kurowicka2003parameterization} the elements in $\boldsymbol{P}_{\mathcal{C}\left(\mathcal{V}_{d}\right)}$ are algebraically independent, i.e.\ for any arbitrary assignment of values in $\left(-1,1\right)$ to the edges of R-vine structure $\mathcal{V}_{d}$ the correlation matrix calculated from $\boldsymbol{P}_{\mathcal{C}\left(\mathcal{V}_{d}\right)}$ using the `inverse' of \eqref{eq:bijectionRVinePCor2Cor} is positive definite with correlation values in $\left(-1,1\right)$ for all off-diagonal elements. An efficient implementation of the bijection $F_{\text{Cor2PCor}}$ and its `inverse' is available in the R-package \texttt{VineCopula} \citep{schepsmeier2014vinecopula}. Pseudo-code is provided in  \cite{joe2014dependence}. Note that while in the derivation of \eqref{eq:PCorsSINGLEMatrixInversion} and in the following explanations we assume the submatrix of standard correlations $\boldsymbol{\Omega}$ to be assorted blockwise with indices $(i,j)$ and $D_{\{i,j\}}$, \cite{joe2014dependence} assorts the indices using the order $D_{\{i,j\}}$ and $(i,j)$.}

\setcounter{bsp}{0}
\begin{bsp}[continued] \textcolor{black}{We illustrate the data transformation based on R-vine structure $\mathcal{V}_6$ in \autoref{Fig:6D_RVine_Example}. As illustrated below, each standard correlation in $\boldsymbol{R}$ on the left-hand side is specified in the partial correlation vine corresponding to $\mathcal{V}_6$ through a (partial) correlation of order $\ell - 1$ modeled in tree $\mathcal{T}_{\ell}$ ($\ell=1,\ldots,5$)
		, i.e.\ }
	
	\textcolor{black}{\begin{align*}\small
		\boldsymbol{R}  = \left(\begin{array}{c c c c c c}  & \rho_{1,2} & \rho_{1,3} & \rho_{1,4} & \rho_{1,5} & \rho_{1,6}\\ 
		&  & \rho_{2,3} & \rho_{2,4} & \rho_{2,5} & \rho_{2,6}\\
		&  &  & \rho_{3,4} & \rho_{3,5} & \rho_{3,6}\\
		&  &  &  & \rho_{4,5} & \rho_{4,6}\\
		&  &  &  &  & \rho_{5,6}\\
		&  &  &  &  & \\
		\end{array} \right) \rightleftarrows \left(\begin{array}{c c c c c c}  & \rho_{1,2} & \rho_{1,3;2} & \rho_{1,4;2,3} & \rho_{1,5;2,3,4} & \rho_{1,6;2}\\ 
		&  & \rho_{2,3} & \rho_{2,4} & \rho_{2,5;3} & \rho_{2,6}\\
		&  &  & \rho_{3,4;2} & \rho_{3,5} & \rho_{3,6;1,2}\\
		&  &  &  & \rho_{4,5;2,3} & \rho_{4,6;1,2,3}\\
		&  &  &  &  & \rho_{5,6;1,2,3,4}\\
		&  &  &  &  & \\
		\end{array} \right)
		\end{align*}}
	
	%	\renewcommand{\arraystretch}{0.65}
	%	\begin{tabular}{p{1.25cm}p{.0025cm}p{0.75cm}p{.0025cm}p{.75cm}p{.0025cm}p{0.75cm}p{.0025cm}p{0.75cm}p{.0025cm}p{0.75cm}}
	%		& \cellcolor{TUMblue} & {\small $\mathcal{T}_1$} & \cellcolor{TUMgreen} & {\small $\mathcal{T}_2$} & \cellcolor{TUMorange} & {\small $\mathcal{T}_3$} & \cellcolor{lightgray} & {\small $\mathcal{T}_4$} & \cellcolor{lightgray!25} & {\small $\mathcal{T}_5$}
	%		\\\vspace*{.25cm}
	%	\end{tabular}
	%	\renewcommand{\arraystretch}{1.0}
	
	\noindent
	\textcolor{black}{\textbf{Transformation of correlation matrix}. First, we derive from $\boldsymbol{R}$ the partial correlations corresponding to  $\mathcal{V}_6$. %, which are also listed in \autoref{Fig:6D_RVine_Example}. 
		Thus, proceeding in the illustration of the above matrices is from left to right. While the standard correlations in $\mathcal{T}_1$ can simply be taken from the correlation matrix $\boldsymbol{R}$, the first order partial correlations in $\mathcal{T}_2$ can be calculated using recursion formula \eqref{eq:recursionCor2Pcor}, e.g.\ } 
	
	\textcolor{black}{\begin{align*}
		\rho_{1,6;2} = \frac{\rho_{1,6} - \rho_{1,2}\rho_{2,6}}{\sqrt{1-\rho_{1,2}^2}\sqrt{1-\rho_{2,6}^2}}.
		\end{align*}}
	
	\textcolor{black}{From tree level $\ell = 3$ on, we rely on formula \eqref{eq:PCorsSINGLEMatrixInversion} and elementwise calculate the partial correlations specified by $\mathcal{T}_3$ to $\mathcal{T}_5$. For example, for $\rho_{3,6;1,2}$ we set} 
	
	\textcolor{black}{\begin{align*}\small
		\boldsymbol{\Omega}_{1,1} = \left(\begin{matrix} 1 & \rho_{3,6} \\ 
		\\ \rho_{3,6} & 1
		\end{matrix} \right), \ \boldsymbol{\Omega}_{1,2} = \left(\begin{matrix} \rho_{1,3} & \rho_{2,3} \\ 
		\\ \rho_{1,6} & \rho_{2,6}
		\end{matrix} \right), \ \boldsymbol{\Omega}_{2,1} = \left(\begin{matrix} \rho_{1,3} & \rho_{1,6} \\ 
		\\ \rho_{2,3} & \rho_{2,6}
		\end{matrix} \right) \text{\normalsize \ and \ } \boldsymbol{\Omega}_{2,2} = \left(\begin{matrix} 1 & \rho_{1,2} \\ 
		\\ \rho_{1,2} & 1
		\end{matrix} \right)
		\end{align*}\normalsize
		and evaluate $\left(\begin{matrix} \tilde{p}_{1,1} & \tilde{p}_{1,2} \\ 
		\\ \tilde{p}_{1,2} & \tilde{p}_{2,2}
		\end{matrix} \right) = \boldsymbol{\Omega}_{1,1} - \boldsymbol{\Omega}_{1,2}\boldsymbol{\Omega}_{2,2}^{-1}\boldsymbol{\Omega}_{2,1}$. Then, we calculate}
	
	\textcolor{black}{\begin{align*}
		\rho_{3,6;1,2} \stackrel{\eqref{eq:PCorsSINGLEMatrixInversion}}{=}
		\frac{\tilde{p}_{1,2}}{\sqrt{\tilde{p}_{1,1}\tilde{p}_{2,2}}}.
		\end{align*}\noindent
		\textbf{Back-transformation to correlation matrix}. Now, proceeding in the illustration of the above matrices is from right to left. We proceed treewise. The standard correlations from $\mathcal{T}_1$ can directly be taken. To calculate the standard correlations that correspond to the conditioned sets of the first order partial correlations available in $\mathcal{T}_2$, we use recursion formula \eqref{eq:recursionCor2Pcor}, e.g.\ } 
	
	\textcolor{black}{\begin{align*}
		\rho_{1,6} = \rho_{1,6;2}\sqrt{1-\rho_{1,2}^2}\sqrt{1-\rho_{2,6}^2} + \rho_{1,2}\rho_{2,6}.
		\end{align*}
		Note that due to the proximity condition of an R-vine structure all standard correlations needed for this evaluation are available from the previous step.}	 
	
	\textcolor{black}{From tree level $\ell = 3$ on, we rely on formula \eqref{eq:PCorsSINGLEMatrixInversion} to calculate the standard correlations that correspond to the conditioned sets of the partial correlations specified in $\mathcal{T}_3$ to $\mathcal{T}_5$. For example, to obtain $\rho_{3,6}$ we set $\boldsymbol{\Omega}_{1,2}$, $\boldsymbol{\Omega}_{2,2}$ and $\boldsymbol{\Omega}_{2,1}$ as above. Due to the proximity condition all standard correlations to do so are available from previous steps $\ell = 1,2$. We calculate}
	
	\textcolor{black}{\begin{align*}
		\left(\begin{matrix} q_{1,1} & q_{1,2} \\ 
		\\ q_{1,2} & q_{2,2}
		\end{matrix} \right) =  \boldsymbol{\Omega}_{1,2}\boldsymbol{\Omega}_{2,2}^{-1}\boldsymbol{\Omega}_{2,1}
		\end{align*} such that $\tilde{p}_{1,1} = 1- q_{1,1}$, $\tilde{p}_{1,2} = \tilde{p}_{2,1} = \rho_{3,6} - q_{1,2}$, $\tilde{p}_{2,2} = 1- q_{2,2}$ and obtain}
	
	\textcolor{black}{\begin{align*}
		\rho_{3,6} \stackrel{\eqref{eq:PCorsSINGLEMatrixInversion}}{=} \rho_{3,6;1,2}\sqrt{(1-q_{1,1})(1-q_{2,2})} + q_{1,2}.
		\end{align*}}
\end{bsp}    
To conclude, the set of all standard correlations $\mathcal{C}_d$ can be determined from any set $\mathcal{C}\left(\mathcal{V}_{d}\right)$ specified by the partial correlation vine with R-vine structure $\mathcal{V}_{d}$. In particular, positive definiteness of the correlation matrix is always guaranteed. \autoref{fig:FlowchartData2} provides a summary overview of the relationships between the sets $\mathcal{C}_d$, $\mathcal{C}_d^{\text{p}}$ and $\mathcal{C}\left(\mathcal{V}_{d}\right)$.

\tikzstyle{ClassicalNode} = [rectangle, fill = lightgray!43, draw = black, text = black, align = center]
\tikzstyle{TreeLabels} = [draw = none, fill = none, text = black, font = \bf]
\tikzstyle{DummyNode}  = [draw = none, fill = none, text = white]
\tikzset{
	block/.style = {rectangle, draw, text width=5cm,
		line width=.5pt,minimum height=1.37cm},
	block_half/.style = {block, text width=1.6cm, draw = white},
	connector/.style={
		-latex,
		font=\scriptsize
	},
	rectangle connector/.style={
		connector,
		to path={(\tikztostart) -- ++(#1,0pt) \tikztonodes |- (\tikztotarget) },
		pos=0.5
	},
	rectangle connector/.default=-2cm,
}
\renewcommand{\yshift}{-1cm}
\renewcommand{\yshiftlabel}{-0.1cm}
\renewcommand{\labelsize}{\small}
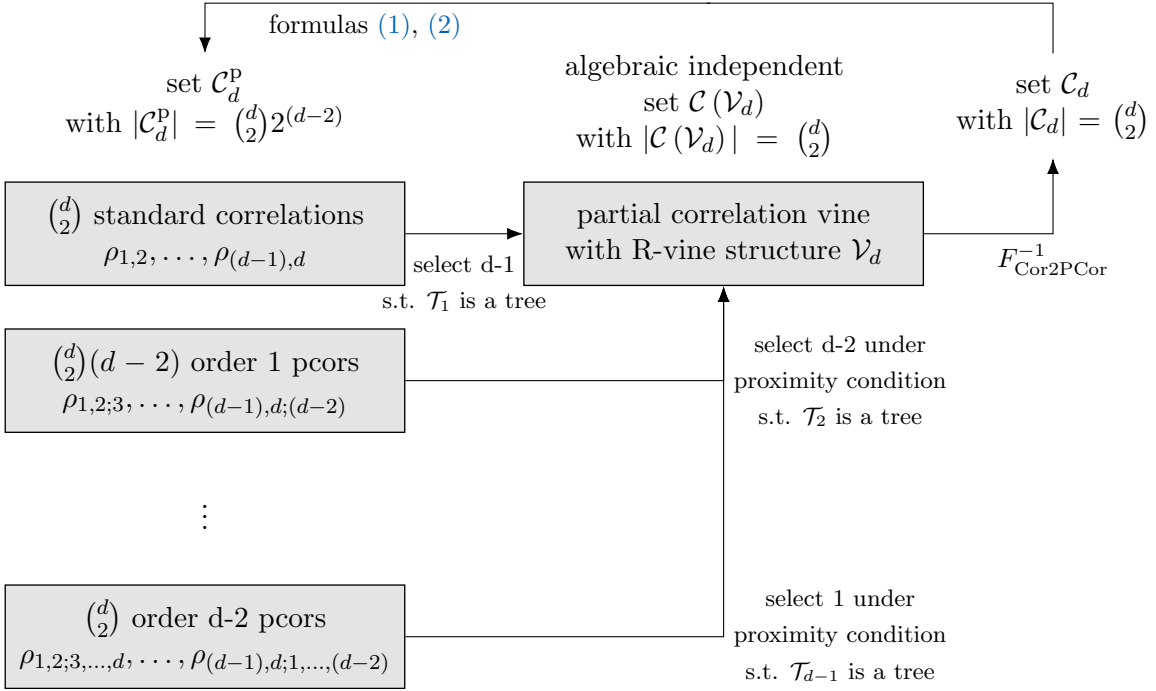
\begin{figure}[h!]
	\centering
	\begin{tikzpicture}	[every node/.style = ClassicalNode, node distance = .31cm, font = \normalsize]
	\node[block, anchor=north east, draw = white, fill = white] (setPCors) {set $\mathcal{C}_d^{\text{p}}$\\ with $|\mathcal{C}_d^\text{p}| = \binom{d}{2}2^{(d-2)}$};
	\node [block, below =  of setPCors] (order0Values) {
		$\binom{d}{2}$ standard correlations\\
		$\rho_{1,2}, \ldots, \rho_{(d-1),d}$};
	%	\node [block_half, left = -.05cm of order0Values] (order0) {order 0};	
	\node [block, below = of order0Values, yshift = -.25cm] (order1Values) {
		$\binom{d}{2}(d-2)$ order 1 pcors \\
		$\rho_{1,2;3}, \ldots, \rho_{(d-1),d;(d-2)}$};
	%	\node [block_half, left = of order1Values] (order1) {order 1};
	%	\node [block, below = of order1Values] (order2Values) {
	%		$\binom{d}{2}\binom{d-2}{2}$ partial correlations\\
	%		$\rho_{1,2;3,4}, \ldots, \rho_{(d-1),d;(d-3),(d-2)}$};
	%	\node [block_half, left = of order2Values] (order2) {order 2};
	\node [block_half,below = of order1Values, fill = white] (dots1) {$\vdots$};
	%	\node [block_half,below = of order1] (dots2) {$\vdots$};	
	\node [block, below = of dots1] (orderLastValues) {
		$\binom{d}{2}$ order d-2 pcors\\
		$\rho_{1,2;3,\ldots,d}, \ldots, \rho_{(d-1),d;1,\ldots,(d-2)}$};
	%	\node [block_half, left = of orderLastValues] (order d-2) {order d-2};
	
	\node [block, right = 1.3 cm of setPCors, draw = white, fill = white] (RVineStructure) {algebraic independent\\ set $\mathcal{C}\left(\mathcal{V}_{d}\right)$\\ with $|\mathcal{C}\left(\mathcal{V}_{d}\right)| = \binom{d}{2}$};
	\node [block, right = 1.55 cm of order0Values] (PCorVine) {partial correlation vine\\ with R-vine structure $\mathcal{V}_{d}$};
	
	\node [block, right = of RVineStructure, draw = white, text width=3.0cm, fill = white] (StandardCors) {set $\mathcal{C}_d$\\ with $|\mathcal{C}_d| = \binom{d}{2}$};
	
	\node [block, above = -.15cm of RVineStructure, draw = white, fill = white, text width=-2cm, text height=-2cm] (dummy1) {};
	\node [block, right = .05cm of dummy1, draw = white, fill = white, text width=-2cm, text height=-2cm] (dummy2) {};
	
	\draw[-{Latex[scale=1.25]}] (order0Values) -- node[draw=none, fill = none, below, yshift = -0.15cm] {{\footnotesize select d-1}\\ \footnotesize{s.t. $\mathcal{T}_1$ is a tree}} (PCorVine);
	\draw[-{Latex[scale=1.25]}] (order1Values) -| node[draw=none, fill = none, right] {{\footnotesize select d-2 under}\\ \footnotesize{proximity condition}\\ \footnotesize{s.t. $\mathcal{T}_2$ is a tree}} (PCorVine);
	%	\draw[->] (Data) to node[draw=none, fill = none] {} (CorMatrix);
	%	\draw[->] (CorMatrix) to node[draw=none, fill = none] {} (vecStandardCors);
	\draw[-{Latex[scale=1.25]}] (orderLastValues) -| node[draw=none, fill = none, right] {{\footnotesize select 1 under}\\ \footnotesize{proximity condition}\\ \footnotesize{s.t. $\mathcal{T}_{d-1}$ is a tree}} (PCorVine);
	\draw[-{Latex[scale=1.25]}] (PCorVine) -| node[draw=none, fill = none, below] {\small{$F^{-1}_{\text{Cor2PCor}}$}} (StandardCors);
	\draw[-] (StandardCors) |-  node[draw=none, fill = none] {} (dummy1);
	\draw[-{Latex[scale=1.25]}] (dummy2) -| node[draw=none, fill = none, below right] {\small{\quad \quad formulas \eqref{eq:recursionCor2Pcor}, \eqref{eq:PCorsMatrixInversion}}} (setPCors);
	\end{tikzpicture}
	\caption{Illustration of the transformation of the set of standard correlations $\mathcal{C}_d$ through a partial correlation vine, which consists of a subset of algebraic independent (partial) correlations $\mathcal{C}\left(\mathcal{V}_{d}\right) \subset \mathcal{C}_d^\text{p}$ from all standard and partial correlations. The abbreviation ``pcor'' is used for partial correlation.}
	\label{fig:FlowchartData2}
	%\vspace*{-2cm}
\end{figure}

\section{General setting and benchmark models}\label{Sec:DataSettingBenchmark}
\textcolor{black}{In the following, partial correlation vine based data transformation will be used to model and forecast multivariate volatility time-series. To do so, we introduce the general data setting first. For the daily price series $\boldsymbol{S}_t \in \mathbb{R}^d$, $t = 1, \ldots, T$, of $d$ assets let
	%\begin{align*}
	$\boldsymbol{r}_{t} = \log\left(\boldsymbol{S}_{t}\right) - \log\left(\boldsymbol{S}_{t-1}\right)$
	%\end{align*}
	be the $d \times 1$ vector of daily log-returns. The process $\boldsymbol{r}_{t}$ can be written as}

\textcolor{black}{\begin{align*}
	\boldsymbol{r}_{t} = \mathbb{E}[\boldsymbol{r}_t|\mathcal{F}_{t-1}] + \boldsymbol{\epsilon}_t,
	\end{align*}
	where $\mathcal{F}_{t-1}$ is the information set containing all information up to and including time point $t-1$. For the innovation term $\boldsymbol{\epsilon}_t$, we suppose that $\boldsymbol{\epsilon}_t = \boldsymbol{\Sigma}_t^{1/2}\boldsymbol{\eta}_t$, where $\boldsymbol{\Sigma}_t = \text{Var}[\boldsymbol{r}_t|\mathcal{F}_{t-1}]$ is the $\left(d \times d\right)$-dimensional symmetric and positive definite conditional covariance matrix. For the i.i.d.\ vector $\boldsymbol{\eta}_t \in \mathbb{R}^{d}$ it holds that $\mathbb{E}[\boldsymbol{\eta}_t] = 0$ and $\text{Var}[\boldsymbol{\eta}_t] = I_d$. Interest is in modeling and forecasting the series of daily conditional covariance matrices $\boldsymbol{\Sigma}_t$, $t=1,\ldots,T$, which however are naturally latent variables and therefore are unobservable. Still, as proposed by \cite{Barndorff2004ApproxRCOV} $\boldsymbol{\Sigma}_t$, $t=1,\ldots,T$, can be specified nonparametrically using the realized covariance matrices as consistent estimates. Considering $M$ intra-day periods per day $t$, the latter are calculated from high-frequency intra-day log-returns $\boldsymbol{r}_{\ell,t}= \log\left(\boldsymbol{S}_{t-1+\ell/M}\right) - \log\left(\boldsymbol{S}_{t-1+(\ell-1)/M}\right)$ based on the price series $\boldsymbol{S}_{\ell, t} \in \mathbb{R}^d$, $\ell = 1,\ldots,M$. The modeling and forecasting framework is then based on the matrix valued time-series of realized covariance matrices} 

\textcolor{black}{\begin{align}
	\boldsymbol{Y}_t = \sum_{l=1}^{M}\boldsymbol{r}_{\ell,t}\boldsymbol{r}_{\ell,t}', \quad t=1,\dots,T.
	\label{Eq:RCOV}
	\end{align}
	Since for the matrix forecasts symmetry and positive definiteness have to be ensured, algebraic restrictions are imposed on time-series models. Thus, popular modeling strategies avoid direct modeling of the realized covariance matrices considering transformed data instead. Then, the modeling approach basically consists of three consecutive steps: (S1) data transformation of the realized covariance matrices; (S2) multivariate time-series modeling and prediction based on the transformed data; (S3) back-transformation of the transformed data to obtain predictions for the realized covariance matrices, which are proxies for the future conditional covariance matrices.}

\textcolor{black}{The novelty in this paper lies in the use of partial correlation vines for data transformation in steps (S1) and (S3). By modeling and forecasting the time-series of partial correlation vines, we obtain forecasts for the transformed data, which do not underly any algebraic restrictions. On the contrary, due to the algebraic independence of the model components positive definiteness of the corresponding predicted correlation matrices is always guaranteed. In literature, besides the matrix log transformation suggested by \cite{Bauer2011} data transformation based on the Cholesky factorization is one of the most commonly used approaches.} 

\textcolor{black}{Here, the series of realized covariance matrices 
	$\boldsymbol{Y}_t$, $t=1,\ldots,T$, is decomposed such that $\boldsymbol{Y}_t = \boldsymbol{C}'_t\boldsymbol{C}_t$, where $\boldsymbol{C}'_t$ is a lower triangular matrix with positive diagonal elements. The Cholesky elements $c_{i,j;t}$ ($i,j = 1,\ldots,d$) are recursively calculated by}

\textcolor{black}{\begin{align}\label{eq:CholeskyRecursion}
	c_{i,j;t} = \left\{ \begin{matrix}
	\frac{1}{c_{i,i;t}}\left(y_{i,j;t} - \sum_{k=1}^{i-1}c_{k,i;t}c_{k,j;t}\right) & \text{ for } i<j,\\ \sqrt{y_{j,j;t} - \sum_{k=1}^{j-1} c_{k,j;t}^2} & \text{ for } i=j, \\ 0 & \text{ for } i > j.
	\end{matrix}\right. 
	\end{align}
	By modeling and forecasting the Cholesky elements in step (S2) no parameter restrictions need to be imposed on the multivariate time-series models. Symmetry and positive definiteness of the predicted covariance matrices $\hat{\boldsymbol{Y}}_t$, $t=T+1, T+2, \ldots$, are automatically guaranteed through the back-transformation}

\textcolor{black}{\begin{align}\label{eq:CholeskyBackRecursion}
	\hat{y}_{i,j;t} = \sum_{k = 1}^{\min\{i,j\}}\hat{c}_{k,i;t}\hat{c}_{k,j;t}.
	\end{align}
	\citet{ChiriacVoev2011ForecastingMultVol} use the Cholesky decomposition in steps (S1) and (S3) and apply a parsimonious VARFIMA model to the multivariate time-series of the Cholesky components in step (S2). In their detailed analysis, they show the superiority of their approach over a variety of competitor models. The comparison includes the above mentioned matrix log transformation used in steps (S1) and (S3) for data transformation combined with VARFIMA and vector HAR models in step (S2). Further, the Wishart autoregressive model of \cite{gourieroux2009wishart} as well as the multivariate GARCH model with dynamic conditional correlations of \cite{Engle2002} and its fractionally integrated version proposed by \cite{baillie1996fractionally} are considered. \cite{Brechmann2015Cholesky} refine the Cholesky-VARFIMA model of \cite{ChiriacVoev2011ForecastingMultVol} and allow for more flexible modeling of the multivariate time-series in step (S2). They take account of challenging data characteristics in the Cholesky elements by modeling the univariate marginal time-series with elaborate HAR and ARFIMA models including GARCH-augmentations for the residuals. The possibly complex dependence between the Cholesky components is captured by a copula. Given these profound model reviews and comparisons already existing in literature, models based on the Cholesky decomposition will be our main benchmarks.}

\textcolor{black}{\cite{ChiriacVoev2011ForecastingMultVol} and \cite{Brechmann2015Cholesky} both consider high frequency data from the NYSE TAQ database containing tick-by-tick bid and ask quotes on six stocks listed on the New York Stock Exchange (NYSE), American Stock Exchange (AMEX) and the National Association of Security Dealers Automated Quotation System (NASDAQ). The original raw data were processed by \cite{ChiriacVoev2011ForecastingMultVol}, who provide detailed information on the employed data preparation. Data of the six stocks American Express Inc. (AXP), Citigroup (C), General Electric (GE), Home Depot Inc. (HD), International Business Machines (IBM) and JPMorgan Chase \& Co (JPM) were sampled from 9:30 until 16:00 for the period January 1, 2000, until July 30, 2008, i.e.\ for 2156 trading days. While in \eqref{Eq:RCOV} a single realized covariance matrix is computed from $M$ intra-day log-returns, \cite{ChiriacVoev2011ForecastingMultVol} obtained for each day a refined subsampled realized covariance matrix, which is more robust to market microstructure noise. For each day $t$, a 5-minute spaced time grid, i.e.\ $M = 78$, was shifted by 10 seconds, resulting in 30 distinct sets of realized covariance matrices calculated from 78 intra-day log-returns. By taking the average of these sets, the subsampled realized covariance matrix for day $t$ was calculated. Although the data are less recent, we consider the same data for comparison reasons. Further, the data cover interesting periods of financial turmoil such as the aftermath of the dotcom bubble in 2000 and the beginning of the financial crisis in 2008. Since focus in this paper is on the novel data transformation defined by partial correlation vines, this will provide new interesting insights about the data.}

\section{Partial correlation vine data transformation approach}\label{Sec:PCVapproach}
\textcolor{black}{In this section, we outline -- supported by real data characteristics -- steps (S1) to (S3) for the proposed modeling strategy based on partial correlation vines.}

\subsection{Data characteristics}\label{Subsec:DataCharacteristics}
\textcolor{black}{Time-series of realized co(variances) typically exhibit long-memory behavior detectable by high autocorrelations, which decay at a slow rate (see e.g.\ \cite{andersen1997heterogeneous, andersen2001distribution}). \cite{ChiriacVoev2011ForecastingMultVol} find that the time-series of Cholesky components obtained through data transformation inherit this data feature. Further, according to \cite{Brechmann2015Cholesky} appropriate time-series models need to capture non-Gaussianity and volatility clustering of the residuals extracted from the series of Cholesky elements.}

\textcolor{black}{In order to also appropriately setup the partial correlation vine data transformation model it is essential to understand the properties of the corresponding model components, namely realized variances and realized (partial) correlations. The latter are specified through the realized covariance matrix via $\boldsymbol{Y}_t = \boldsymbol{D}_t^{1/2}\boldsymbol{R}_t\boldsymbol{D}_t^{1/2}$, $t=1,\ldots,T$. For day $t$, $\boldsymbol{D}_t = \text{diag}\left(y_{1,1;t},\ldots,y_{d,d;t}\right)$ contains the realized variances and $\boldsymbol{R}_t$ is the realized correlation matrix. Realized partial correlations can easily be obtained either using recursion formula \eqref{eq:recursionCor2Pcor} or through simultaneous calculation using \eqref{eq:PCorsMatrixInversion}. For reasonable time-series modeling later in step (S2), for all model components data on the real line are needed. Thus, we log-transform the all positive realized variance time-series and apply the Fisher z-transformation to the series of (partial) correlations, i.e.\ for $\rho_t$ being an arbitrary (partial) correlation at day $t$}%elements $\rho_{(j);t} \in \left(-1,1\right)$ of $\boldsymbol{P}_{\mathcal{C}_d^\text{p},t}$  with $(j) \in \lbrace \left(1,2\right), \ldots, \left(d-1,d\right), \left(1,2;3\right),\ldots, \left(d-1,d;1,\ldots,d-2\right) \rbrace$, i.e.\
%\begin{align}\label{Eq:Fisher-z}
%\vspace*{-5cm}
%z\left(\rho_{(j);t}\right) = \frac{1}{2}\log\left(\frac{1+\rho_{(j);t}}{1-\rho_{(j);t}}\right).
%\end{align} 

\textcolor{black}{\begin{align}\label{Eq:Fisher-z}
	\vspace*{-5cm}
	z\left(\rho_t\right) = \frac{1}{2}\log\left(\frac{1+\rho_t}{1-\rho_t}\right), \quad t=1,\ldots,T.
	\end{align}}

\begin{figure}[h!]
	%	\vspace{-1.5cm}
	\centering
	\hspace*{1.75cm}	
	{\footnotesize original scale}	\hspace*{3.35cm} {\footnotesize log-/Fisher z-transformed data}
	\includegraphics[width=.5\linewidth]{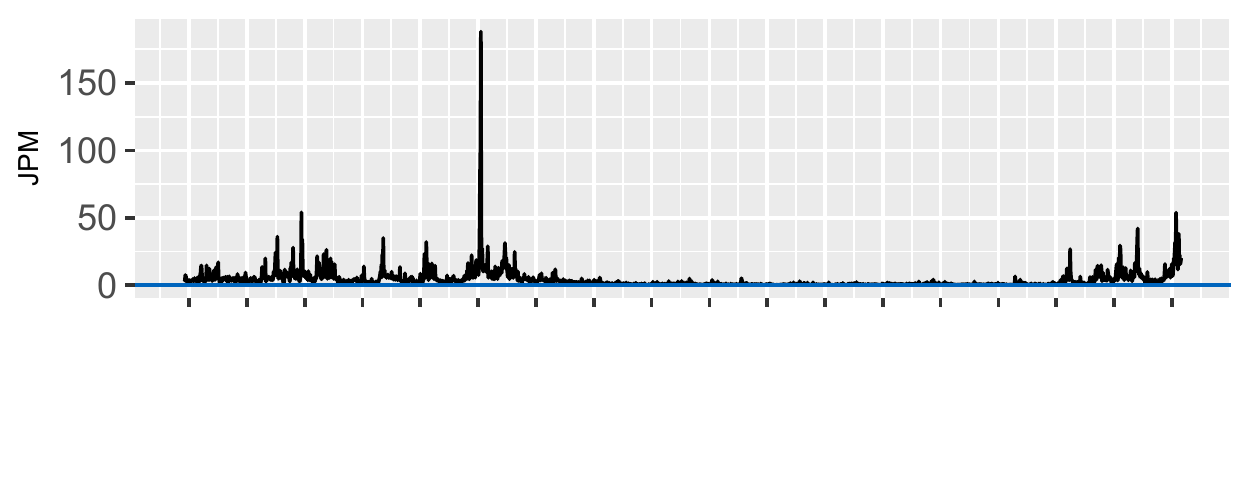}\includegraphics[width=.5\linewidth]{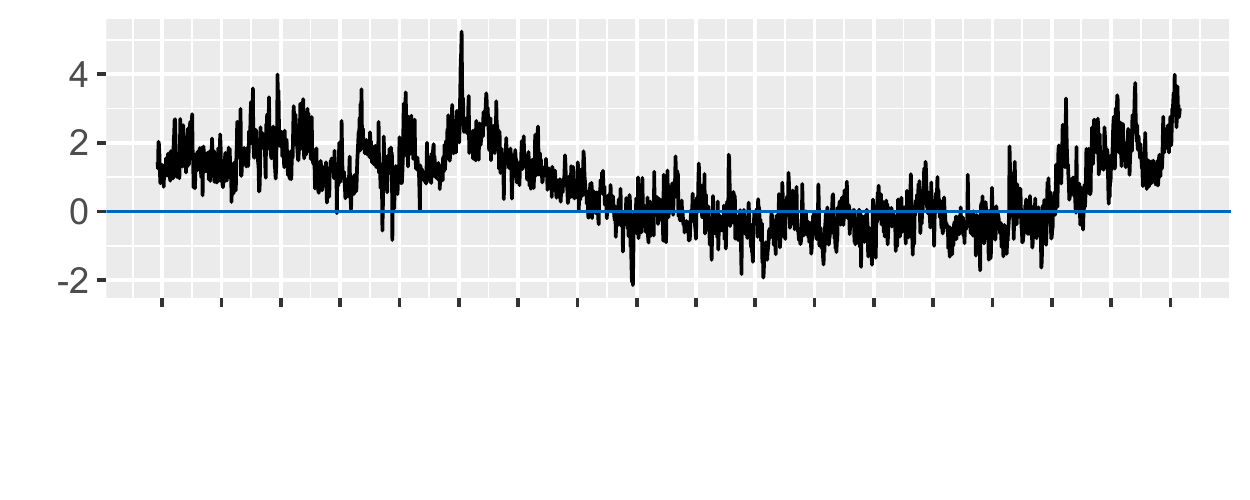}\vspace{-1.15cm}
	\includegraphics[width=.5\linewidth]{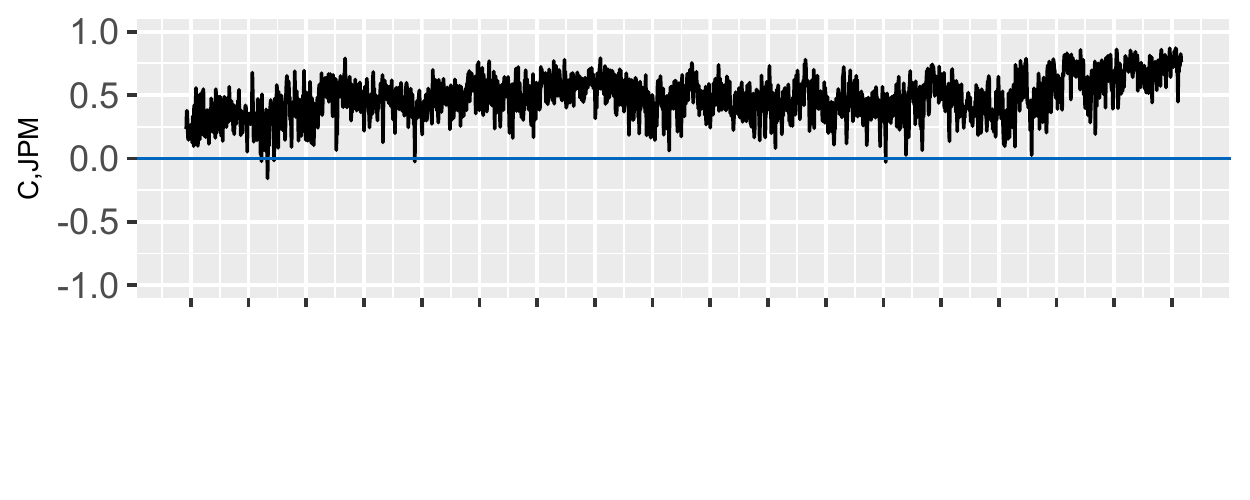}\includegraphics[width=.5\linewidth]{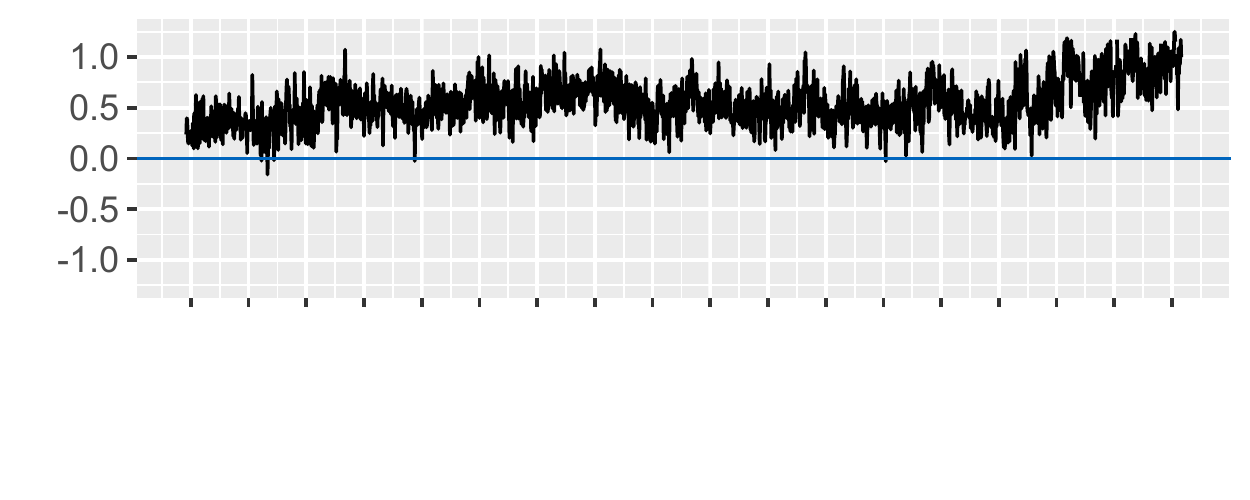}\vspace{-1.15cm}
	\includegraphics[width=.5\linewidth]{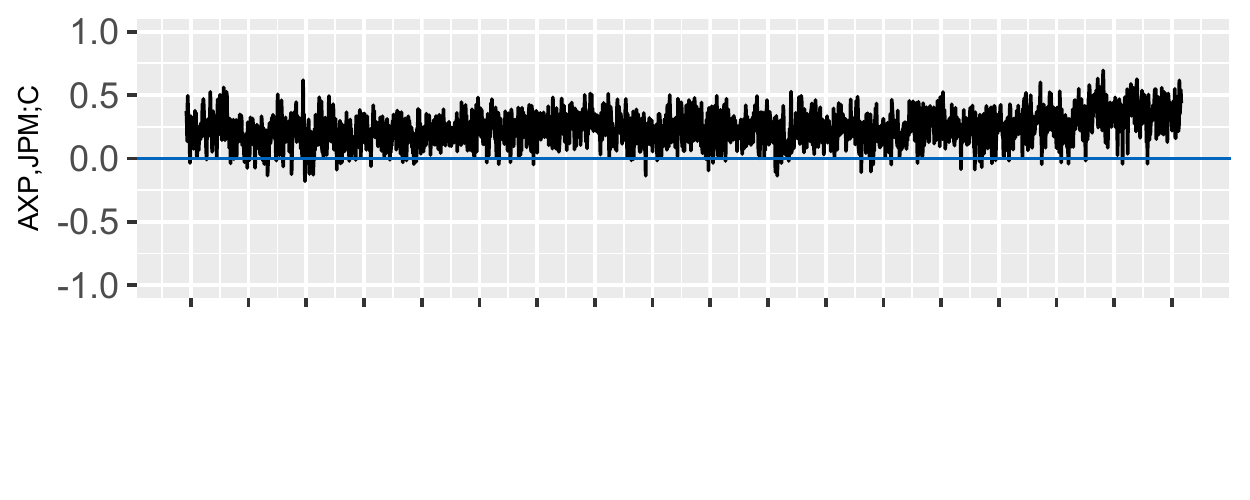}\includegraphics[width=.5\linewidth]{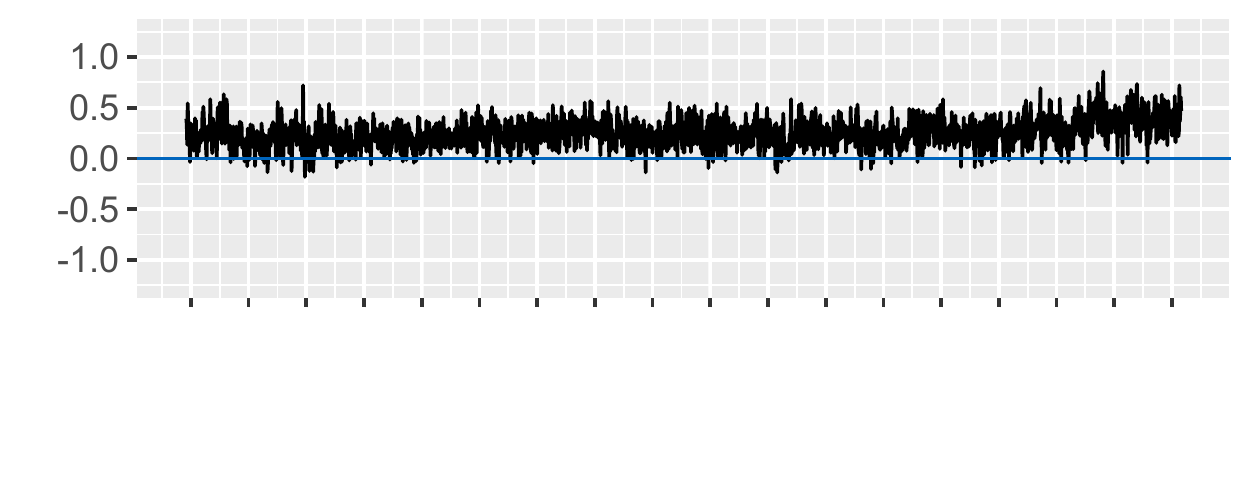}\vspace{-1.15cm}
	\includegraphics[width=.5\linewidth]{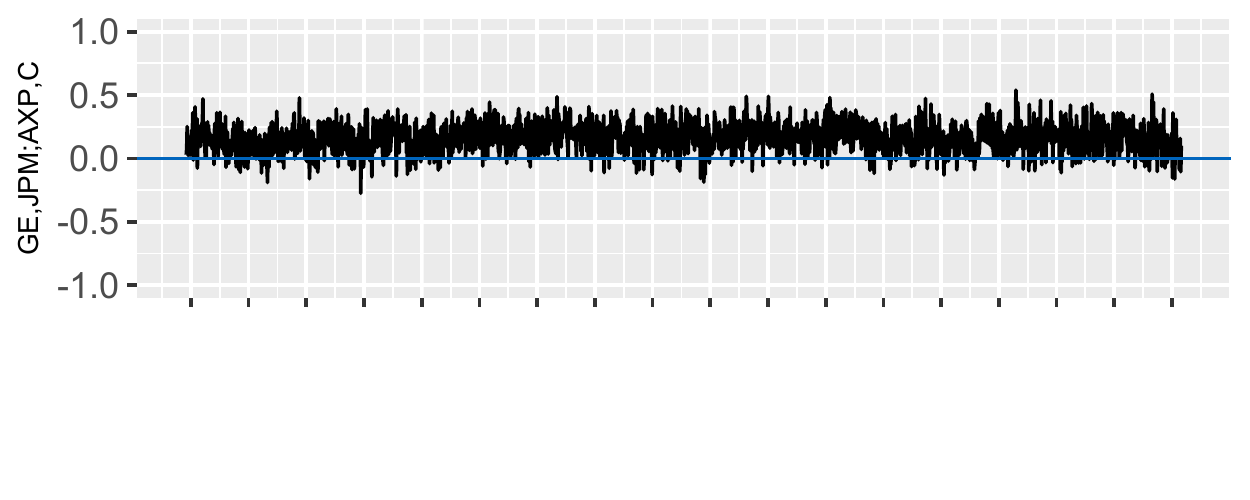}\includegraphics[width=.5\linewidth]{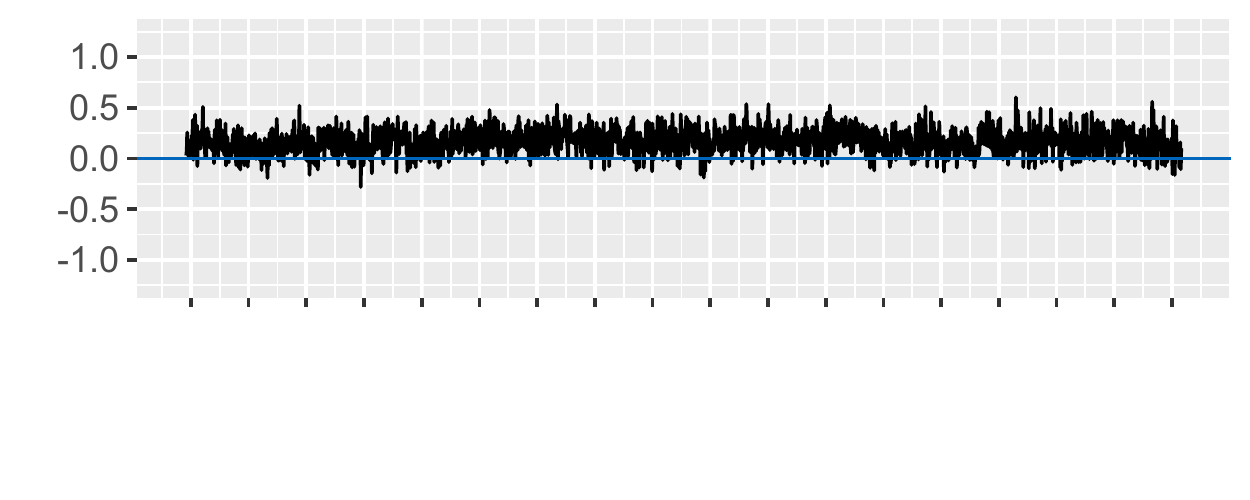}\vspace{-1.15cm}
	\includegraphics[width=.5\linewidth]{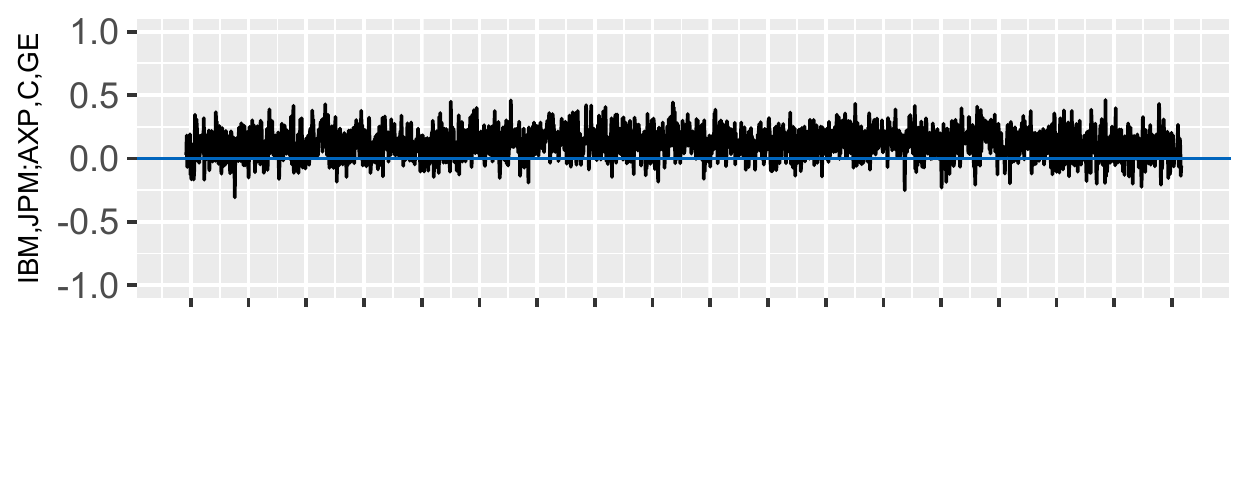}\includegraphics[width=.5\linewidth]{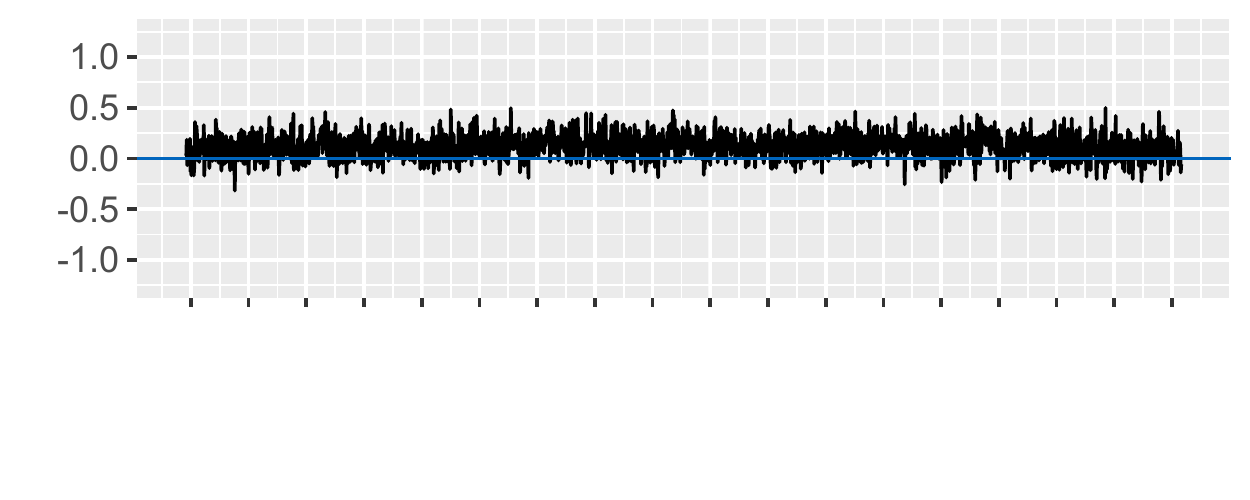}\vspace{-1.15cm}
	\includegraphics[width=.5\linewidth]{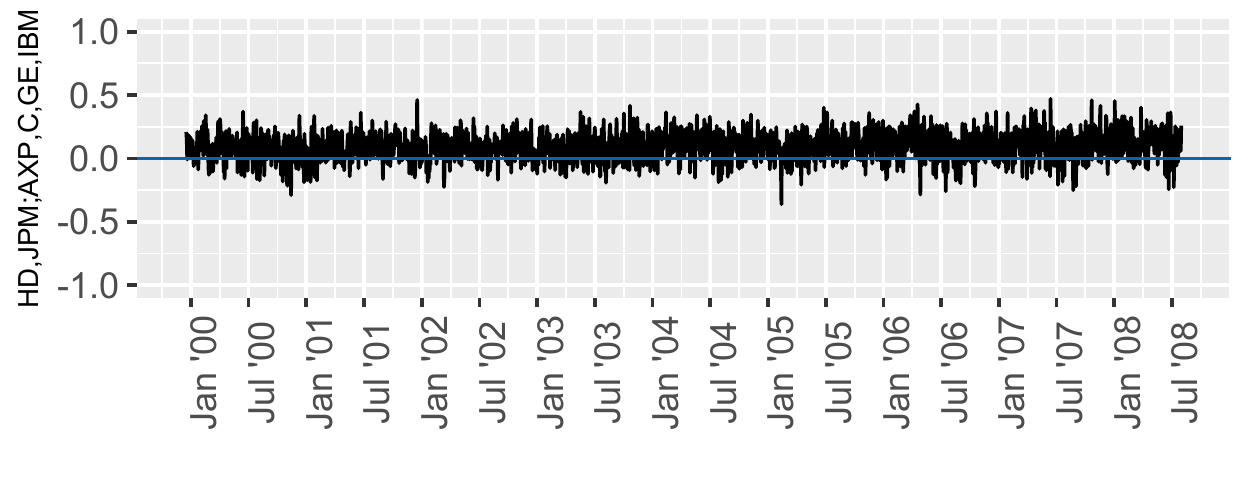}\includegraphics[width=.5\linewidth]{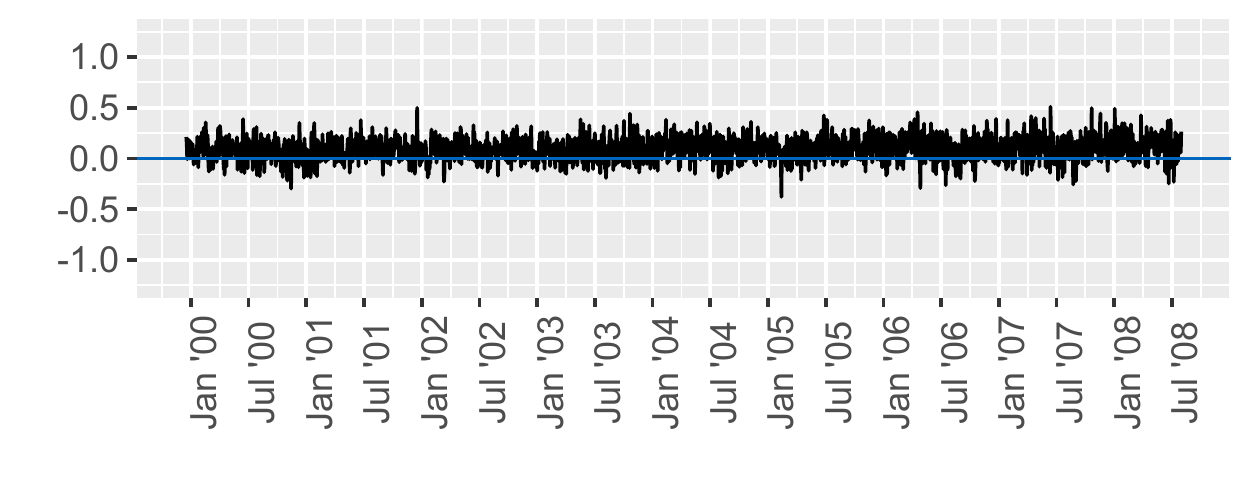}
	\caption{Daily realized variance series (1st row) and daily realized (partial) correlation series (2nd -- 6th row). Original data are shown in the left panel, log-transformed and Fisher z-transformed data, respectively, are shown in the right panel.}
	\label{fig:timeSeries_JPM}
	%\vspace{-1.5cm}
\end{figure} 

\begin{figure}[h!]
	%\vspace{-.35cm}	
	\small
	\centering		
	\begin{tabular}{c c} 
		\multicolumn{2}{c}{$\lbrace\log\left(y_{JPM,JPM;t}\right)\rbrace$}\\
		\includegraphics[width=0.43\linewidth]{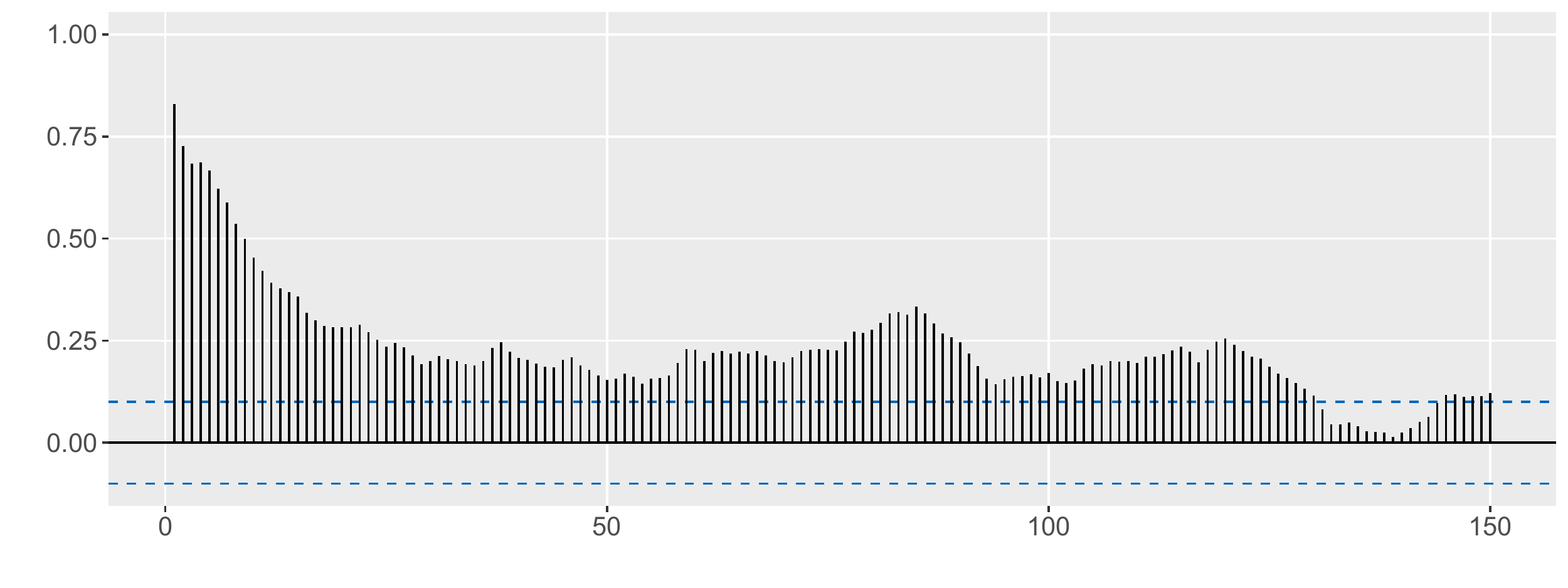}&\includegraphics[width=0.43\linewidth]{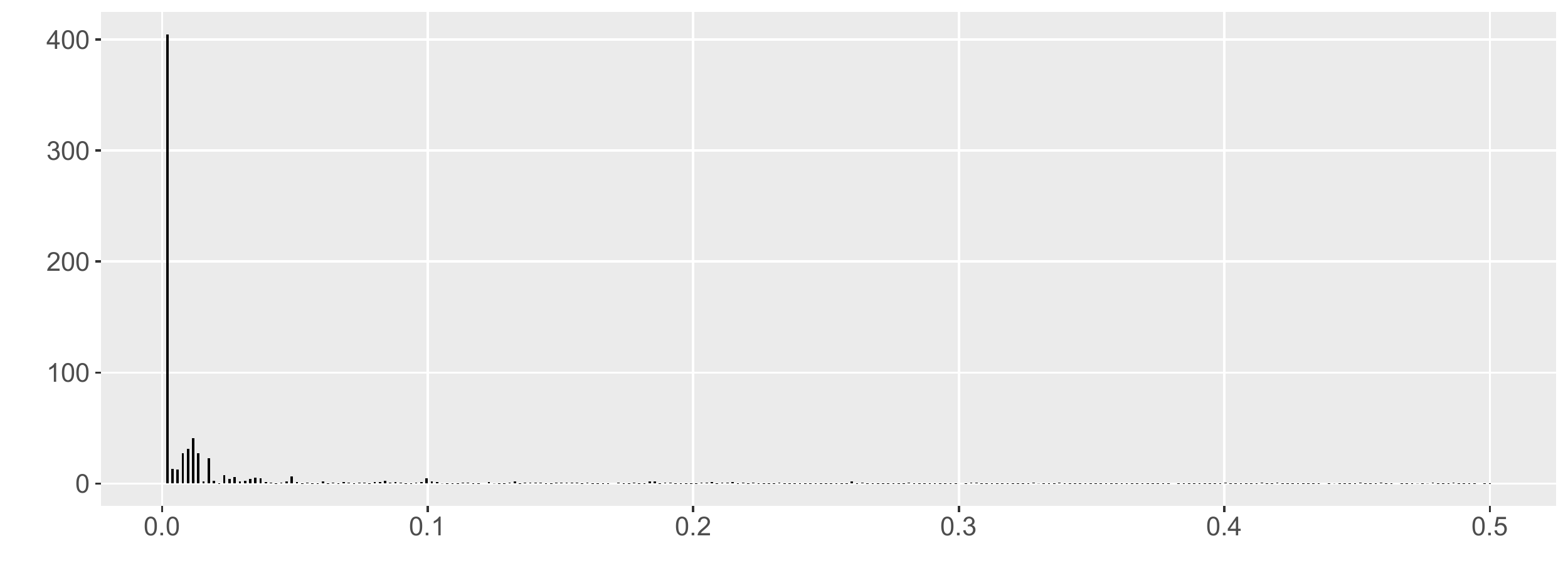}\\
		\multicolumn{2}{c}{$\mathcal{T}_1: \ \ \lbrace z\left(\rho_{C,JPM;t}\right) \rbrace$}\\   \includegraphics[width=0.43\linewidth]{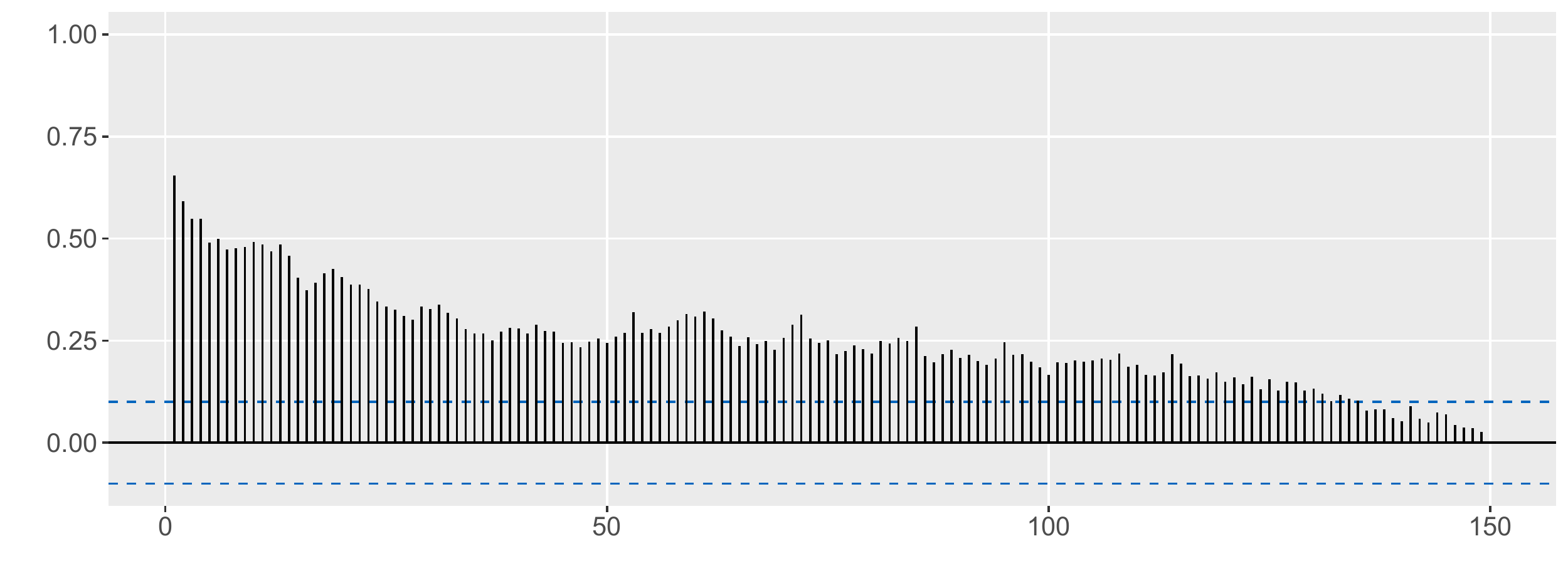}&\includegraphics[width=0.43\linewidth]{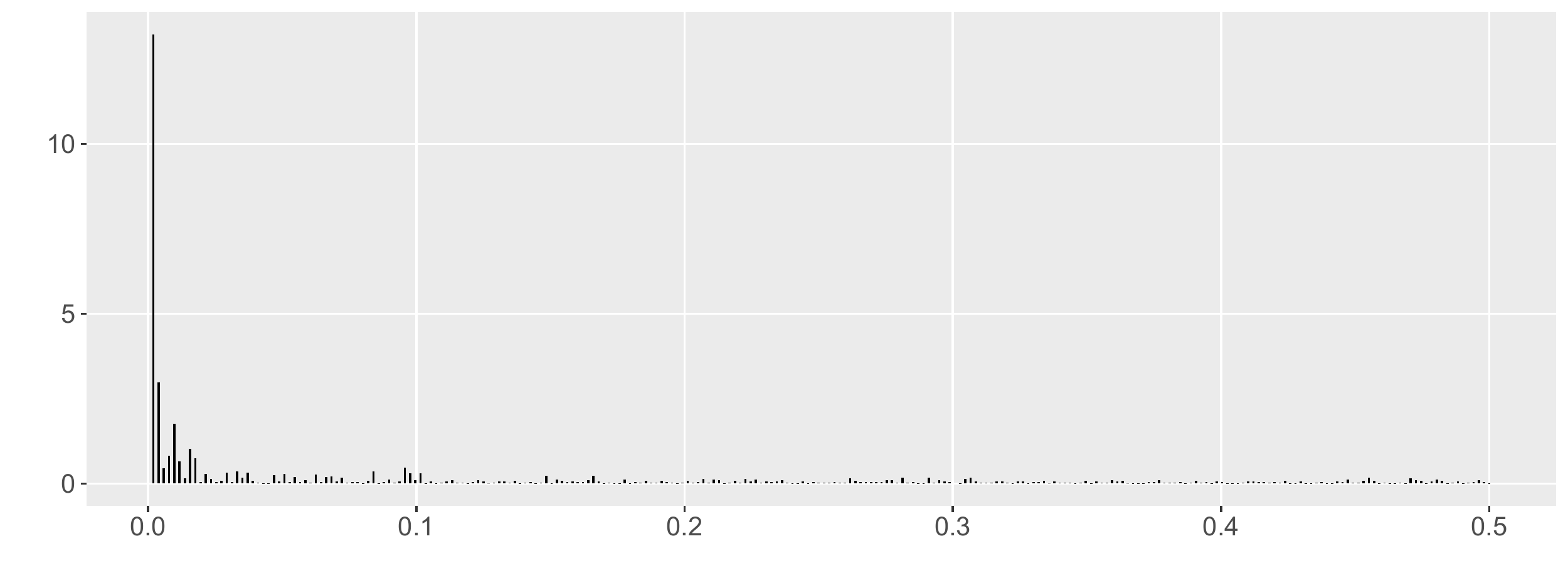}\\
		\multicolumn{2}{c}{$\mathcal{T}_4: \ \ \lbrace z\left(\rho_{IBM,JPM;AXP,C,GE;t}\right) \rbrace$}\\	\includegraphics[width=0.43\linewidth]{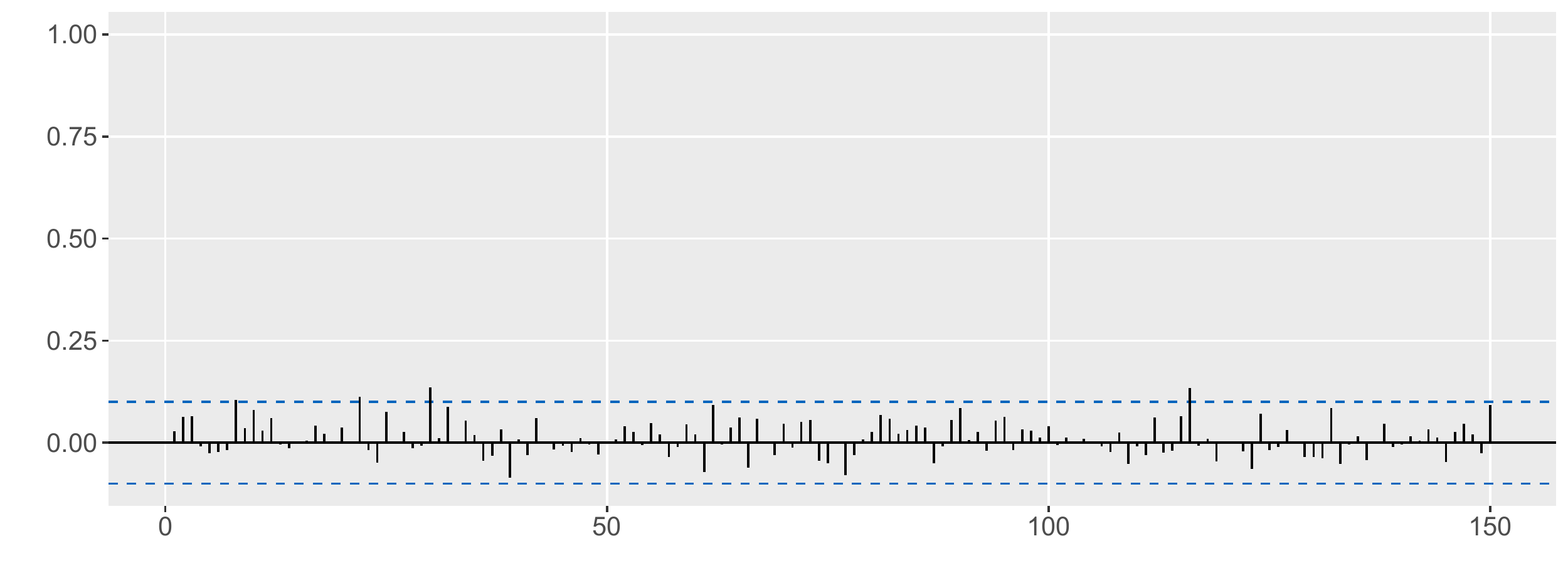}&\includegraphics[width=0.43\linewidth]{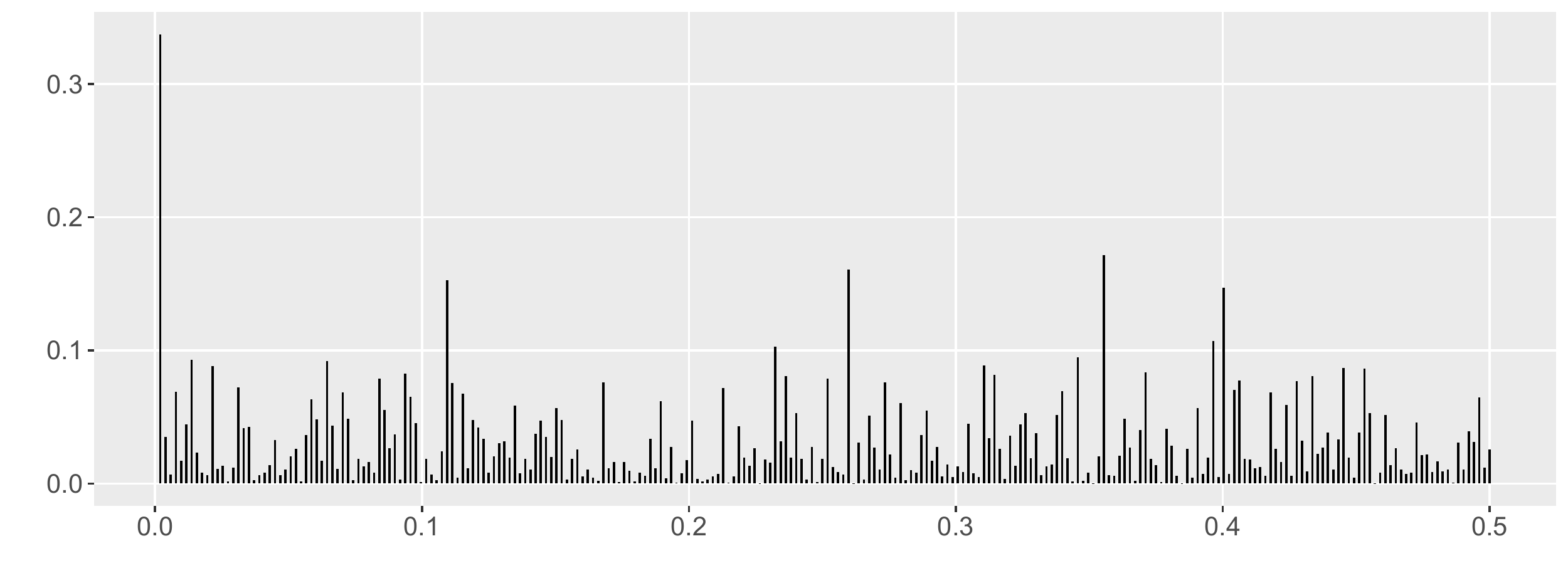}\\
		\multicolumn{2}{c}{$\mathcal{T}_5: \ \ \lbrace z\left(\rho_{C,JPM;AXP,GE,HD,IBM;t}\right) \rbrace$}\\	 \includegraphics[width=0.43\linewidth]{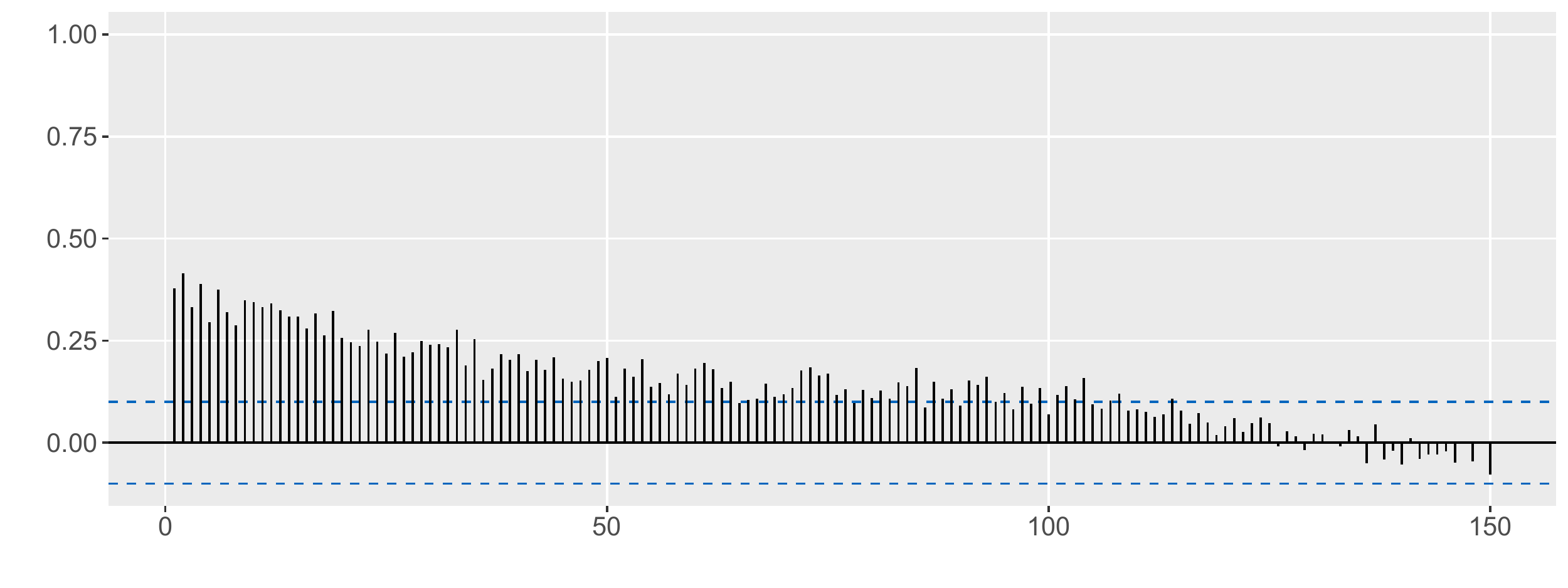} & \includegraphics[width=0.43\linewidth]{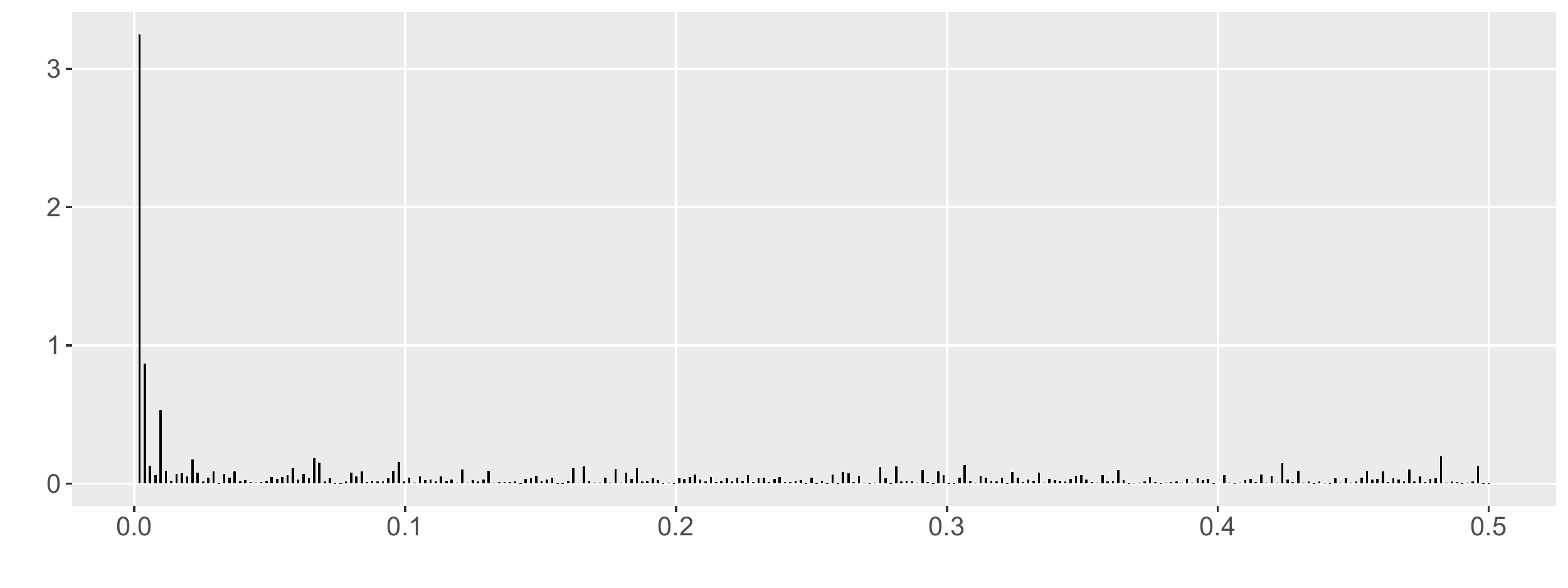}\\	
		\vspace*{-0.20cm}
	\end{tabular}
	\caption{Illustration of autocorrelation functions of squared data (left panel) and corresponding log-periodograms (right panel) based on data from July 1, 2006, to June 30, 2008. In the first row, the log-transformed realized variance time-series of JPM  is considered. In rows 2 to 4 exemplary Fisher z-transformed realized (partial) correlation time-series of increasing order are considered. }
	\label{fig:characteristicsTimeSeries}
	\vspace*{-.5cm}
\end{figure}

For the considered real data, \autoref{fig:timeSeries_JPM} shows a selection of time-series both on the original (left) and the transformed (right) scale. The first panel illustrates for JPM the daily realized variance series. Striking is the highly volatile behavior particularly during periods of financial turmoil such as the aftermath of the dotcom bubble and the beginning of the financial crisis in August 2007. Panels 2 to 6 show selected daily time-series of realized (partial) correlations with increasing order, \textcolor{black}{e.g.\ the last panel illustrates the time-series of the fourth order realized partial correlation between HD and JPM. For each day, the latter is a proxy of the conditional (with respect to the information set) correlation between the log-returns of HD and JPM given the four remaining stocks.} With increasing order the realized partial correlation time-series become more stable while still exhibiting highly volatile behavior. \textcolor{black}{Further, in \autoref{fig:characteristicsTimeSeries} data characteristics of four time-series are illustrated. The figures in the top row correspond to the realized variance time-series of JPM, which together with the remaining five realized variance series always will be a model component. The long hyperbolic decay of the autocorrelation function of the squared data on the left confirms the long-memory behavior and the presence of volatility clustering. The log-periodogram shows higher peaks only for short frequencies as expected for self-similar processes. In the second and third row,  exemplary time-series, which would appear in tree level $\mathcal{T}_1$ and $\mathcal{T}_4$, respectively, of an R-vine structure are shown. Interestingly, while the realized standard correlation time-series corresponding to tree level $\mathcal{T}_1$ inherits the data characteristics of the realized variance time-series, the latter are less pronounced for the realized third order partial correlation time-series in $\mathcal{T}_4$.}

\normalsize

\textcolor{black}{To gain a deeper understanding of this last observation, recall that in a partial correlation vine each variable pair $(i,j)$ ($i,j \in \lbrace 1,\ldots,d \rbrace$, $i \neq j$) forms exactly once the conditioned set of an edge. Thus, depending on the tree level $\ell$ the proxy for the conditional (with respect to the information set) correlation between the log-returns of stocks $i$ and $j$ is either represented by the realized standard correlation (if $\left(i,j\right)$ occurs as conditioned set in $\mathcal{T}_1$) or through a ($\ell-1$)-th order realized partial correlation (if $\left(i,j\right)$ occurs as conditioned set in $\mathcal{T}_\ell$ ($\ell=2,\ldots,d-1$)). In the latter case, the linear effect of the $\ell-1$ stocks forming the conditioning set is removed. Clearly, for some pairs the realized standard correlations might mainly be driven by other variables. Once this influence is removed data features such as long-memory behavior weaken and the corresponding realized partial correlation time-series behave more and more like noise. On the other hand, this effect is not observable for pairs, which truly are strongly correlated such as the log-returns of the two financial stocks C and JPM. The corresponding realized fourth order partial correlation time-series, which would occur in the highest tree level, i.e.\ $\mathcal{T}_5$, of an R-vine structure, underlies the two figures in the last row of \autoref{fig:characteristicsTimeSeries}. It shows similar data characteristics as the realized variance time-series in the top row.}

\textcolor{black}{The above analysis clearly stresses the practical interpretability of the model components in the partial correlation vine data transformation approach, namely realized variances and realized (partial) correlations. In the following, the detected inhomogeneous data complexity motivates a specific choice for the R-vine structure used for data transformation in step (S1).}

\subsection{Step (S1): R-vine structure selection for data transformation}\label{Sec:R-VineStructureSelection}
\textcolor{black}{In $d$ dimensions there exist $d!/2\cdot d^{(d-2)(d-3)/2}$ valid R-vines \citep{morales2010number}. It is important to note that, in general, each of these tree structures allows a valid transformation of the realized correlation matrices. To allow for model parsimony later in step (S2), when modeling the dynamics of the realized variance series and the realized (partial) correlation time-series selected by the R-vine structure, we refer to the data characteristics detected in \autoref{Subsec:DataCharacteristics} and propose in this section an algorithm for R-vine structure selection.} 

\textcolor{black}{We know that when transforming a series of realized correlation matrices based on the same R-vine structure, to each edge in this R-vine a univariate time-series of realized standard or partial correlations is assigned.} Therefore, each edge can be characterized by a weight derived from sample properties of the corresponding time-series. We decide for the average (partial) correlation strength: we consider the average correlation matrix $\bar{\boldsymbol{R}} = \left(\bar{\rho}_{i,j}\right)_{i,j=1,\ldots,d}$ (which is positive definite) calculated from $\boldsymbol{R}_t$, $t=1,\ldots,T$. \textcolor{black}{Then, we find the maximum spanning tree $\mathcal{T}_1$ \citep{KatohAlgorithm1981} with edge weights set to  $\bar{\rho}_{C_{e,a},C_{e,b}}$. To construct tree $\mathcal{T}_2$, we calculate all average first order partial correlations $\bar{\rho}_{C_{e,a},C_{e,b};D_e}$, i.e.\ $|D_e| = 1$, where $(C_{e,a},C_{e,b};D_e)$ satisfies the proximity condition given $\mathcal{T}_1$. Based on these weights, we find the maximum spanning tree $\mathcal{T}_2$.} In general, we construct the R-vine structure $\mathcal{V}_d$ within a top-down procedure and find tree by tree ($\ell=1,\ldots,d-1$) the maximum spanning tree $\mathcal{T}_\ell$ with edge weights set to  $\bar{\rho}_{C_{e,a},C_{e,b};D_e}$, where $|D_e| = \ell-1$ and $(C_{e,a},C_{e,b};D_e)$ satisfies the proximity condition given $\mathcal{T}_1$ to $\mathcal{T}_{\ell-1}$. By doing so, we equip based on historical information the R-vine structure with the highest realized (partial) correlation means.

The correlation matrix $\bar{\boldsymbol{R}}$ can be obtained in various ways depending on how the average is calculated. Considering for each pair $(i,j)$ ($i,j \in \lbrace 1,\ldots,d \rbrace$, $i \neq j$) the empirical mean $\bar{\rho}_{i,j} = \frac{1}{T}\sum_{t=1}^{T} \rho_{i,j;t}$ assigns to each day's value $\rho_{i,j;t}$ the same influence $1/T$ irrespective of how far it lies in the past. By using e.g.\ an exponentially weighted moving average (EWMA) more influence can be assigned to values of more recent days. Here, the exact weights are controlled by the smoothing parameter $\lambda \in \ ]0,1[$ and are defined as $w_t = (1-\lambda)\lambda^{T-t}$, $t=1,\ldots,T$. Thus, for decreasing $\lambda$ the impact of more recent days increases and therewith the sensitivity of the selected R-vine structure with respect to market changes.

\tikzstyle{ClassicalVineNode} = [rectangle, fill = white, draw = black, text = black, align = center, minimum width = 0.55cm]
\tikzstyle{TimeAxisNode} = [fill = none, draw = none, text = black, align = center, minimum height = 0.55cm, minimum width = 0.55cm]
\tikzstyle{TreeLabels} = [draw = none, fill = none, text = black, font = \bf]
\tikzstyle{DummyNode}  = [draw = none, fill = none, text = white]
\renewcommand{\yshift}{.25cm}
\newcommand{\xshift}{2cm}
\renewcommand{\yshiftlabel}{-0.08cm}
\renewcommand{\labelsize}{\tiny}
\renewcommand{\arraystretch}{1.61}
\begin{table}[h!]
%	\vspace{-1.95cm}	
	\small
	\centering
	\caption{Illustration of the R-vine structure selection method for the real data example considering all available data points, i.e.\ the mean values $\bar{\rho}_{\mathcal{C}_{e,a},\mathcal{C}_{e,b};\mathcal{D}_e}$ are based on  $t=1,\ldots,2156$.}
	\label{Table:SelectionMethodRealData}
	\hspace*{-.5cm}\begin{tabular}{l l c c p{0.075cm}} \cmidrule{1-5}
		\multicolumn{3}{c}{pairs allowed by proximity condition} & \multicolumn{2}{c}{\multirow{2}{*}{selected tree}} \\
		$\mathcal{D}_e$ &  $\mathcal{C}_{e,a},\mathcal{C}_{e,b}$  & $\bar{\rho}_{\mathcal{C}_{e,a},\mathcal{C}_{e,b};\mathcal{D}_e}$ & & \\
		\cmidrule{1-5}\morecmidrules\cmidrule{1-5}
		$\emptyset$ & \cellcolor{gray!25}C,JPM &\cellcolor{gray!25} 0.547 & \multirow{15}{*}{\begin{tikzpicture}	[every node/.style = ClassicalVineNode, node distance =1.5cm, font = \scriptsize]
			\node[very thick] (IBM)        {IBM}			node[very thick] (AXP) [below left  of = IBM, xshift = -0.5*\xshift]{AXP}
			node[very thick]             (HD)         [below of = AXP, yshift = -1cm] {HD}
			node[very thick](JPM)         [below right of = HD, xshift = 0.5*\xshift] {JPM}	
			node[very thick]             (C)         [above right of = JPM, xshift = 0.5*\xshift] {C}																				    							node[very thick]             (GE)         [above of = C, yshift = 1cm] {GE}	
			;						
			\draw (IBM) to node[draw=none, text = black, fill = none, font = \labelsize, above, rotate = 25, yshift = \yshiftlabel] {} (AXP);
			\draw (IBM) to node[pos = .275, draw=none, text = black, fill = none, font = \labelsize, above, rotate = 48, yshift = \yshiftlabel] {} (HD);
			\draw (IBM) to node[draw=none, pos = 0.175, text = black, fill = none, font = \labelsize, left, xshift = -0.7*\yshiftlabel] {} (JPM);
			\draw (IBM) to node[draw=none, pos = .275, text = black, fill = none, font = \labelsize, above, rotate = -48, yshift = \yshiftlabel] {} (C);
			\draw[very thick] (IBM) to node[draw=none, text = black, fill = none, font = \labelsize, above, rotate = -29, yshift = \yshiftlabel] {0.400} (GE);
			\draw[very thick] (IBM) to node[draw=none, text = black, fill = none, font = \labelsize, below, rotate = -29, yshift = -0.75*\yshiftlabel] {GE,IBM} (GE);					
			\draw (AXP) to node[draw=none, text = black, fill = none, font = \labelsize, above, rotate = 90, yshift = \yshiftlabel] {} (HD);
			\draw (AXP) to node[draw=none, text = black, fill = none, font = \labelsize, above, pos = .15, rotate = -48, yshift = \yshiftlabel] {} (JPM);
			\draw[very thick] (AXP) to node[draw=none, text = black, fill = none, font = \labelsize, above, pos = .65, rotate = -31, yshift = \yshiftlabel] {0.456} (C);
			\draw[very thick] (AXP) to node[draw=none, text = black, fill = none, font = \labelsize, below, pos = .65, rotate = -31, yshift = -0.75*\yshiftlabel] {AXP,C} (C);
			\draw (AXP) to node[draw=none, pos = .225, text = black, fill = none, font = \labelsize, left, yshift = -\yshiftlabel] {} (GE);						
			\draw (HD) to node[draw=none, text = black, fill = none, font = \labelsize, below, rotate = -24, yshift = -\yshiftlabel] {} (JPM);
			\draw[very thick] (HD) to node[draw=none, pos = .15, text = black, fill = none, font = \labelsize, below, yshift = -\yshiftlabel] {0.355} (C);
			\draw[very thick] (HD) to node[draw=none, pos = .15, text = black, fill = none, font = \labelsize, above, yshift = \yshiftlabel] {C,HD} (C);
			\draw (HD) to node[draw=none, text = black, fill = none, font = \labelsize, above, pos = .35, rotate = 17.5, yshift = \yshiftlabel] {} (GE);						
			\draw[very thick] (JPM) to node[draw=none, text = black, fill = none, font = \labelsize, below, rotate = 30, yshift = -\yshiftlabel] {0.547} (C);
			\draw[very thick] (JPM) to node[draw=none, text = black, fill = none, font = \labelsize, above, rotate = 30, yshift = \yshiftlabel] {C,JPM} (C);
			\draw (JPM) to node[draw=none, text = black, fill = none, font = \labelsize, above, pos = .85, rotate = 48, yshift = \yshiftlabel] {} (GE);						
			\draw[very thick] (C) to node[draw=none, text = black, fill = none, font = \labelsize, above, rotate = 90, yshift = \yshiftlabel] {0.437} (GE);
			\draw[very thick] (C) to node[draw=none, text = black, fill = none, font = \labelsize, below, rotate = 90, yshift = -0.75*\yshiftlabel] {C,GE} (GE);						
			\end{tikzpicture}						
		} &	\multirow{15}{*}{$\mathcal{T}_1$} \\
		$\emptyset$	&\cellcolor{gray!25}AXP,C &\cellcolor{gray!25} 0.456 &&\\
		$\emptyset$	&\cellcolor{gray!25}C,GE &\cellcolor{gray!25} 0.437 &&\\
		$\emptyset$	&\cellcolor{white}AXP,JPM &\cellcolor{white} 0.433 &&\\
		$\emptyset$	&\cellcolor{gray!25}GE,IBM &\cellcolor{gray!25} 0.400 &&\\
		$\emptyset$	&\cellcolor{white}AXP,GE &\cellcolor{white} 0.394 &&\\
		$\emptyset$	&\cellcolor{white}GE,JPM &\cellcolor{white} 0.393 &&\\
		$\emptyset$	&\cellcolor{white}C,IBM &\cellcolor{white} 0.390 &&\\
		$\emptyset$	&\cellcolor{white}IBM,JPM &\cellcolor{white} 0.362 &&\\
		$\emptyset$	&\cellcolor{white}AXP,IBM &\cellcolor{white} 0.358 &&\\
		$\emptyset$	&\cellcolor{gray!25}C,HD &\cellcolor{gray!25} 0.355 &&\\
		$\emptyset$	&\cellcolor{white}GE,HD &\cellcolor{white} 0.352 &&\\
		$\emptyset$	&\cellcolor{white}AXP,HD &\cellcolor{white} 0.333 &&\\
		$\emptyset$	&\cellcolor{white}HD,JPM &\cellcolor{white} 0.333 &&\\
		$\emptyset$	&\cellcolor{white}HD,IBM &\cellcolor{white} 0.330 &&\\
		\cmidrule{1-5}
		GE	&\cellcolor{gray!25}C,IBM &\cellcolor{gray!25} 0.253 & \multirow{7}{*}{
			\begin{tikzpicture}	[every node/.style = ClassicalVineNode, node distance = 2.25cm, font = \scriptsize]
			%%%%% Tree 2 %%%%%
			\node[very thick] (GEIBM)         {GE,IBM}
			node[very thick]  (CGE)         [above right of = GEIBM, yshift = \yshift]{C,GE}															
			node[DummyNode] (Dummy1)    [right of = CGE, xshift = -1cm] {}
			node[very thick]  (AXPC)        [right of = Dummy1, xshift = -1cm] {AXP,C}
			node[very thick]  (CJPM)         [below right of = AXPC, yshift = -\yshift] {C,JPM}	
			node[very thick]  (CHD)         [below of = Dummy1, yshift = -3*\yshift] {C,HD}																	    							;					
			\draw[very thick] (CGE) to node[draw=none, text = black, fill = none, font = \labelsize, below, rotate = 49, yshift = -\yshiftlabel] {0.253} node[draw=none, text = black, fill = none, font = \labelsize, above, rotate = 49, yshift = \yshiftlabel] {C,IBM;GE}(GEIBM);
			\draw[very thick] (CGE) to node[draw=none, text = black, fill = none, font = \labelsize, below, pos = .31, rotate = -66, yshift = -\yshiftlabel]  {0.229} node[draw=none, text = black, fill = none, font = \labelsize, above, pos = .31, rotate = -66, yshift = \yshiftlabel]  {GE,HD;C}(CHD);
			\draw (CGE) to node[draw=none, text = black, fill = none, font = \labelsize, above, pos = 0.21, rotate = -22, yshift = \yshiftlabel] {} (CJPM);
			\draw[very thick] (CGE) to node[draw=none, text = black, fill = none, font = \labelsize, above, yshift = \yshiftlabel]{AXP,GE;C}node[draw=none, text = black, fill = none, font = \labelsize, below, yshift = -\yshiftlabel]{0.241}  (AXPC);					
			\draw (CHD) to node[draw=none, text = black, fill = none, font = \labelsize, above, pos = 0.31, rotate = 21, yshift = \yshiftlabel] {} (CJPM);
			\draw (CHD) to node[draw=none, text = black, fill = none, font = \labelsize, above, pos = 0.31, rotate = 66, yshift = \yshiftlabel] {} (AXPC);					
			\draw[very thick] (CJPM) to node[draw=none, text = black, fill = none, font = \labelsize, below, rotate = -49, yshift = -\yshiftlabel] {0.247}node[draw=none, text = black, fill = none, font = \labelsize, above, rotate = -49, yshift = \yshiftlabel] {AXP,JPM;C} (AXPC);			
			\end{tikzpicture}
		} & \multirow{7}{*}{$\mathcal{T}_2$}\\
		C &\cellcolor{gray!25}AXP,JPM &\cellcolor{gray!25} 0.247 & &\\
		C	&\cellcolor{gray!25}AXP,GE &\cellcolor{gray!25} 0.241 & &\\
		C	&\cellcolor{gray!25}GE,HD &\cellcolor{gray!25} 0.229 & &\\
		C	&\cellcolor{white}GE,JPM &\cellcolor{white} 0.214 & &\\
		C	&\cellcolor{white}AXP,HD &\cellcolor{white} 0.203 & &\\
		C	&\cellcolor{white}HD,JPM &\cellcolor{white} 0.182 & &\\
		\cmidrule{1-5}
		C,GE &\cellcolor{gray!25}HD,IBM &\cellcolor{gray!25} 0.164 & \multirow{4}{*}{			
			\begin{tikzpicture}	[every node/.style = ClassicalVineNode, node distance = 2.55cm, font = \scriptsize]
			%%%%% Tree 3 %%%%%
			\node[very thick] (CIBMGE)       {C,IBM;GE}
			node[very thick]  (GEHDC)         [below right of = CIBMGE]{GE,HD;C}											
			node[very thick] (AXPGEC)    [above right of = GEHDC] {AXP,GE;C}
			node[very thick]  (AXPJPMC)        [below right of = AXPGEC] {AXP,JPM;C}																    							;					
			\draw[very thick] (CIBMGE) to node[draw=none, text = black, fill = none, font = \labelsize, below, rotate = -45.5, yshift = -\yshiftlabel] {0.164} node[draw=none, text = black, fill = none, font = \labelsize, above, rotate = -45.5, yshift = \yshiftlabel] {\ \ \ \ HD,IBM;C,GE}(GEHDC);
			\draw[very thick] (CIBMGE) to node[draw=none, text = black, fill = none, font = \labelsize, below, rotate = 0, yshift = -\yshiftlabel] {0.163} node[draw=none, text = black, fill = none, font = \labelsize, above, rotate = 0, yshift = \yshiftlabel] {AXP,IBM;C,GE}(AXPGEC);
			\draw (GEHDC) to node[draw=none, text = black, fill = none, font = \labelsize, below, rotate = 0, yshift = -\yshiftlabel] {} node[draw=none, text = black, fill = none, font = \labelsize, above, rotate = 0, yshift = \yshiftlabel] {}(AXPGEC);
			\draw[very thick] (AXPGEC) to node[draw=none, text = black, fill = none, font = \labelsize, below, rotate = -45.5, yshift = -\yshiftlabel] {0.163} node[draw=none, text = black, fill = none, font = \labelsize, above, rotate = -45.5, yshift = \yshiftlabel] {\ \ \ \ GE,JPM;AXP,C}(AXPJPMC);		
			\end{tikzpicture}
		} & \multirow{4}{*}{$\mathcal{T}_3$}\\
		C,GE	&\cellcolor{gray!25}AXP,IBM &\cellcolor{gray!25} 0.163 & &\\
		AXP,C	&\cellcolor{gray!25}GE,JPM &\cellcolor{gray!25} 0.163 & &\\
		C,GE	&\cellcolor{white}AXP,HD &\cellcolor{white} 0.154 & &\\
		\cmidrule{1-5}
		C,GE,IBM	&\cellcolor{gray!25}AXP,HD &\cellcolor{gray!25} 0.129 & \multirow{2}{*}{
			\begin{tikzpicture}	[every node/.style = ClassicalVineNode, node distance = 2.65cm, font = \scriptsize]
			%%%%% Tree 4 %%%%%
			\node[very thick] (HDIBMCGE)         {HD,IBM;C,GE}
			node[very thick]  (AXPIBMCGE)         [right of = HDIBMCGE]{AXP,IBM;C,GE}	
			node[very thick]  (GEJPMAXPC)         [right of = AXPIBMCGE]{GE,JPM;AXP,C}					 				;
			;					
			\draw[very thick] (HDIBMCGE) to node[draw=none, text = black, fill = none, font = \labelsize, below, rotate = 0, yshift = -.2cm] {0.129} node[draw=none, text = black, fill = none, font = \labelsize, above, rotate = 0, yshift = .15cm] {AXP,HD;C,GE,IBM}(AXPIBMCGE);
			\draw[very thick] (AXPIBMCGE) to node[draw=none, text = black, fill = none, font = \labelsize, below, rotate = 0, yshift = -.2cm]  {0.121} node[draw=none, text = black, fill = none, font = \labelsize, above, rotate = 0, yshift = .15cm]  {IBM,JPM;AXP,C,GE}(GEJPMAXPC);		
			\end{tikzpicture}
		}&\multirow{2}{*}{$\mathcal{T}_4$}\\
		AXP,C,GE &\cellcolor{gray!25}IBM,JPM &\cellcolor{gray!25}0.121 & & \\
		\midrule
		AXP,C,GE,IBM	& \cellcolor{gray!25}HD,JPM &\cellcolor{gray!25} 0.093 &
		\multirow{2}{*}{\begin{tikzpicture}	[every node/.style = ClassicalVineNode, node distance = 3.75cm, font = \scriptsize]
			%%%%% Tree 5 %%%%%
			\node[very thick] (AXPHDCGEIBM)         {AXP,HD;C,GE,IBM}
			node[very thick]  (IBMJPMAXPCGE)         [right of = HDIBMCGE]{IBM,JPM;AXP,C,GE}					 				;
			;					
			\draw[very thick] (AXPHDCGEIBM) to node[draw=none, text = black, fill = none, font = \labelsize, below, rotate = 0, yshift = -.15cm] {0.093} node[draw=none, text = black, fill = none, font = \labelsize, above, rotate = 0, , yshift = .15cm] {HD,JPM;AXP,C,GE,IBM}(IBMJPMAXPCGE);	
			\end{tikzpicture}}
		&\multirow{2}{*}{$\mathcal{T}_5$}\\
		& & & & \\
		\hline\hline
	\end{tabular}%\vspace*{-1cm}
\end{table}
\renewcommand{\arraystretch}{1}
\normalsize

For the real data example, the proposed R-vine structure selection method is illustrated in \autoref{Table:SelectionMethodRealData}. Recall that the data include three market participants of financial sectors, namely AXP, C and JPM, IBM as an IT service, HD representing building materials trade and the diversified industrial corporation GE. As edge weights the empirical means of the realized (partial) correlation series based on all data points, i.e.\ $t=1,\ldots,2156$ (January 1, 2000 - July 30, 2008), are chosen. In $\mathcal{T}_1$, we start with a full graph, i.e.\ all edges are allowed to be chosen. Edge by edge a tree, i.e.\ a connected and acyclic graph, is built adding edges with the highest possible correlation mean. Including e.g.\ the pair (AXP,JPM) would result in a cycle and is thus not allowed. For $\mathcal{T}_2$ only edges satisfying the proximity condition given $\mathcal{T}_1$ are allowed. A tree is constructed by the four edges with the highest mean values of realized first order partial correlations, etc. The resulting R-vine structure captures the strong pairwise realized correlations between the three financial services in trees $\mathcal{T}_1$ and $\mathcal{T}_2$. From $\mathcal{T}_3$ on only realized partial correlations corresponding to stocks from different market sectors are modeled such as the one investigated in the last row of \autoref{fig:characteristicsTimeSeries}. Consequently, using in step (S1) an R-vine structure selected by the proposed algorithm will likely result in higher order realized partial correlation series, which allow for a parsimonious time-series model specification in step (S2).

\subsection{Step (S2): Multivariate time-series modeling and forecasting}\label{Subsec:Step2TimeSeries}
\textcolor{black}{After transforming the series of realized covariance matrices in step (S1), multivariate time-series models in step (S2) can be applied to the transformed data without imposing any parameter restrictions. Except for the considered modeling components this step does not differ from Cholesky decomposition based benchmark models. In both approaches, there are $d(d+1)/2$ model components after data transformation. In particular, the time-series of log-transformed realized variances and Fisher z-transformed realized (partial) correlations could be modeled using a VARFIMA model as suggested for the Cholesky elements in \cite{ChiriacVoev2011ForecastingMultVol}. Compared with this, copula based time-series modeling as applied in \cite{Brechmann2015Cholesky} showed superior results especially for economic applications.}

\textcolor{black}{A $\tilde{d}$-dimensional copula is a multivariate distribution function on $[0,1]^{\tilde{d}}$ with uniformly distributed margins. Since data are required to be approximately i.i.d., the copula model usually is not directly applied to the observed time-series, but to the corresponding standardized residuals $\left(\varepsilon_{1;t},\ldots,\varepsilon_{\tilde{d};t}\right)$, $t=1,\ldots,T$. The latter are extracted after fitting appropriate univariate time-series models to the original marginal data.  While no longer being subject to temporal dependence, the residuals inherit the cross-sectional dependence between the time-series components. According to \cite{Sklar59}, their joint distribution function $F$ can be expressed in terms of its marginal distributions $F_j$ ($j=1,\ldots,\tilde{d}$) and its corresponding copula, i.e.\ $F\left(\varepsilon_1,\ldots,\varepsilon_{\tilde{d}}\right) = \mathbb{C}\{F_1\left(\varepsilon_1\right),\ldots,F_{\tilde{d}}\left(\varepsilon_{\tilde{d}}\right)\}$. Consequently, in a copula based time-series model the individual behavior of the time-series components and their dependence are modeled separately. This allows us to deepen the analysis of the realized variance and (partial) correlation series.}

\subsubsection*{Marginal time-series modeling}  

\textcolor{black}{As discussed in \autoref{Subsec:DataCharacteristics} specific univariate time-series models are needed to reproduce the long-memory property of the Cholesky components as well as of the realized variance and some of the realized (partial) correlation series. HAR \citep{corsi2009simple} and ARFIMA \citep{andersen2003modeling} models are popular models capable of doing so.}

Let $\eta_t$ denote the variable of interest, i.e.\ a log-transformed realized variance, a Fisher z-transformed realized (partial) correlation or a Cholesky element, at time $t$. A basic HAR model accounts for different time horizons by incorporating one day ($d = 1$), one week ($w = 5$) and one month ($m = 22$) averages $\eta_{t-1}$, $\eta_{t-1}^{(w)}$ and $\eta_{t-1}^{(m)}$ as regressors for $\eta_t$: 
\begin{align*}
\eta_{t} = \alpha_{0} + \alpha_{1}\eta_{t-1} + \alpha_{2}\eta_{t-1}^{(5)} + \alpha_{3}\eta_{t-1}^{(22)} + \epsilon_{t}.
\end{align*}
The error term $\epsilon_{t}$ is usually assumed to be Gaussian white noise. While showing very good modeling and prediction performance given complex data features, the basic HAR model describes an easy to estimate restricted autoregressive process. 

\textcolor{black}{The ARFIMA(p, D, q) model for the time-series $\eta_t$, $t=1,\ldots,T$, is specified by}

\textcolor{black}{\begin{align*}
	\phi\left(L\right)\left(1-L\right)^D\left(\eta_{t} - \mu\right) = \psi\left(L\right)\epsilon_{t},
	\end{align*}
	where $\phi\left(L\right)=1-\phi_{1}L-\ldots-\phi_{p}L^{p}$ and $\psi\left(L\right)=1+\psi_{1}L+\ldots+\psi_{q}L^{q}$ are lag polynomials for $p, q \in \mathbb{N}$. $D$ is the parameter of fractional differencing. We choose $D \in \left(0, 0.5\right)$ to guarantee stationarity of the process. Gaussian white noise is usually assumed for the error term $\epsilon_{t}$.}

\textcolor{black}{In these basic models, the volatility $h$ of the error term $\epsilon_{t} = h\varepsilon_t$ with $\varepsilon_t \sim \mathcal{N}\left(0,1\right)$ is assumed to be constant. Given the presence of volatility clustering in the Cholesky series, \cite{Brechmann2015Cholesky} include a $\text{GARCH}\left(1, 1\right)$ component, i.e.\  $\epsilon_{t} = h_{t}\varepsilon_{t}$ with $h_{t}^{2} = \omega + \beta_{1}\epsilon_{t-1}^{2} + \beta_{2}h_{t-1}^{2}$. Usually, the innovation terms $\varepsilon_t$ are standard normally distributed, i.e.\ $\varepsilon_{t}\sim \mathcal{N}\left(0,1\right)$. To additionally capture possible high kurtosis and skewness, \cite{Brechmann2015Cholesky} further allow the innovations to follow a skewed generalized error distribution, i.e.\ $\varepsilon_{t} \sim \text{SGED}\left(\mu, \sigma, \nu, \xi \right)$ \citep{bai2003kurtosis,corsi2008volatility,fernandez1998bayesian}. A specification of the skewed generalized error distribution is provided in \autoref{Sec:AppSGED}.}

\subsubsection*{Dependence modeling}

\textcolor{black}{After fitting one of the above univariate time-series models to each of the model components, the sample of i.i.d.\ standardized residuals $\left(\varepsilon_{1;t}, \ldots, \varepsilon_{d(d+1)/2;t}\right)$, $t=1,\ldots,T$, can be extracted. To them, usually a two-stage proceeding is applied, called inference for margins methods \citep{joe1996IFM,joe2005asymptotic}. First, the probability integral transform, $\hat{u}_{j;t} = \hat{F}_j\left(\varepsilon_{j;t}\right)$, is applied to each residual component ($j=1,\ldots,d(d+1)/2$) to obtain pseudo copula data  $\left(\hat{u}_{1;t},\ldots,\hat{u}_{d(d+1)/2;t}\right)$, $t=1,\ldots,T$. The marginal estimates $\hat{F}_j$ are specified through the corresponding marginal time-series fit. In case of e.g.\ a basic HAR or ARFIMA model, $\hat{F}_j$ is a normal distribution with sample mean (approximately 0) and sample standard deviation (approximately 1).} 

\textcolor{black}{Second, a copula is fitted to the pseudo copula data. To do so, we consider R-vine copulas. Note that while in step (S1) and (S3) of the partial correlation vine data transformation approach, R-vines are exclusively used as a graph theoretical tool for data transformation, they are now the cornerstones for this flexible copula class. Given the limited choice of copula families for more than two variables, \cite{Joe1996} proposed to recursively decompose the multivariate copula density into a cascade of (conditional) bivariate copula densities through conditioning. Through R-vines the variety of possible pair-copula constructions can be organized. To each edge in an R-vine structure bivariate unconditional copulas in tree $\mathcal{T}_1$ and bivariate conditional copulas in trees $\mathcal{T}_2$ to $\mathcal{T}_{d-1}$ with arbitrary type and strength of dependence are assigned. This is allows to capture complex asymmetric and nonlinear dependence patterns. A detailed introduction to R-vine copulas is provided in \autoref{Sec:DepModelingVines}. Fitting an R-vine copula to the sample $\left(\hat{u}_{1;t},\ldots,\hat{u}_{d(d+1)/2;t}\right)$, $t=1,\ldots,T$, finalizes the model specification based on in-sample data.}

\subsubsection*{Forecasting of the model components}
\textcolor{black}{A one-day-ahead out-of-sample forecast $\hat{\boldsymbol{Y}}_{T+1}$ is now generated in multiple steps. First, innovations on the copula scale $\left(\hat{u}_{1;T+1},\ldots,\hat{u}_{d(d+1)/2;T+1}\right)$ are sampled from the R-vine copula fit. The corresponding innovations on their original scale are obtained using the inverse probability integral transform, i.e.\ $\hat{\varepsilon}_{j;T+1} = \hat{F}_j^{-1}\left(\hat{u}_{j;T+1}\right)$ ($j=1,\ldots,d(d+1)/2$). Then, based on the corresponding time-series fit forecasts for the model components, which involve the corresponding simulated innovations, are calculated. In the Cholesky decomposition based model, this results in a predicted upper triangular matrix $\hat{\boldsymbol{C}}_{T+1}$. In the partial correlation vine data transformation approach, the log-transformation of the realized variances and the Fisher z-transformation of the realized (partial) correlations have to be reversed first. This results in the predicted realized partial correlation vine stored in  $\hat{\boldsymbol{P}}_{\mathcal{C}\left(\mathcal{V}_{d}\right);T+1}$ and the corresponding predicted realized variance vector $\left(\hat{y}_{1,1;T+1},\ldots,\hat{y}_{d,d;T+1}\right)$.} 

\subsection{Step (S3): Back-transformation}\label{Sec:Step3Back}
\textcolor{black}{Finally, based on $\hat{\boldsymbol{C}}_{T+1}$ back-transformation \eqref{eq:CholeskyBackRecursion} is applied for the Cholesky decomposition based approach. Likewise,  $\hat{\boldsymbol{P}}_{\mathcal{C}\left(\mathcal{V}_{d}\right);T+1}$ is back-transformed to a symmetric and positive definite correlation matrix (\autoref{Sec:PartialCorrelationVines}) based on the R-vine structure selected in step (S1) (\autoref{Sec:R-VineStructureSelection}). Combined with the predicted realized variances a forecast for the realized covariance matrix is obtained. Given that both backward procedures involve nonlinear transformations of the copula-distributed innovation terms, the underlying dependence pattern has an explicit effect on the matrix forecast. Clearly, in practice this simulation based procedure is to be replicated several times. The final point-forecast  $\hat{\boldsymbol{Y}}_{T+1}$, which is considered as a proxy for the conditional covariance matrix, is obtained as the mean of the simulation based matrix forecasts.}

\textcolor{black}{In both modeling approaches the predictions of $\hat{\boldsymbol{Y}}_{T+1}$ are obtained after inverting a nonlinear data transformation. Consequently, while the prediction errors of the model components have zero mean, the nonlinear back-transformation induces a bias. Even though \cite{ChiriacVoev2011ForecastingMultVol} derive the theoretical bias correction for the Cholesky decomposition based model, they stress that the theoretical formula crucially depends on the considered time-series model and thus, has to be estimated in practice. However, given that in a copula based time-series model the marginal time-series are estimated independently of each other, consistent estimation of the covariance matrix of the forecast errors in \cite{Brechmann2015Cholesky} is not feasible. Against this background, \cite{ChiriacVoev2011ForecastingMultVol} and \cite{Brechmann2015Cholesky} both advocate a data-driven bias correction. In the partial correlation vine data transformation approach, the forecast bias of the variable pair $(i,j)$ in $\hat{\boldsymbol{Y}}_{T+1}$ depends not only on the underlying time-series model but also on the R-vine structure used for data transformation, making a theoretical correction practically infeasible. We therefore as well opt for the heuristic data-driven bias correction proposed in \cite{ChiriacVoev2011ForecastingMultVol}. The basic idea is to match the level of the observed volatilities by scaling the predicted volatilities $\sqrt{\hat{y}_{j,j;T+1}}$ ($j=1,\ldots,d$) by the corresponding mean $\frac{1}{T-s + 1}\sum_{t=s}^{T} \frac{\sqrt{y_{j,j;t}}}{\sqrt{\hat{y}_{j,j;t}}}$, where $s$ controls the number of past days included for level matching. Note that this proceeding has no influence on the predicted correlation structure. Thus, in the partial correlation vine data transformation approach only the nonlinear inversion of the log-transformation can be corrected.}

\subsection{Modeling approach at a glance}\label{Subsec:Guide}
\autoref{fig:FlowchartData3} summarizes the partial correlation vine data transformation approach discussed in the previous sections.
%\newpage

\tikzstyle{ClassicalNode} = [rectangle, fill = lightgray!43, draw = black, text = black, align = center, minimum height=1.00cm]
\renewcommand{\yshift}{-1cm}
\renewcommand{\yshiftlabel}{-0.1cm}
\renewcommand{\labelsize}{\small}
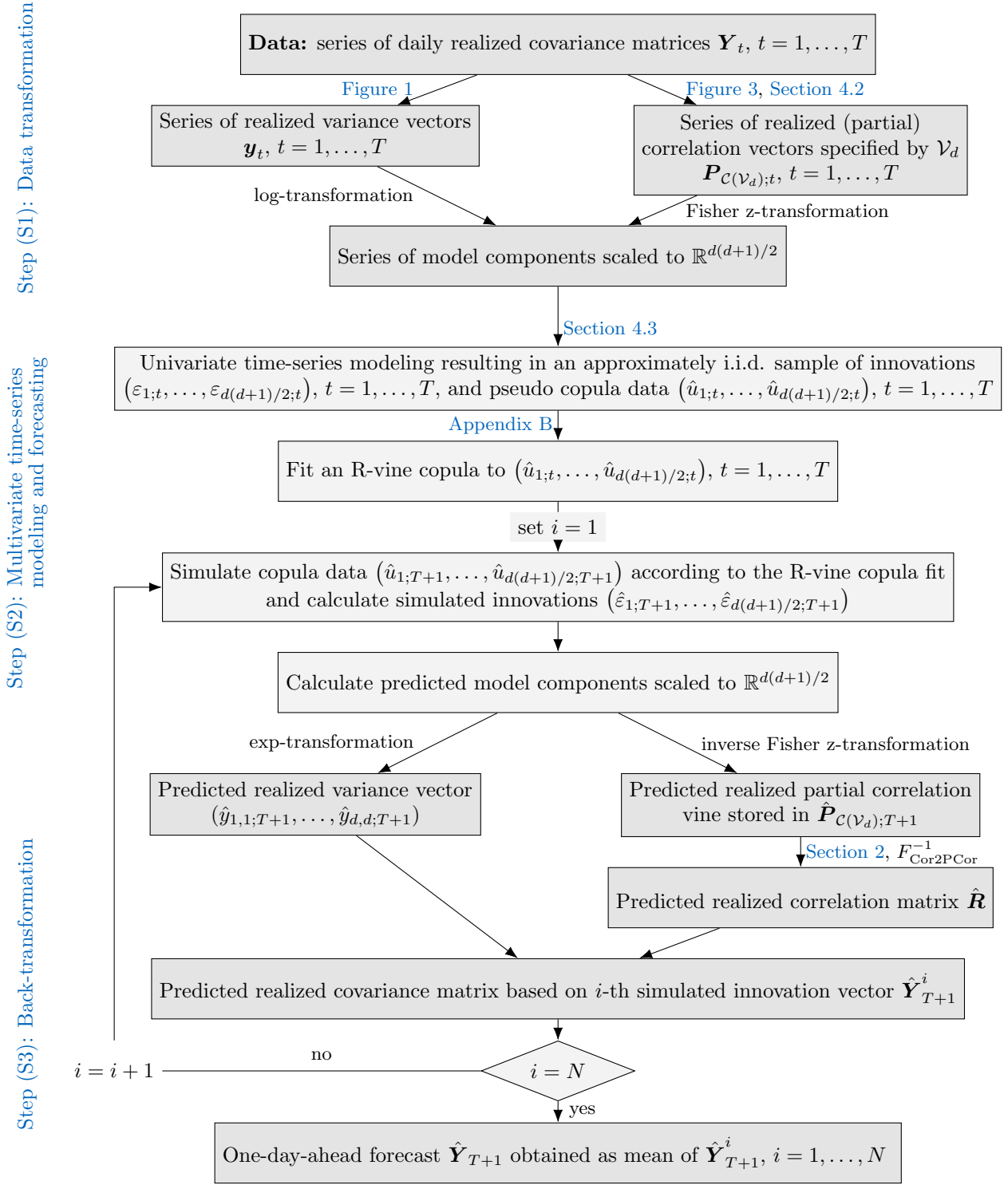
\begin{figure}[p!]
	\centering\vspace*{-2.50cm}
	\hspace*{-1.25cm}\begin{tikzpicture}	[every node/.style = ClassicalNode, node distance = 1cm, font = \small]
	\node (Data) {\textbf{Data:} series of daily realized covariance matrices
		$\boldsymbol{Y}_t$, $t=1,\ldots,T$};
	\node [below =  of Data, xshift = -4cm, yshift = .5cm] (varVectors) {
		Series of realized variance vectors\\ $\boldsymbol{y}_t$, $t=1,\ldots,T$};
	\node[fill = white, draw = none, text = TUMblue] (StepS1)        [left of = varVectors, xshift = -3.75cm, yshift = -0.20cm] {\begin{sideways} Step (S1): Data transformation \end{sideways}};
	\node[fill = white, draw = none, text = TUMblue] (StepS2)        [below of = StepS1, yshift = -5.225cm] {\begin{sideways} Step (S2): Multivariate time-series \end{sideways}\newline \begin{sideways} \phantom{Step (S2):} modeling and forecasting \end{sideways}};
	\node[fill = white, draw = none, text = TUMblue] (StepS3)        [below of = StepS2, yshift = -6.67cm] {\begin{sideways} Step (S3): Back-transformation \end{sideways}};
	\node [below =  of Data, xshift = 4cm, yshift = .5cm] (PCorVineVectors) {
		Series of realized (partial)\\ correlation vectors specified by $\mathcal{V}_{d}$\\
		$\boldsymbol{P}_{\mathcal{C}\left(\mathcal{V}_{d}\right);t}$, $t=1,\ldots,T$};
	\node [below =  of PCorVineVectors, xshift = -4cm, yshift = .5cm] (jointData) {
		Series of model components scaled to $\mathbb{R}^{d(d+1)/2}$};
	%$\boldsymbol{P}'_t \coloneqq \big(p_{1;t}, \ldots, p_{d(d+1)/2;t}\big)'$, $t=1,\ldots,T$};	
	\node [fill = lightgray!19, below =  of jointData] (margins) {
		Univariate time-series modeling resulting in an approximately i.i.d. sample of innovations\\ $\left(\varepsilon_{1;t},\ldots,\varepsilon_{d(d+1)/2;t}\right)$, $t=1,\ldots,T$, and pseudo copula data $\left(\hat{u}_{1;t},\ldots,\hat{u}_{d(d+1)/2;t}\right)$, $t=1,\ldots,T$
	};
	\node [fill = lightgray!19,below =  of margins, yshift = .5cm] (dep) {Fit an R-vine copula to $\left(\hat{u}_{1;t},\ldots,\hat{u}_{d(d+1)/2;t}\right)$, $t=1,\ldots,T$};
	\node [draw = none, fill = white, fill = lightgray!19,below =  of dep, yshift = .83cm,minimum height=0.00cm] (sim) {set $i = 1$};
	\node [fill = lightgray!19, below =  of sim, yshift = .83cm] (forecast1) {
		Simulate copula data $\left(\hat{u}_{1;T+1},\ldots,\hat{u}_{d(d+1)/2;T+1}\right)$ according to the R-vine copula fit\\ and
		calculate simulated innovations $\left(\hat{\varepsilon}_{1;T+1},\ldots,\hat{\varepsilon}_{d(d+1)/2;T+1}\right)$};
	\node [fill = lightgray!19, below =  of forecast1, yshift = .5cm] (forecast2) {
		Calculate predicted model components scaled to $\mathbb{R}^{d(d+1)/2}$};	
	\node [below =  of forecast2, xshift = -4cm] (varFC) {
		Predicted realized variance vector\\ 
		$\left(\hat{y}_{1,1;T+1}, \ldots, \hat{y}_{d,d;T+1}\right)$};
	\node [below =  of forecast2, xshift = +4cm] (PCorVineFC) {
		Predicted realized partial correlation\\ vine stored in $\hat{\boldsymbol{P}}_{\mathcal{C}\left(\mathcal{V}_{d}\right);T+1}$};	
	\node [below =  of PCorVineFC, yshift = .5cm] (CorFC) {
		Predicted realized correlation matrix $\hat{\boldsymbol{R}}$};
	\node [below =  of CorFC, xshift = -4cm, yshift = .5cm] (VarCovFC) {
		Predicted realized covariance matrix based on $i$-th simulated innovation vector  
		$\hat{\boldsymbol{Y}}^i_{T+1}$};	
	\node [diamond, fill = lightgray!19, aspect = 2.75, below = of VarCovFC, yshift = .65cm](done){$i = N$};
	\node [draw = none, fill = white, left = of done, xshift = -4.25cm](increment){$i = i + 1$};
	\node [below =  of done, yshift = .65cm] (finalFC) {
		One-day-ahead forecast $\hat{\boldsymbol{Y}}_{T+1}$ obtained as mean of $\hat{\boldsymbol{Y}}^i_{T+1}$, $i=1,\ldots,N$
	};
	
	\draw[-{Latex[scale=1.25]}] (Data) -- node[draw=none, fill = none, left, xshift = -.25cm] {{\footnotesize \autoref{fig:FlowchartData1}}} (varVectors);
	\draw[-{Latex[scale=1.25]}] (Data) -- node[draw=none, fill = none, right, xshift = +.25cm] {{\footnotesize \autoref{fig:FlowchartData2}, \autoref{Sec:R-VineStructureSelection}}} (PCorVineVectors);
	\draw[-{Latex[scale=1.25]}] (varVectors) -- node[draw=none, fill = none, left, xshift = -.25cm] {{\footnotesize log-transformation}} (jointData);
	\draw[-{Latex[scale=1.25]}] (PCorVineVectors) -- node[draw=none, fill = none, right, xshift = +.25cm] {{\footnotesize Fisher z-transformation}} (jointData);
	\draw[-{Latex[scale=1.25]}] (jointData) -- node[draw=none, fill = none, right, xshift = -.05cm, yshift = -.19cm] {{\footnotesize \autoref{Subsec:Step2TimeSeries}}} (margins);
	\draw[-{Latex[scale=1.25]}] (margins) -- node[draw=none, fill = none, left, xshift = .05cm] {{\footnotesize \autoref{Sec:DepModelingVines}}} (dep);
	\draw (dep) -- node[draw=none, fill = none, left] {} (sim);
	\draw[-{Latex[scale=1.25]}] (sim) -- node[draw=none, fill = none, left] {} (forecast1);
	\draw[-{Latex[scale=1.25]}] (forecast1) -- node[draw=none, fill = none, left] {} (forecast2);
	\draw[-{Latex[scale=1.25]}] (forecast2) -- node[draw=none, fill = none, left, xshift = -.25cm] {{\footnotesize exp-transformation}} (varFC);
	\draw[-{Latex[scale=1.25]}] (forecast2) -- node[draw=none, fill = none, right, xshift = +.25cm] {{\footnotesize inverse Fisher z-transformation}} (PCorVineFC);
	\draw[-{Latex[scale=1.25]}] (PCorVineFC) -- node[draw=none, fill = none, right, xshift = -.05cm] {{\footnotesize \autoref{Sec:PartialCorrelationVines}, $F_{\text{Cor2PCor}}^{-1}$}} (CorFC);
	\draw[-{Latex[scale=1.25]}] (varFC) -- node[draw=none, fill = none, right] {} (VarCovFC);
	\draw[-{Latex[scale=1.25]}] (CorFC) -- node[draw=none, fill = none, right] {} (VarCovFC);
	\draw[-{Latex[scale=1.25]}] (VarCovFC) -- node[draw=none, fill = none, above, yshift = -.25cm] {} (done);
	\draw (done) -- node[draw=none, fill = none, above, yshift = -.25cm] {{\footnotesize no}} (increment);
	\draw[-{Latex[scale=1.25]}] (increment) |- node[draw=none, fill = none, left] {} (forecast1);
	\draw[-{Latex[scale=1.25]}] (done) -- node[draw=none, fill = none, right, xshift = +.05cm] {{\footnotesize yes}} (finalFC);
	\end{tikzpicture}
	\caption{Modeling and forecasting approach using partial correlation vine based data transformation of the series of realized covariance matrices in step (S1) and an R-vine copula based time-series model in step (S2). The one-day-ahead forecast $\hat{\boldsymbol{Y}}_{T+1}$ is obtained as mean of $N$ simulation based matrix forecasts.}
	\label{fig:FlowchartData3}
	\vspace*{-1.5cm}
\end{figure} 

\section{Empirical study}\label{Sec:EmpiricalStudy}	
\textcolor{black}{The real data example introduced in \autoref{Sec:DataSettingBenchmark} and \autoref{Sec:PCVapproach} will now be investigated in more detail. Based on the model specifications in \autoref{Sec:EmpStepS1} and \autoref{Sec:EmpStepS2}, the out-of-sample forecasting performance of the partial correlation vine data transformation approach and Cholesky decomposition based benchmark models will be evaluated both with respect to statistical precision and mean-variance trade-off in portfolio optimization strategies.}

\textcolor{black}{It is crucial to keep in mind that the realized covariance matrices are proxies for the unobservable true conditional covariance matrices, which we aim to predict. As a consequence, when comparing the performance of different forecasting models, loss functions have to satisfy the condition to deliver the same ranking whether the evaluation is based on the unbiased proxy, i.e.\ the realized covariance matrix, or the true conditional covariance matrix. We will therefore rely on loss functions, which according to \cite{patton2011volatility} and \cite{laurent2013loss} are robust to noise in the volatility proxies. Further, numerous different models for prediction will be compared. To avoid pairwise comparison of loss functions we apply the model confidence set (MCS) approach developed by \cite{hansen2011model}. It allows to select a set of superior models, which contains the best one with a specified level of confidence.}

\subsection{Moving window approach}
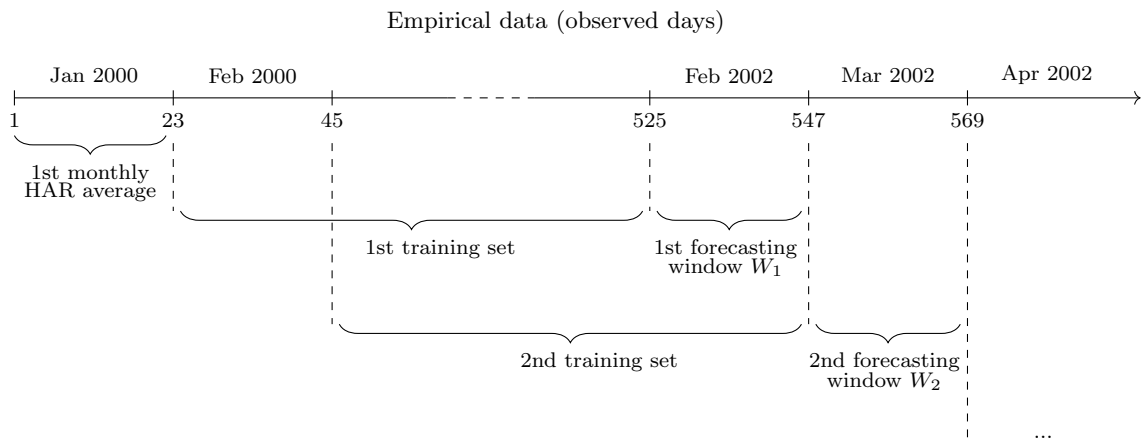
\begin{figure}[h]
	\tikzstyle{letty} = [line width=0.25mm, TUMblue]
	\begin{tikzpicture}[x=.095cm, y = .05cm,]
	\def \shift {6}	
	\def \shiftTik {2}	
	
	\def \orx {0} 
	\def \endx {150}

	\def \ory {0}
	\def \endy {-100}
	
	\def \facTime {3}
	\def \daysInMonth {22}
	
	% vertical lines
	\draw[-] ($(\orx,\ory)$) to ($(\orx+2.5*22+5,\ory)$);
	\draw[-, dashed] ($(\orx+2.5*22+5,\ory)$) to ($(\orx+3.5*22-5,\ory)$);
	\draw[->] ($(\orx+3.5*22-5,\ory)$) to ($(\endx+\shift,\ory)$);
	%\node at (\orx - 20, \ory+4)[font=\normalsize] {Empirical}; 
	%\node at (\orx - 20, \ory-4)[font=\normalsize] {data}; 
	\node at (75, \ory + 5*4)[font=\footnotesize] {Empirical data (observed days)};
	
	%%% HAR
	\node at (\orx, \ory-\shift)[font=\scriptsize] {1};
	\draw[-] (\orx, \ory-\shiftTik) to (\orx, \ory+\shiftTik);
	%\node at (21, \ory-\shift)[font=\scriptsize] {22};

	%%% block 1
	\node at (22, \ory-\shift)[font=\scriptsize] {23};
	\draw[-] (22, \ory-\shiftTik) to (22, \ory+\shiftTik);
	\node at (4*22, \ory-\shift)[font=\scriptsize] {525};
	\draw[-] (4*22, \ory-\shiftTik) to (4*22, \ory+\shiftTik);
	\node at (5*22, \ory-\shift)[font=\scriptsize] {547};
	\draw[-] (5*22, \ory-\shiftTik) to (5*22, \ory+\shiftTik);
	\node at (6*22, \ory-\shift)[font=\scriptsize] {569};
	\draw[-] (6*22, \ory-\shiftTik) to (6*22, \ory+\shiftTik);
	
	\node at (0*22 + 11, \ory+\shift)[font=\scriptsize] {Jan 2000};
	\node at (1*22 + 11, \ory+\shift)[font=\scriptsize] {Feb 2000};
	\node at (4*22 + 11, \ory+\shift)[font=\scriptsize] {Feb 2002};
	\node at (5*22 + 11, \ory+\shift)[font=\scriptsize] {Mar 2002};
	\node at (6*22 + 11, \ory+\shift)[font=\scriptsize] {Apr 2002};
	
	%%% block 2
	\node at (2*22, \ory-\shift)[font=\scriptsize] {45};
	\draw[-] (2*22, \ory-\shiftTik) to (2*22, \ory+\shiftTik);
	%\node at (3*22, \ory-\shift)[font=\scriptsize] {67};
	%\draw[-] (3*22, \ory-\shiftTik) to (3*22, \ory+\shiftTik);
	
	%\node at ($(22+\facTime*\daysInMonth + 1*\daysInMonth, \ory-\shift)$)[font=\scriptsize] {525};
	%\node at ($(22+\facTime*\daysInMonth + 2*\daysInMonth, \ory-\shift)$)[font=\scriptsize] {547};

	%\textsc{
	
	% dashed lines  
	\draw[-, dashed] (22, \ory-2*\shift) to (22, - 30);	
	\draw[-, dashed] (4*22, \ory-2*\shift) to (22+\facTime*\daysInMonth, - 30);
	\draw[-, dashed] (5*22, \ory-2*\shift) to (5*22, - 60);
	\draw[-, dashed] (2*22, \ory-2*\shift) to (2*22, - 60);
	\draw[-, dashed] (6*22, \ory-2*\shift) to (6*22, - 90);

	\draw [decorate, decoration = {brace, amplitude = 7pt, mirror}](\orx, -10)  to (21,-10);
	\node at (10.5, -20)[font=\scriptsize] {1st monthly};
	\node at (10.5, -25)[font=\scriptsize] {HAR average};
	
	\draw[decorate, decoration = {brace, amplitude = 7pt, mirror}](23, -30)  to (4*22-1,-30);
	\node at (2*22+15, -40)[font=\scriptsize] {1st training set};
	\draw[decorate, decoration = {brace, amplitude = 7pt, mirror}](4*22+1, -30)  to (5*22-1,-30);
	\node at (4*22+10.5, -40)[font=\scriptsize] {1st forecasting};
	\node at (4*22+10.5, -45)[font=\scriptsize] {window $W_1$};
	\draw[decorate, decoration = {brace, amplitude = 7pt, mirror}](2*22+1, -60)  to (5*22-1,-60);
	\node at (3*22+15, -70)[font=\scriptsize] {2nd training set};
	\draw[decorate, decoration = {brace, amplitude = 7pt, mirror}](5*22+1, -60)  to (6*22-1,-60);
	\node at (5*22+10.5, -70)[font=\scriptsize] {2nd forecasting};
	\node at (5*22+10.5, -75)[font=\scriptsize] {window $W_2$};
	
	\node at (6*22+10.5, -90)[font=\scriptsize] {...};
	
	% dashed lines  
	%\foreach \facx/\lbl/\x in {.6575//-.4,.6975//.4} {
	%	\draw[-, TUMorange, dashed] ($(\orxr+\endx*\facx, \ory)$) to ($(\orxr+\endx*\facx, \endy)$);	
	%}
	\end{tikzpicture}
	\caption{Moving window approach illustrated for the considered real data example.}
	\label{fig:MovingWindowApproach}
\end{figure}

In the following, we proceed in a moving window approach. Data for the period from January 1, 2000, until June 30, 2008, are available, i.e.\ for 2156 days. For each time window 502 days (about two years) are used as training set and 22 days (about one month) constitute the test set for which one-day-ahead forecasts are made. Since in case of HAR based time-series models a monthly (22 days) average of the data is involved, the first forecast is obtained for day 525. In total, there are 75 time windows. \autoref{fig:MovingWindowApproach} illustrates the moving window approach.

\subsection{Step (S1): Dynamic data transformation}\label{Sec:EmpStepS1}
\textcolor{black}{For each time window $W_i$ ($i=1,\ldots,75$) the realized covariance matrices of the corresponding training set are transformed in step (S1). Clearly, application of the R-vine structure selection algorithm proposed in \autoref{Sec:R-VineStructureSelection} can lead to varying R-vine structures among time windows. Thus, data transformation in the partial correlation vine data transformation approach may dynamically change over time. Depending on how the average correlation matrix used for R-vine structure selection is calculated, the selected R-vine structure is more or less sensitive to market developments.} 
\newpage

\begin{figure}[h!]
	\centering\vspace*{.25cm}
	{\footnotesize edge weights based on empirical means}
	\includegraphics[width=.85\linewidth]{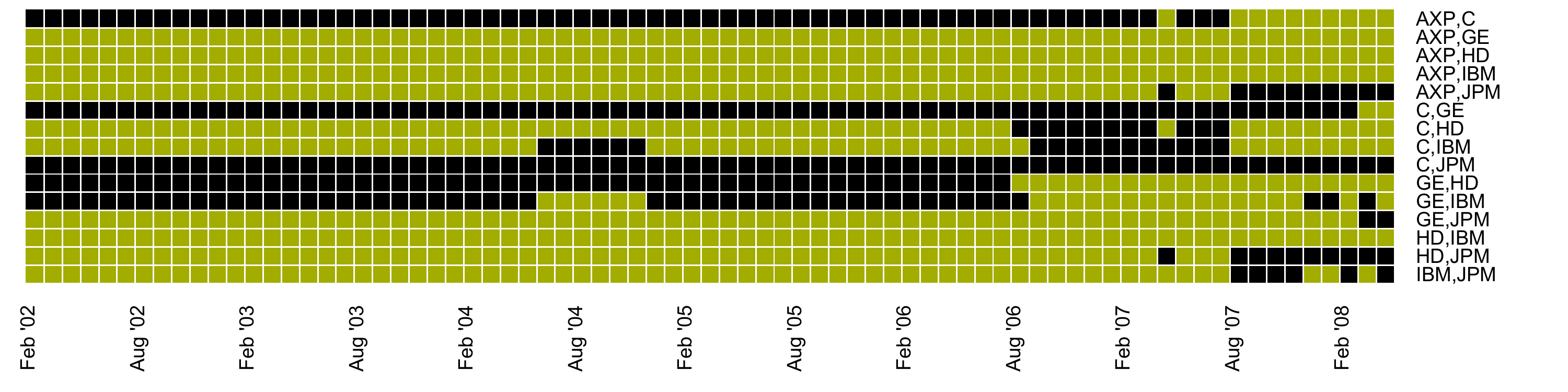}\\
	{\footnotesize edge weights based on EWMA with $\lambda = 0.995$}
	\includegraphics[width=.85\linewidth]{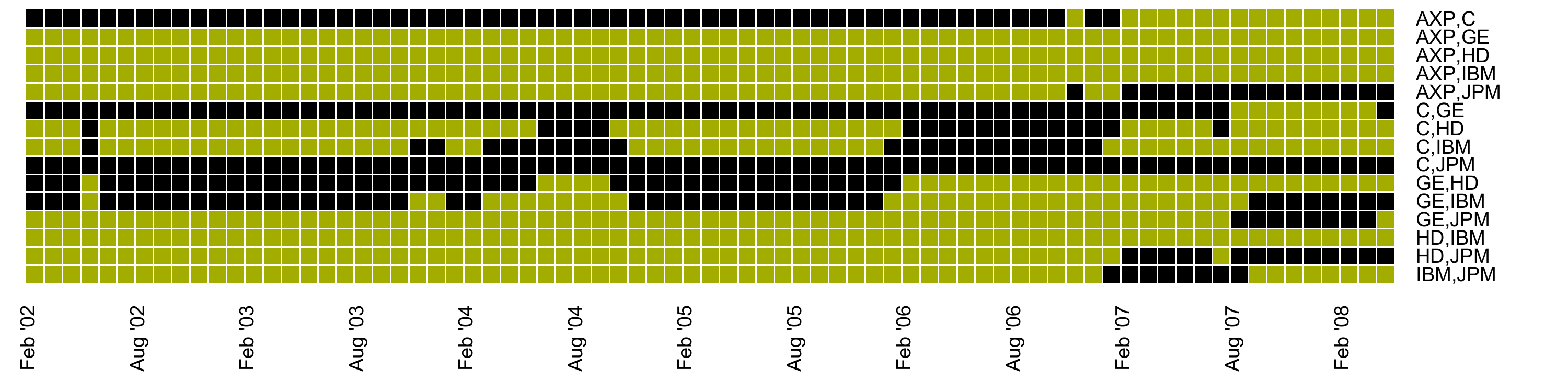}\\
	{\footnotesize edge weights based on EWMA with $\lambda = 0.98$}
	\includegraphics[width=.85\linewidth]{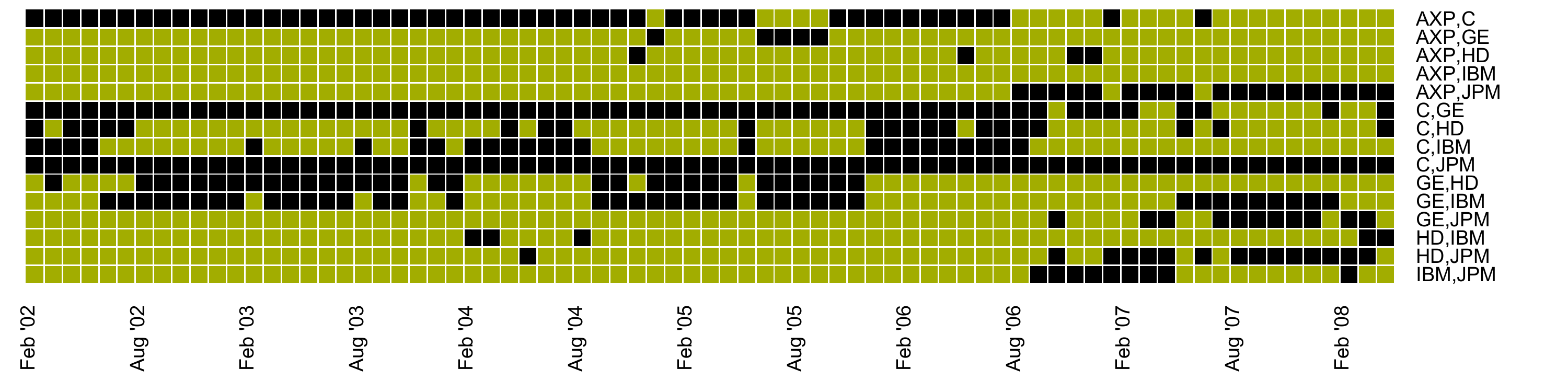}
	\renewcommand{\arraystretch}{0.75}
	\begin{tabular}{p{.5cm}p{.001cm}p{3.5cm}p{.001cm}p{3cm}}
		& \cellcolor{TUMgreen} & {\scriptsize allowed but not selected}
		& \cellcolor{black} & {\scriptsize selected}\\
		\vspace{-3.00cm}
	\end{tabular}	
	\renewcommand{\arraystretch}{1}
	\caption{Illustration of the dynamically changing R-vine structures used for transforming the series of realized correlation matrices in each of the 75 time windows. The horizontal time axis states the prediction month. Model components selected in tree $\mathcal{T}_1$ are indicated by black squares. Green squares indicate selectable components allowed by the proximity condition (which does not trigger in $\mathcal{T}_1$). In the first panel, the average correlations used for R-vine structure selection are the empirical means of the training set data. In the second and third row, they are exponentially weighted moving averages with $\lambda = 0.995$ and $\lambda = 0.98$, respectively.}
	\label{fig:selected251t1}
	%	\vspace*{-1cm}
\end{figure}

In \autoref{fig:selected251t1}, the first trees of the R-vine structures selected in each of the 75 time windows are shown indicating the included model components by a black square. In the first row, empirical means of the realized standard correlations are considered. In the second and third row, exponentially weighted moving averages based on $\lambda = 0.995$ and $\lambda = 0.98$, respectively, are used. While in case of empirical means all days of the two training years are of equal weight, for $\lambda = 0.995$ and $\lambda = 0.98$ the six most recent months and one and a half months, respectively, already contribute half of the information for average calculation. Thus, in the latter case R-vine structure selection is most sensitive to market changes resulting in more frequent variations of the selected model components. Changes e.g.\ for the prediction months in mid 2004 or at the beginning of the financial crisis are observed earliest. Nevertheless, for all three setups the selected first tree is quite stable and we may identify three distinct periods:  February 2002--August 2006, September 2006--July 2007 and August 2007--July 2008. For these periods, \autoref{fig:predominantStructuresData} illustrates the first tree $\mathcal{T}_1$ of the predominantly chosen R-vine structures. Until August 2006, pairwise correlations between the log-returns including Citigroup (C) and General Electric (GE) seem to be most pronounced. While C plays a key role within the financial sector, GE as a diversified industrial corporation connects the representatives of the financial sector with the two non-financial stocks. During the period from September 2006 to July 2007 C becomes the first root node in a C-vine, i.e.\ the node in $\mathcal{T}_1$ with the highest possible number of edges attached to it. At the beginning of the financial crisis in August 2007, the correlations between JP Morgan (JPM) and the other market participants seem to tighten. This results in a predominantly chosen R-vine structure, where except for GE all pairwise correlations including JPM are modeled. Note that in 2007 JPM replaced C as the biggest US-bank in terms of revenues.

\textcolor{black}{To conclude, selecting the R-vine structure for data transformation as proposed in \autoref{Sec:R-VineStructureSelection} gives interesting insights into market activities over time. In addition, we already know about the resulting inhomogeneous data complexity of the corresponding time-series, which will be further analyzed in the next section. There, the R-vine structures will be selected using EWMA based edge weights with $\lambda = 0.995$. Recall, however, that any R-vine structure could be used for data transformation. To demonstrate the general adequacy of the partial correlation vine data transformation approach irrespective of the R-vine structure used for data transformation, we will consider two alternative ways of R-vine structure selection as well. First, we reverse the idea of inducing model parsimony and select for each time window a C-vine, where in each tree level the root node induces the on average lowest correlation strength. Thus, the effect of decreasing data complexity as for the proposed R-vine structure selection should be eliminated. Second, an R-vine structure on 6 elements is randomly sampled in each time window \citep{joe2010regular}.}

\noindent

\tikzstyle{ClassicalVineNode} = [rectangle, fill = lightgray!43, draw = black, text = black, align = center, minimum height = 0.75cm, minimum width = 0.75cm]
\tikzstyle{TreeLabels} = [draw = none, fill = none, text = black, font = \bfseries]
\tikzstyle{DummyNode}  = [draw = none, fill = none, text = white]
\renewcommand{\yshiftTreeLabel}{.8cm}
\renewcommand{\labelsizeRVine}{\footnotesize}
\renewcommand{\shiftlabel}{-.05cm}  
\renewcommand{\dummyxShift}{.25cm}					

\begin{figure}[h!]\vspace*{1.00cm}	
	\centering
	\begin{minipage}{.32\linewidth}
		\begin{center}
			{\small	Feb '02 -- Aug '06\vspace{.1cm}}	
			
			\begin{tikzpicture}	[every node/.style = ClassicalVineNode, node distance = 1.75cm, font = \small]
			
			%%%%% Tree 1 %%%%%
			\node (C){C}
			node             (AXP)         [above left of = C] {AXP}			
			node             (JPM)         [above right of = C] {JPM}
			node             (GE)         [below left of = C] {GE}
			node             (HD)         [below right of = GE] {HD}
			node             (IBM)        [above right of =HD] {IBM}
			%node[TreeLabels] (T1)        [above  of = C, yshift = -\yshiftTreeLabel] {$\mathcal{T}_1$}
			;
			
			%		\draw (C) to node[draw=none, text = black, fill = none, font = \labelsizeRVine, left, xshift = \shiftlabel] {$\lbrace{\rho_{AXP,C;t}}\rbrace$} (AXP);
			%		\draw (C) to node[draw=none, text = black, fill = none, font = \labelsizeRVine, right, xshift = -1*\shiftlabel] {$\lbrace{\rho_{C,JPM;t}}\rbrace$} (JPM);
			%		\draw (C) to node[draw=none, text = black, fill = none, font = \labelsizeRVine, left, xshift = \shiftlabel] {$\lbrace{\rho_{C,GE;t}}\rbrace$} (GE);
			%		\draw (GE) to node[draw=none, text = black, fill = none, font = \labelsizeRVine, left, xshift = \shiftlabel] {$\lbrace{\rho_{GE,HD;t}}\rbrace$} (HD);
			%		\draw (GE) to node[draw=none, text = black, fill = none, font = \labelsizeRVine, above, yshift = -\shiftlabel] {$\lbrace{\rho_{GE,IBM;t}}\rbrace$} (IBM); 
			\draw (C) to node[draw=none, text = black, fill = none, font = \labelsizeRVine, left, xshift = \shiftlabel] {} (AXP);
			\draw (C) to node[draw=none, text = black, fill = none, font = \labelsizeRVine, right, xshift = -1*\shiftlabel] {} (JPM);
			\draw (C) to node[draw=none, text = black, fill = none, font = \labelsizeRVine, left, xshift = \shiftlabel] {} (GE);
			\draw (GE) to node[draw=none, text = black, fill = none, font = \labelsizeRVine, left, xshift = \shiftlabel] {} (HD);
			\draw (GE) to node[draw=none, text = black, fill = none, font = \labelsizeRVine, above, yshift = -\shiftlabel] {} (IBM); 
			\end{tikzpicture}
		\end{center}
	\end{minipage}
	%	\hfill
	\begin{minipage}{.32\linewidth}	
		\begin{center}
			{\small	Sep '06 -- Jul '07\vspace{.1cm}}	
			
			\begin{tikzpicture}	[every node/.style = ClassicalVineNode, node distance = 1.75cm, font = \small]
			
			%%%%% Tree 1 %%%%%
			\node (C){C}
			node             (AXP)         [above left of = C] {AXP}			
			node             (JPM)         [above right of = C] {JPM}
			node             (GE)         [below left of = C] {GE}
			node             (HD)         [below right of = GE] {HD}
			node             (IBM)         [above right of = HD] {IBM}
			%node[TreeLabels] (T1)        [above  of = C, yshift = -\yshiftTreeLabel] {$\mathcal{T}_1$}
			;
			
			%		\draw (C) to node[draw=none, text = black, fill = none, font = \labelsizeRVine, left, xshift = \shiftlabel] {$\lbrace{\rho_{AXP,C;t}}\rbrace$} (AXP);
			%		\draw (C) to node[draw=none, text = black, fill = none, font = \labelsizeRVine, right, xshift = -\shiftlabel] {$\lbrace{\rho_{C,JPM;t}}\rbrace$} (JPM);
			%		\draw (C) to node[draw=none, text = black, fill = none, font = \labelsizeRVine, left, xshift = \shiftlabel] {$\lbrace{\rho_{C,GE;t}}\rbrace$} (GE);
			%		\draw (C) to node[draw=none, text = black, fill = none, font = \labelsizeRVine, right, xshift = -1.5*\shiftlabel, yshift = -3*\shiftlabel] {$\lbrace{\rho_{C,HD;t}}\rbrace$} (HD);
			%		\draw (C) to node[draw=none, text = black, fill = none, font = \labelsizeRVine, right, xshift = -\shiftlabel] {$\lbrace{\rho_{C,IBM;t}}\rbrace$} (IBM); 
			\draw (C) to node[draw=none, text = black, fill = none, font = \labelsizeRVine, left, xshift = \shiftlabel] {} (AXP);
			\draw (C) to node[draw=none, text = black, fill = none, font = \labelsizeRVine, right, xshift = -\shiftlabel] {} (JPM);
			\draw (C) to node[draw=none, text = black, fill = none, font = \labelsizeRVine, left, xshift = \shiftlabel] {} (GE);
			\draw (C) to node[draw=none, text = black, fill = none, font = \labelsizeRVine, right, xshift = -1.5*\shiftlabel, yshift = -3*\shiftlabel] {} (HD);
			\draw (C) to node[draw=none, text = black, fill = none, font = \labelsizeRVine, right, xshift = -\shiftlabel] {} (IBM); 
			
			\end{tikzpicture}
		\end{center}
	\end{minipage}
	%\hfill
	\begin{minipage}{.32\linewidth}
		\begin{center}
			{\small	Aug '07 -- Jul '08\vspace*{-.3cm}}	
			
			\begin{tikzpicture}	[every node/.style = ClassicalVineNode, node distance = 1.75cm, font = \small]
			
			%%%%% Tree 1 %%%%%
			\node (C){C}
			node             (AXP)         [above left of = C] {AXP}			
			node             (JPM)         [above right of = C] {JPM}
			node             (GE)         [below left of = C] {GE}
			node             (HD)         [below right of = GE] {HD}
			node             (IBM)         [above right of = HD] {IBM}
			%node[TreeLabels] (T1)        [above  of = C, yshift = -\yshiftTreeLabel] {$\mathcal{T}_1$}
			;
			
			%		\draw (JPM) to node[draw=none, text = black, fill = none, font = \labelsizeRVine, above, yshift = -\shiftlabel] {$\lbrace{\rho_{AXP,JPM;t}}\rbrace$} (AXP);
			%		\draw (C) to node[draw=none, text = black, fill = none, font = \labelsizeRVine, left, xshift = 1.5*\shiftlabel] {$\lbrace{\rho_{C,JPM;t}}\rbrace$} (JPM);
			%		\draw (C) to node[draw=none, text = black, fill = none, font = \labelsizeRVine, left, xshift = 1*\shiftlabel] {$\lbrace{\rho_{C,GE;t}}\rbrace$} (GE);
			%		\draw (JPM) to node[draw=none, text = black, fill = none, font = \labelsizeRVine, left, yshift = -6*\shiftlabel] {$\lbrace{\rho_{HD,JPM;t}}\rbrace$} (HD);
			%		\draw (JPM) to node[draw=none, text = black, fill = none, font = \labelsizeRVine, right, xshift = -2*\shiftlabel] {$\lbrace{\rho_{IBM,JPM;t}}\rbrace$} (IBM); 
			\draw (JPM) to node[draw=none, text = black, fill = none, font = \labelsizeRVine, above, yshift = -\shiftlabel] {} (AXP);
			\draw (C) to node[draw=none, text = black, fill = none, font = \labelsizeRVine, left, xshift = 1.5*\shiftlabel] {} (JPM);
			\draw (C) to node[draw=none, text = black, fill = none, font = \labelsizeRVine, left, xshift = 1*\shiftlabel] {} (GE);
			\draw (JPM) to node[draw=none, text = black, fill = none, font = \labelsizeRVine, left, yshift = -6*\shiftlabel] {} (HD);
			\draw (JPM) to node[draw=none, text = black, fill = none, font = \labelsizeRVine, right, xshift = -2*\shiftlabel] {} (IBM); 		
			\end{tikzpicture}
		\end{center}
	\end{minipage}
	\caption{Predominantly selected first tree of the R-vine structure used for data transformation during the periods February 2002 to August 2006, September 2006 to July 2007 and August 2007 to July 2008.}			\label{fig:predominantStructuresData}			
\end{figure}
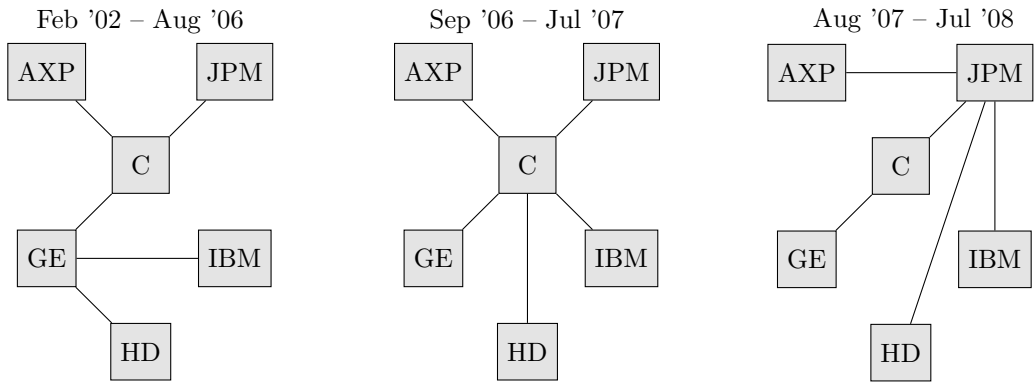\vspace*{0.79cm}
\newpage

\textcolor{black}{For the Cholesky decomposition, the model components depend on the ordering of the assets. However, as opposed to the data transformation based on partial correlation vines there is no justifiable rule to decide `on the fly' for a specific order. Thus, the ordering has to be set upfront. In this sense, the Cholesky decomposition based data transformation is static. As the only way to proceed, \cite{Brechmann2015Cholesky} test all possible permutations for the underlying data. Coincidently, the alphabetic ordering of the six stocks performs best. We therefore, choose the latter for all time windows.}

\subsection{Step (S2): Multivariate time-series modeling}\label{Sec:EmpStepS2}
\textcolor{black}{As explained in \autoref{Subsec:Step2TimeSeries}, to the sample of model components obtained after transforming the series of realized covariance matrices, marginal time-series models need to be applied first.}

\subsubsection*{Marginal time-series modeling}
Given the proposed R-vine structure selection method, we know that with increasing tree level the data complexity decreases such that less elaborate time-series models might already be sufficient for accurate in-sample estimation and out-of-sample forecasting. To support this presumption, for each period within the moving window approach time-series models of different complexity are fitted to the log-transformed realized variance time-series and to the Fisher z-transformed realized (partial) correlations specified by the R-vine structure found in \autoref{Sec:R-VineStructureSelection}. \textcolor{black}{For comparison, also the time-series specified by the C-vine, of which in each tree level the root node induces the on average lowest correlation strength, are investigated.}  

\textcolor{black}{Besides simply considering the mean value over time, basic univariate HAR and ARFIMA models as well as HAR and ARFIMA models including a $\text{GARCH(1,1)}$ component with normal innovations (abbreviated as HN and AN) and with SGED innovations (abbreviated as HSGED and ASGED) are fitted.
	To evaluate the statistical precision we use the root mean squared error (RMSE), which according to \cite{patton2011volatility} is robust to noise in the volatility proxies. \autoref{table:timeSeriesPerformancePCV} shows the out-of-sample RMSE for all time-series model components under consideration. In each row, the set of superior models based on the MCS approach of \cite{hansen2011model} with a confidence level of 10$\%$ is highlighted in gray. The model with the lowest RMSE, which is the last one that would be rejected from the model confidence set, is highlighted in bold. In general, ARFIMA based models show a superior prediction performance with respect to the RMSE criterion. However, especially within the variations of the two base models the RMSE values often are very close to each other. For the realized variance time-series and the realized standard correlation time-series in $\mathcal{T}_1$ of the selected R-vine structure, the best model usually includes a $\text{GARCH}(1,1)$ augmentation. For tree level $\mathcal{T}_2$ and $\mathcal{T}_3$, there is a shift to basic ARFIMA models, while for tree level $\mathcal{T}_4$ and $\mathcal{T}_5$ even simply using the mean realized partial correlation value as forecast is included in the model confidence set at a confidence level of 10\%. This confirms the presumption that given the proposed R-vine structure selection method with increasing tree level more parsimonious time-series models already are sufficient. This hierarchical pattern is not observed for the considered C-vine. Here, base models including a $\text{GARCH}(1,1)$ augmentation with normal or SGED innovations most often would be the last ones to be eliminated from the model confidence set. In particular, a simple mean forecast clearly is insufficient even in high tree levels. Similar results are detected for the Cholesky elements and are given in \autoref{table:timeSeriesPerformanceCholesky} in \autoref{Sec:AddOnEmpiricalStudy}.} 

\renewcommand{\arraystretch}{1.00}
\begin{table}[h!]
	\centering
	\small
	\caption{RMSE with respect to the complete out-of-sample forecasting horizon (1632 days) for the model components in the partial correlation vine data transformation approach. Two different R-vine structures for data transformation are considered. The set of superior models according to the MCS approach at a confidence level of 10\% is highlighted in gray. The lowest RMSE is highlighted in bold.}
	\label{table:timeSeriesPerformancePCV}
	%	\footnotesize
	\begin{tabular}{c c c c c c c c c} 
		\midrule
		& & mean & HAR & HN & HSGED & ARFIMA & AN & ASGED \\
		\midrule
		\midrule
		&	{\tiny AXP} & 1.0429 & 0.4711 & 0.4715 & 0.4719 & \cellcolor{lightgray}0.4684 & \cellcolor{lightgray}\textbf{0.4668} & \cellcolor{lightgray}0.4680\\		
		&	{\tiny C} & 1.0154 & \cellcolor{lightgray}0.4469 & \cellcolor{lightgray}0.4479 & 0.4510 & \cellcolor{lightgray}0.4465 & \cellcolor{lightgray}\textbf{0.4455} & \cellcolor{lightgray}0.4483\\	
		&	{\tiny GE} & 0.8105 & \cellcolor{lightgray}0.4634 & \cellcolor{lightgray}0.4632 & \cellcolor{lightgray}0.4647 & \cellcolor{lightgray}0.4627 & \cellcolor{lightgray}\textbf{0.4625} & \cellcolor{lightgray}0.4627 \\		
		&	{\tiny HD} & 0.7766 & \cellcolor{lightgray}0.4554 & \cellcolor{lightgray}0.4557 & \cellcolor{lightgray}0.4568 & \cellcolor{lightgray}\textbf{0.4540} & \cellcolor{lightgray}0.4543 & \cellcolor{lightgray}0.4543\\
		&	{\tiny IBM} & 0.7242 & \cellcolor{lightgray}0.4320 & \cellcolor{lightgray}0.4322 & \cellcolor{lightgray}0.4323 & \cellcolor{lightgray}\textbf{0.4317} & \cellcolor{lightgray}0.4331 & \cellcolor{lightgray}0.4327 \\		
		&	{\tiny JPM} & 1.0137 & \cellcolor{lightgray}0.4653 & \cellcolor{lightgray}0.4647 & \cellcolor{lightgray}0.4671 & \cellcolor{lightgray}0.4652 & \cellcolor{lightgray}\textbf{0.4641} & \cellcolor{lightgray}0.4655\\			
		\midrule
		\multirow{15}{*}{\begin{sideways} {\small R-vine selection (\autoref{Sec:R-VineStructureSelection})} \end{sideways}} & {\tiny AXP,C} & 0.2183 & \cellcolor{lightgray}0.1572 & \cellcolor{lightgray}0.1573 & 0.1576 & \cellcolor{lightgray}0.1568 & \cellcolor{lightgray}\textbf{0.1568} & \cellcolor{lightgray}0.1571\\		
		&	{\tiny C,GE} & 0.1984 & \cellcolor{lightgray}0.1531 & \cellcolor{lightgray}0.1531 & \cellcolor{lightgray}0.1530 & \cellcolor{lightgray}0.1526 & \cellcolor{lightgray}0.1526 & \cellcolor{lightgray}\textbf{0.1524}\\		
		&	{\tiny C,HD} & 0.1857 & \cellcolor{lightgray}0.1519 & \cellcolor{lightgray}0.1520 & \cellcolor{lightgray}0.1520 & \cellcolor{lightgray}0.1515 & \cellcolor{lightgray}0.1516 & \cellcolor{lightgray}\textbf{0.1512}\\		
		&	{\tiny C,JPM} & 0.2149 & \cellcolor{lightgray}0.1619 & \cellcolor{lightgray}0.1620 & \cellcolor{lightgray}0.1620 & \cellcolor{lightgray}0.1615 & \cellcolor{lightgray}\textbf{0.1615} & \cellcolor{lightgray}\textbf{0.1615}\\		
		&	{\tiny GE,IBM} & 0.1914 & \cellcolor{lightgray}\textbf{0.1489} & \cellcolor{lightgray}0.1490 & \cellcolor{lightgray}0.1490 & \cellcolor{lightgray}0.1490 & \cellcolor{lightgray}0.1490 & \cellcolor{lightgray}0.1490\\		
		&	{\tiny AXP,GE;C} & 0.1395 & \cellcolor{lightgray}0.1317 & \cellcolor{lightgray}0.1317 & \cellcolor{lightgray}0.1316 & \cellcolor{lightgray}\textbf{0.1313} & \cellcolor{lightgray}0.1313 & \cellcolor{lightgray}0.1313\\
		&	{\tiny AXP,JPM;C} & 0.1364 & \cellcolor{lightgray}\textbf{0.1295} & \cellcolor{lightgray}0.1295 & \cellcolor{lightgray}0.1296 & \cellcolor{lightgray}0.1297 & \cellcolor{lightgray}0.1297 & \cellcolor{lightgray}0.1297\\		
		&	{\tiny C,IBM;GE} & 0.1340 & \cellcolor{lightgray}0.1271 & \cellcolor{lightgray}0.1271 & \cellcolor{lightgray}0.1272 & \cellcolor{lightgray}\textbf{0.1270} & \cellcolor{lightgray}0.1270 & \cellcolor{lightgray}0.1270\\
		&	{\tiny GE,HD;C} & 0.1384 & 0.1300 & 0.1301 & 0.1301 & \cellcolor{lightgray}\textbf{0.1292} & \cellcolor{lightgray}0.1292 & \cellcolor{lightgray}0.1292\\		
		&	{\tiny AXP,IBM;C,GE} & 0.1246 & \cellcolor{lightgray}0.1237 & \cellcolor{lightgray}0.1237 & \cellcolor{lightgray}0.1237 & \cellcolor{lightgray}\textbf{0.1231} & \cellcolor{lightgray}0.1231 & \cellcolor{lightgray}0.1232\\
		&	{\tiny GE,JPM;AXP,C} & 0.1211 & 0.1196 & 0.1196 & 0.1196 & \cellcolor{lightgray}\textbf{0.1191} & \cellcolor{lightgray}0.1191 & \cellcolor{lightgray}\textbf{0.1191}\\
		
		&	{\tiny HD,IBM;C,GE} & \cellcolor{lightgray}0.1260 & \cellcolor{lightgray}0.1253 & \cellcolor{lightgray}0.1254 & \cellcolor{lightgray}0.1254 & \cellcolor{lightgray}\textbf{0.1250} & \cellcolor{lightgray}0.1250 & \cellcolor{lightgray}0.1251\\		
		&	{\tiny AXP,HD;C,GE,IBM} & \cellcolor{lightgray}0.1175 & \cellcolor{lightgray}0.1171 & \cellcolor{lightgray}0.1171 & \cellcolor{lightgray}0.1172 & \cellcolor{lightgray}0.1168 & \cellcolor{lightgray}\textbf{0.1168} & \cellcolor{lightgray}0.1170\\		
		&	{\tiny IBM,JPM;AXP,C,GE} & \cellcolor{lightgray}0.1237 & \cellcolor{lightgray}0.1227 & \cellcolor{lightgray}0.1227 & \cellcolor{lightgray}0.1227 & \cellcolor{lightgray}0.1225 & \cellcolor{lightgray}\textbf{0.1223} & \cellcolor{lightgray}0.1225\\
		&	{\tiny HD,JPM;AXP,C,GE,IBM} & \cellcolor{lightgray}0.1177 & \cellcolor{lightgray}\textbf{0.1175} & \cellcolor{lightgray}0.1175 & \cellcolor{lightgray}0.1175 & \cellcolor{lightgray}0.1178 & \cellcolor{lightgray}0.1177 & \cellcolor{lightgray}0.1178\\
		\midrule
		\multirow{15}{*}{\begin{sideways} {\small C-vine} \end{sideways}} &	{\tiny AXP,HD} & 0.1873 & 0.1547 & 0.1547 & 0.1549 & \cellcolor{lightgray}\textbf{0.1538} & \cellcolor{lightgray}0.1538 & \cellcolor{lightgray}0.1539\\
		&	{\tiny C,HD} & 0.1857 & \cellcolor{lightgray}0.1519 & \cellcolor{lightgray}0.1520 & \cellcolor{lightgray}0.1520 & \cellcolor{lightgray}0.1515 & \cellcolor{lightgray}0.1516 & \cellcolor{lightgray}\textbf{0.1512}\\		
		&	{\tiny GE,HD} & 0.1856 & 0.1517 & 0.1517 & 0.1517 & \cellcolor{lightgray}0.1509 & \cellcolor{lightgray}\textbf{0.1509} & \cellcolor{lightgray}0.1509\\		
		&	{\tiny HD,IBM} & 0.1731 & \cellcolor{lightgray}0.1488 & \cellcolor{lightgray}0.1489 & \cellcolor{lightgray}0.1489 & \cellcolor{lightgray}0.1484 & \cellcolor{lightgray}0.1484 & \cellcolor{lightgray}\textbf{0.1483}\\
		&	{\tiny HD,JPM} & 0.1804 & 0.1532 & 0.1533 & 0.1533 & \cellcolor{lightgray}0.1525 & \cellcolor{lightgray}\textbf{0.1523} & 0.1525\\
		&	{\tiny AXP,IBM;HD} & 0.1491 & \cellcolor{lightgray}0.1349 & \cellcolor{lightgray}0.1349 & \cellcolor{lightgray}0.1349 & \cellcolor{lightgray}0.1345 & \cellcolor{lightgray}\textbf{0.1344} & \cellcolor{lightgray}0.1345\\		
		&	{\tiny C,IBM;HD} & 0.1502 & \cellcolor{lightgray}0.1320 & \cellcolor{lightgray}0.1320 & \cellcolor{lightgray}0.1320 & \cellcolor{lightgray}0.1317 & \cellcolor{lightgray}0.1317 & \cellcolor{lightgray}\textbf{0.1315}\\		
		&	{\tiny GE,IBM;HD} & 0.1574 & \cellcolor{lightgray}0.1352 & \cellcolor{lightgray}0.1352 & \cellcolor{lightgray}0.1353 & \cellcolor{lightgray}\textbf{0.1353} & \cellcolor{lightgray}0.1354 & \cellcolor{lightgray}0.1355\\		
		&	{\tiny IBM,JPM;HD} & 0.1486 & 0.1382 & 0.1382 & 0.1383 & \cellcolor{lightgray}0.1376 & \cellcolor{lightgray}0.1376 & \cellcolor{lightgray}\textbf{0.1376}\\		
		&	{\tiny AXP,GE;HD,IBM} & 0.1387 & \cellcolor{lightgray}0.1303 & \cellcolor{lightgray}0.1303 & \cellcolor{lightgray}0.1302 & \cellcolor{lightgray}0.1299 & \cellcolor{lightgray}0.1300 & \cellcolor{lightgray}\textbf{0.1299}\\		
		&	{\tiny C,GE;HD,IBM} & 0.1371 & 0.1270 & 0.1270 & \cellcolor{lightgray}0.1268 & 0.1270 & 0.1270 & \cellcolor{lightgray}\textbf{0.1268}\\		
		&	{\tiny GE,JPM;HD,IBM} & 0.1312 & \cellcolor{lightgray}0.1251 & \cellcolor{lightgray}\textbf{0.1251} & \cellcolor{lightgray}0.1251 & \cellcolor{lightgray}0.1251 & \cellcolor{lightgray}0.1251 & \cellcolor{lightgray}0.1252\\		
		&	{\tiny AXP,C;GE,HD,IBM} & 0.1496 & \cellcolor{lightgray}0.1283 & \cellcolor{lightgray}0.1283 & \cellcolor{lightgray}0.1285 & \cellcolor{lightgray}0.1281 & \cellcolor{lightgray}\textbf{0.1281} & \cellcolor{lightgray}0.1282\\
		&	{\tiny AXP,JPM;GE,HD,IBM} & 0.1526 & \cellcolor{lightgray}0.1321 & \cellcolor{lightgray}0.1321 & \cellcolor{lightgray}0.1321 & \cellcolor{lightgray}0.1317 & \cellcolor{lightgray}\textbf{0.1316} & \cellcolor{lightgray}0.1316\\
		&	{\tiny C,JPM;AXP,GE,HD,IBM} & 0.1398 & 0.1247 & 0.1247 & 0.1248 & \cellcolor{lightgray}0.1241 & \cellcolor{lightgray}\textbf{0.1241} & \cellcolor{lightgray}0.1242\\
		\midrule
		\midrule
	\end{tabular}
	%	\vspace*{-.5cm}
\end{table}	
\renewcommand{\arraystretch}{1}

\textcolor{black}{In the following, we consider two groups of models. One including only HAR based time-series models and the other including only ARFIMA based models. Given the above findings within the partial correlation vine data transformation approach, we use HN and AN models, respectively, for all components in case that a C-vine or a randomly sampled R-vine structure is taken for data transformation. Likewise, we proceed for the Cholesky decomposition based model. In case of R-vine structure selection according to \autoref{Sec:R-VineStructureSelection}, we stepwise increase model parsimony. For model components corresponding to the realized variance and realized standard correlation time-series in $\mathcal{T}_1$, we use HN and AN models, respectively. For the ones in tree level $\mathcal{T}_2$ and $\mathcal{T}_3$ we apply basic HAR and ARFIMA models, respectively. For components in $\mathcal{T}_4$ and $\mathcal{T}_5$ we consider in one setting basic HAR and ARFIMA models, respectively, and take in another setting simply the mean value of the underlying training set as forecast.}
\newpage

\subsubsection*{Dependence modeling}

\begin{figure}[b!]
	\centering
	\includegraphics[width=0.85\linewidth]{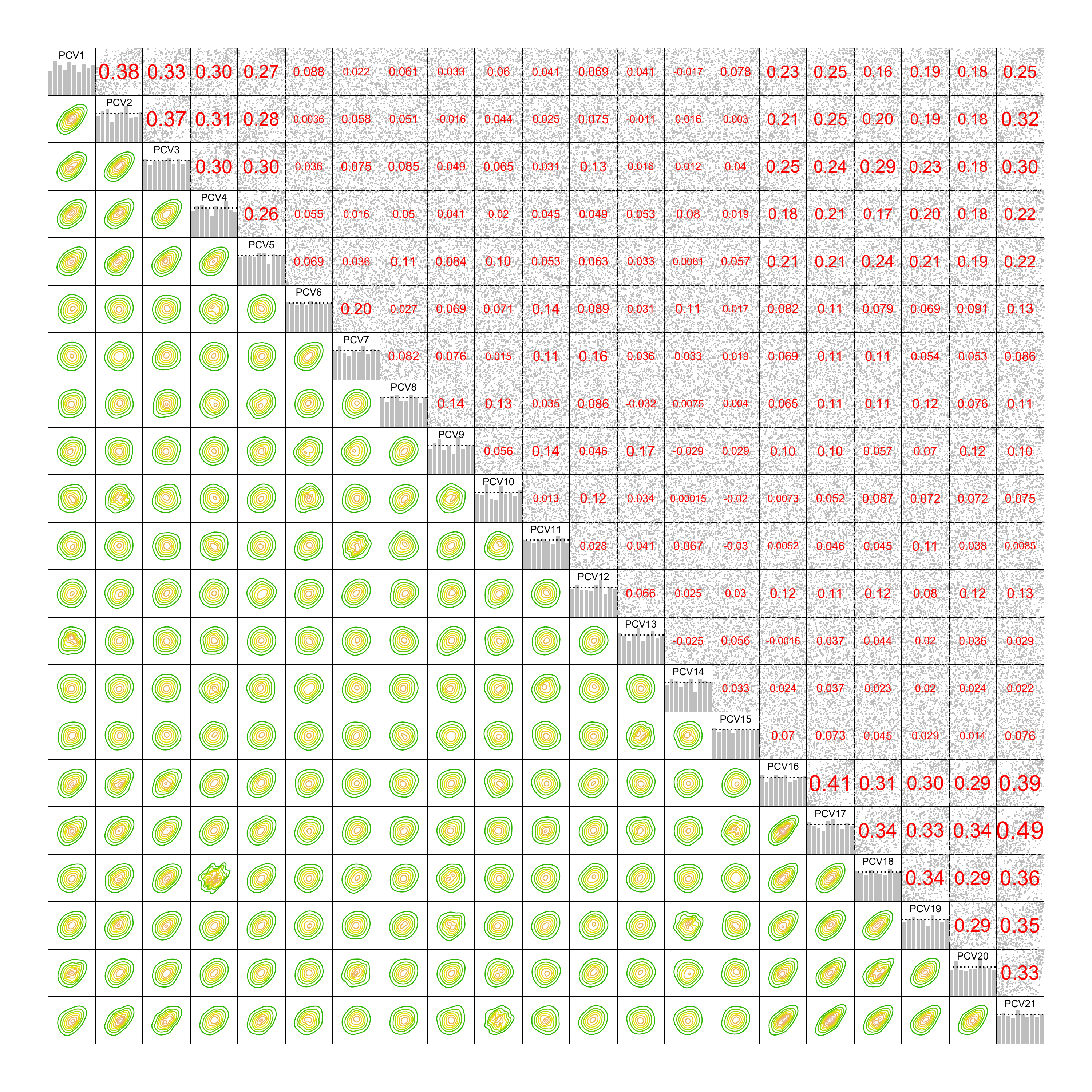}
	\caption{Exploratory data analysis for the pairwise dependencies of the $21$-dimensional pseudo copula data estimated for the period July 2006 to July 2008 using the proposed method for R-vine structure selection (\autoref{Sec:R-VineStructureSelection}) and HAR based time-series modeling. Pairwise contour plots with normalized margins, histograms and pairs plots with empirical Kendall's $\tau$ values are shown. The first five components $\text{PCV}1$ to $\text{PCV}5$ correspond to realized standard correlations in $\mathcal{T}_1$, components $\text{PCV}6$ to $\text{PCV}9$ correspond to realized first order partial correlations in $\mathcal{T}_2$, etc. Variables $\text{PCV}16$ to $\text{PCV}21$ correspond to realized variances.}
	\label{fig:pairsupcvblock74}
\end{figure}

Now, for each time window interest is in the cross-sectional dependence between the model components. Since dependencies between stocks are expected to be most pronounced during financial turmoil, we consider as an example the time window from July 2006 to July 2008. Based on the specified time-series models, the sample of innovations is obtained and transformed to pseudo copula data (\autoref{Subsec:Step2TimeSeries}). \autoref{fig:pairsupcvblock74} shows the resulting data based on R-vine structure selection according to \autoref{Sec:R-VineStructureSelection} and HAR based time-series modeling. It illustrates the corresponding histograms on its diagonal, pairwise contour plots with standard normal margins in the lower left corner and pairs plots with corresponding Kendall's $\tau$ values in the upper right corner. Only dependencies between model components corresponding to realized variances (last six components) and realized standard correlations (first 5 components) are significant with Kendall's $\tau$ values ranging from 0.2 to 0.5. Dependencies including components, which correspond to partial correlations, are rather small and close to zero for higher tree levels. \textcolor{black}{Based on these findings, we subsequently consider five different R-vine copula settings. First, independence for all pairs is assumed. Second, a 21-dimensional R-vine copula is fitted to capture dependence between all model components. Third, a reduced structured dependence is imposed, where a 11-dimensional R-vine copula is fitted only to the components corresponding to realized variances and realized standard correlations. The components corresponding to realized partial correlations are assumed to be independent. Both in case of full and reduced structured R-vine copula based dependence modeling, we allow as a first setting the pair-copulas to stem from various copula families such as Clayton, Gumbel, Frank, etc. including their reflected forms. Thus, possible asymmetric and nonlinear dependence patterns can be detected. Given the primarily elliptical shapes in \autoref{fig:pairsupcvblock74} we also consider an R-vine copula exclusively built from bivariate (conditional) Gaussian copulas, i.e.\ a Gaussian vine. Except for the structured dependence, the same settings for the copula models are taken in the Cholesky decomposition based benchmarks. To fit an R-vine copula model we rely on the \texttt{R}-package \texttt{VineCopula} \citep{schepsmeier2014vinecopula}.}

\subsection{Forecasting performance}\label{Sec:EmpStepS3}
\textcolor{black}{Based on the above model specifications, one-day-ahead forecasts as described in \autoref{Sec:Step3Back} are obtained. In addition to the 36 data transformation based prediction models, we consider three naive benchmarks. First, $\hat{\boldsymbol{Y}}_{T+1}$ is set to the realized covariance matrix at time point $T$, i.e.\ $\hat{\boldsymbol{Y}}_{T+1} = \boldsymbol{Y}_T$. Second, $\hat{\boldsymbol{Y}}_{T+1}$ is calculated as the equally weighted average of the realized covariance matrices in the corresponding training set. Third,  $\hat{\boldsymbol{Y}}_{T+1}$ is obtained as an exponentially weighted moving average, i.e.\ in our setup $\hat{\boldsymbol{Y}}_{T+1} = \lambda\hat{\boldsymbol{Y}}_{T} + \left(1-\lambda\right)\boldsymbol{Y}_{T}$, where the smoothing parameter $\lambda$ is set to 0.94 as commonly suggested in the framework of a RiskMetrics approach \citep{morgan1996riskmetrics}.}

\subsubsection*{Out-of-sample forecasting precision}

\begin{figure}[t!]
	\centering
	\includegraphics[width=.75\linewidth]{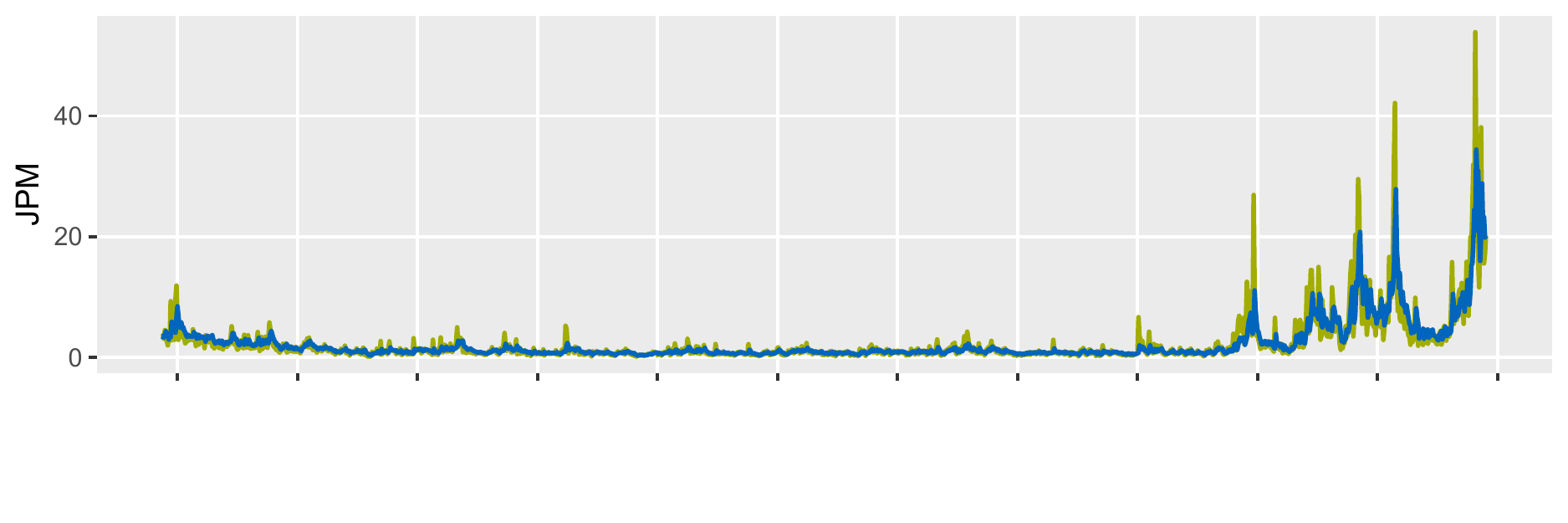}\vspace*{-1.15cm}\\
	\includegraphics[width=.75\linewidth]{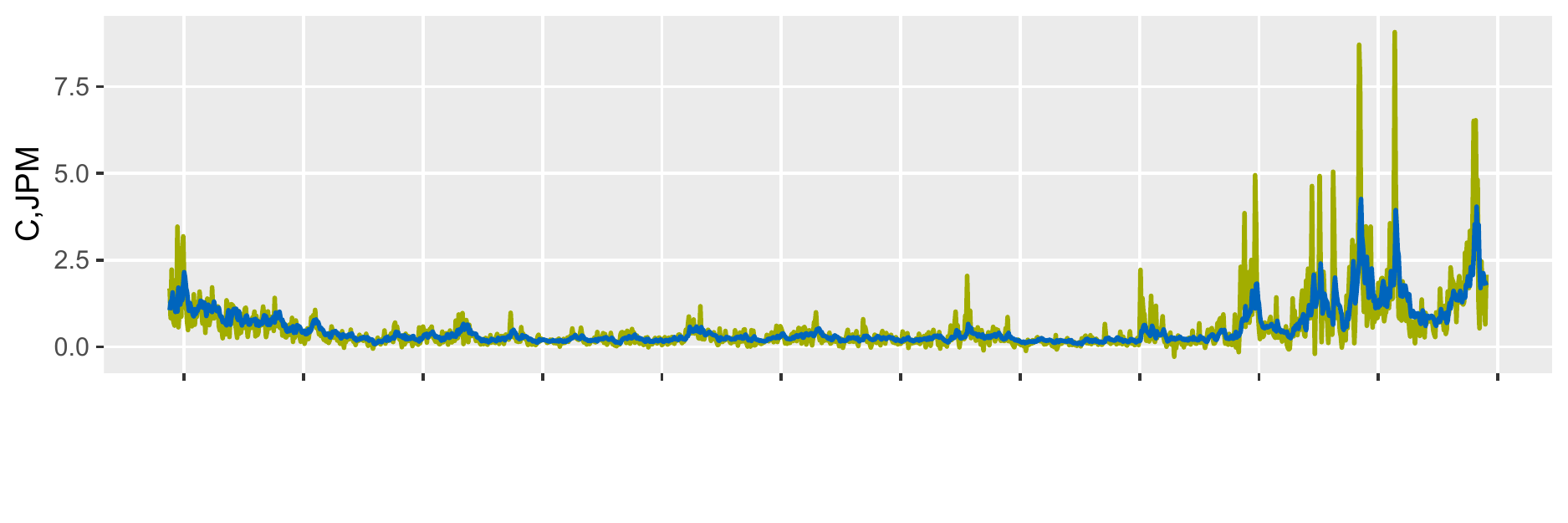}\vspace*{-1.15cm}\\	\includegraphics[width=.75\linewidth]{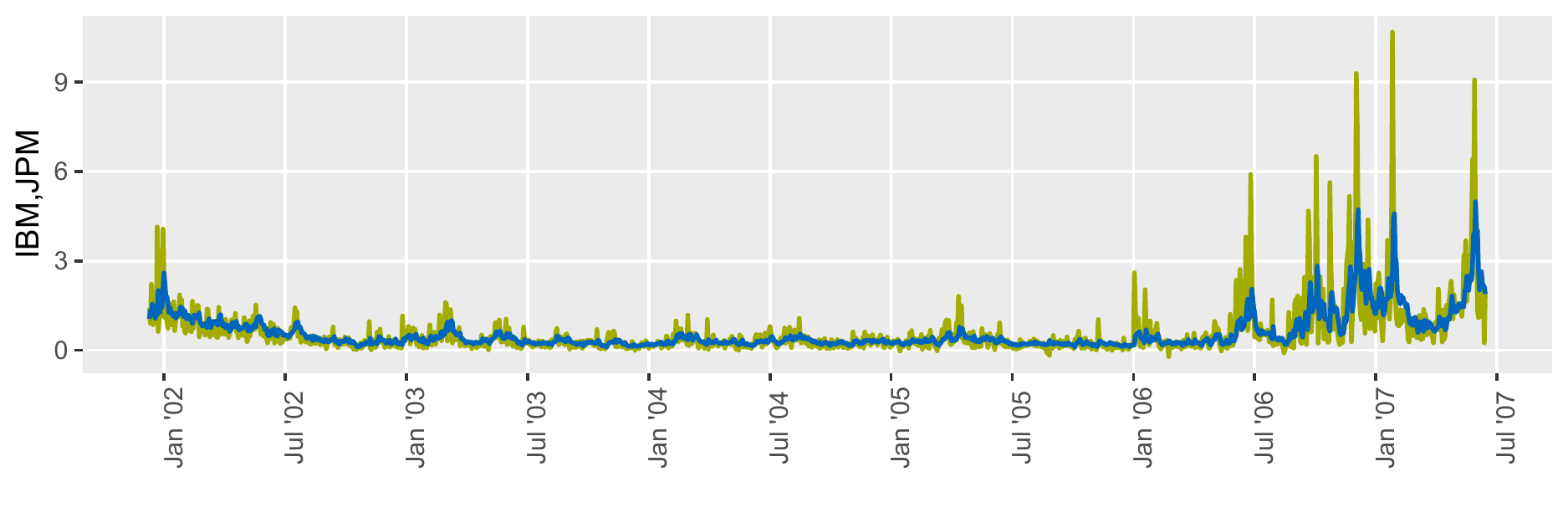}\\
	\renewcommand{\arraystretch}{0.75}
	\begin{tabular}{p{.05cm}lp{.05cm}l}
		\cellcolor{TUMgreen} & {\scriptsize historical data}
		& \cellcolor{TUMblue} & {\scriptsize one-day-ahead forecasts}\\
		\vspace{-3cm}
	\end{tabular}	
	\renewcommand{\arraystretch}{1}
	\caption{Daily realized variance time-series for JPM (1st row) and daily realized covariance time-series (2nd and 3rd row) together with the time-series of the corresponding daily forecasts based on the partial correlation vine data transformation approach with R-vine structure selected according to \autoref{Sec:R-VineStructureSelection}, ARFIMA based time-series modeling and a 21-dimensional Gaussian vine for dependence modeling.}
	\label{fig:tsplotjpm}\vspace*{-.50cm}
\end{figure}

To illustrate that the proposed forecasting approach is on target, \autoref{fig:tsplotjpm} shows for the realized variance time-series of JPM (top panel), the realized covariance time-series of C and JPM (mid panel) as well as IBM and JPM (bottom panel) the historical time-series from January 2002 until July 2008 together with the one-day-ahead forecasts based on the R-vine structure selected according to \autoref{Sec:R-VineStructureSelection}, ARFIMA based time-series modeling and a 21-dimensional Gaussian vine for dependence modeling. Results for all other realized variances and covariance pairs are similar and given in \autoref{fig:tsplot_app3} in \autoref{Sec:AddOnEmpiricalStudy}. The trends in all time-series including high short-term peaks are well detected and modeled. Distances between historical extreme peaks and corresponding forecasts are large. This finding holds true for all prediction models and is due to the high volatility of the realized variances and covariances. The predicted time-series incorporate smoothed long-term information of historical data and thus, are more stable.

To evaluate the statistical precision of the matrix forecasts, \autoref{table:ResultsPerformance} summarizes for all considered models the RMSE based on the Frobenius norm between the realized and the
predicted covariance matrices. This loss function satisfies the conditions in \cite{laurent2013loss} for consistent model ranking. In the right column, the RMSE based on bias corrected (bc) matrix forecasts are shown. We use historical data over the period of one year for level correction reducing the out-of-sample forecasting horizon to 1368 days. As in the previous analysis of the single model components, ARFIMA based models in general have smaller RMSE values compared to HAR based models. All models using partial correlation vine based data transformation and full dependence modeling exhibit a smaller RMSE than Cholesky decomposition based models and show very similar performance among each other. This confirms that any R-vine structure can be used for data transformation in step (S1) of the model approach. Among the partial correlation vine data transformation based models those with a C-vine structure used for data transformation have the highest RMSE. Recall that by construction more complex data features are induced even for high tree levels. For C-vine and random R-vine structures, time-series modeling in step (S2) with independent components and reduced structured dependence between components is clearly improved by models, which capture dependence between all model components. Here, the decreasing data complexity does not trigger. However, for the R-vine structure selected according to \autoref{Sec:R-VineStructureSelection} the performance in case of reduced structured dependence is only slightly improved by full dependence modeling. Thus, also the dependence between the model components allows for model parsimony. In general, using a Gaussian vine for dependence modeling between the model components shows comparable results as using more elaborate copulas allowing e.g.\ for tail-dependence. All discussed prediction models clearly show superior results as compared to the naive benchmarks. Bias correction in step (S3) slightly improves results while maintaining the above observations among the different models.

\begin{figure}[t!]
	\centering
	\includegraphics[width=.91\linewidth]{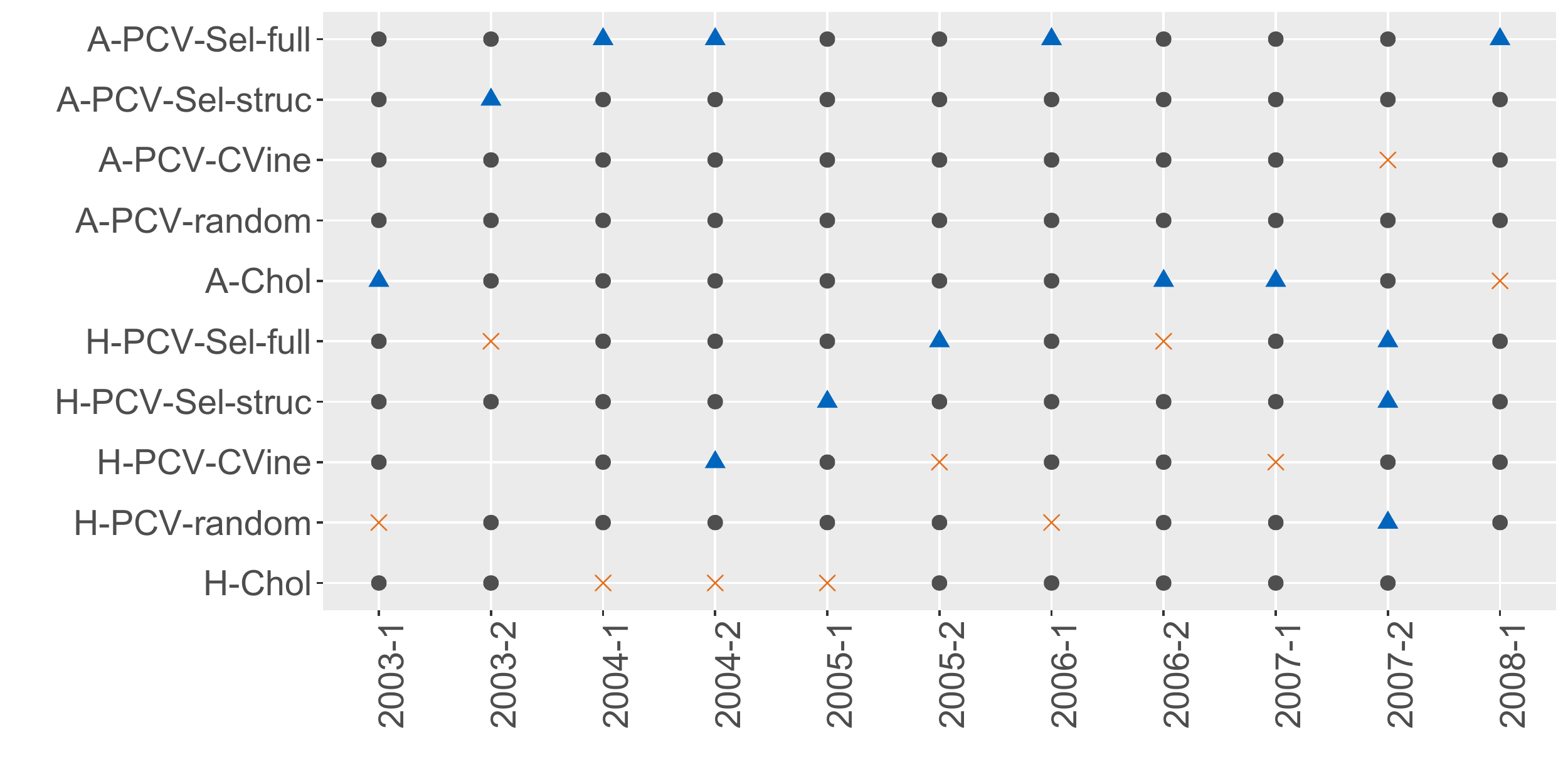}
	\caption{Model confidence sets of \cite{hansen2011model} with confidence level 10\% for all half-year periods of the out-of-sample forecasting horizon. Gray dots indicate selected models, blue triangles and orange crosses indicate the last and the next model, respectively, that would be eliminated from the set of superior models.}
	\label{fig:mcs-forecastingperformance}\vspace*{-.25cm}
\end{figure}

\textcolor{black}{To test the statistical significance of the results, we apply the MCS approach of \cite{hansen2011model}. Based on the above findings, we restrict the analysis to models using a Gaussian vine for dependence modeling between the model components. Only in case of R-vine structure selection according to \autoref{Sec:R-VineStructureSelection} we consider reduced structured dependence modeling in addition to the full one. \autoref{fig:mcs-forecastingperformance} shows for each half-year period of the out-of-sample horizon the set of superior models (indicated by a gray dot), which contains the best model at a confidence level of 10\%. A blue triangle and an orange cross indicate the last and the next model, respectively, that would be eliminated. For almost all periods, all models are selected at the given confidence level showing very close performance of all models. Most often, HAR based models would be eliminated next, while ARFIMA based models usually would be the last ones to be eliminated from the set of superior models. In three out of eleven periods, the ARFIMA-Cholesky model has the smallest RMSE based on the Frobenius norm and therefore automatically would be the last model to be eliminated. All ARFIMA and partial correlation vine data transformation based models show rather robust performance over the out-of-sample forecasting horizon. Especially, the models based on R-vine structure selection according to \autoref{Sec:R-VineStructureSelection} usually are the ones to be eliminated last from the MCS, i.e.\ having the smallest loss.}

\renewcommand{\arraystretch}{1.00}
\begin{table}[p!]
	\centering
	\small
	\caption{RMSE based on the Frobenius norm between the realized and predicted correlation matrices with respect to the complete out-of-sample forecasting horizon (1368 days) for all models.}
	\label{table:ResultsPerformance}
	%	\footnotesize
	\begin{tabular}{c c c l c c} 
		\midrule
		\multirow{2}{*}{Marginals} & \multicolumn{2}{c}{Data transformation} & R-vine copula assumed & \multirow{2}{*}{RMSE} & \multirow{2}{*}{RMSE bc}\\
		& \multicolumn{2}{c}{based on} & for transformed data &  & \\
		\midrule
		\midrule          
		\multirow{18}{*}{\begin{sideways} ARFIMA based \end{sideways}} 
		& \multirow{15}{*}{\begin{sideways} PCV \end{sideways}} &  & independence & 6.6314 & 6.6045 \\
		& & R-vine selection  & full all & 6.5725 & 6.5632 \\
		& & (\autoref{Sec:R-VineStructureSelection}) & full Gauss & 6.5685 & 6.5619 \\
		& & & structured all& 6.5858 & 6.5740 \\
		& & & structured Gauss& 6.5864 & 6.5742 \\
		\cmidrule{3-6}
		& & \multirow{5}{*}{C-vine} & independence & 6.7076 & 6.6733 \\
		& & & full all & 6.5968 & 6.5854 \\
		& & & full Gauss & 6.5967 & 6.5860 \\
		& & & structured all& 6.6436 & 6.6189 \\
		& & & structured Gauss & 6.6410 & 6.6166 \\
		\cmidrule{3-6}
		& & \multirow{5}{*}{random R-vine} & independence & 6.6694 & 6.6393 \\
		& & & full all & 6.5826 & 6.5746 \\
		& & & full Gauss & 6.5886 & 6.5810 \\
		& & & structured all & 6.6155 &  6.5968 \\
		& & & structured Gauss & 6.6138 & 6.5954 \\
		\cmidrule{2-6}
		\cmidrule{2-6}
		& & \multirow{3}{*}{Cholesky } &  independence & 6.6732 & 6.6437 \\
		& & & all & 6.6121 & 6.6001 \\
		& & & Gauss & 6.6193 & 6.6075 \\	
		\midrule
		\multirow{18}{*}{\begin{sideways} HAR based \end{sideways}}  
		& \multirow{15}{*}{\begin{sideways} PCV \end{sideways}} &  & independence & 6.7218 & 6.6566 \\
		& & R-vine selection & full all & 6.6332 & 6.5998 \\
		& & (\autoref{Sec:R-VineStructureSelection}) & full Gauss & 6.6313 & 6.5962 \\
		& & & structured all& 6.6544 & 6.6146 \\
		& & & structured Gauss& 6.6522 & 6.6122 \\
		\cmidrule{3-6}
		& & \multirow{5}{*}{C-vine} & independence & 6.7900 & 6.7085 \\
		& & & full all & 6.6574 & 6.6153 \\
		& & & full Gauss & 6.6575 & 6.6158 \\
		& & & structured all& 6.7117 & 6.6480 \\
		& & & structured Gauss & 6.7094 & 6.6453 \\
		\cmidrule{3-6}
		& & \multirow{5}{*}{random R-vine} & independence & 6.7527 & 6.6830 \\
		& & & full all & 6.6432 & 6.6117 \\
		& & & full Gauss & 6.6474 & 6.6158 \\
		& & & structured all & 6.6821 & 6.6334 \\
		& & & structured Gauss & 6.6796 & 6.6310 \\
		\cmidrule{2-6}
		\cmidrule{2-6}
		& & \multirow{3}{*}{Cholesky } &  independence & 6.7400 & 6.6866 \\
		& & & all & 6.6863 & 6.6621 \\
		& & & Gauss & 6.6841 & 6.6603 \\
		\midrule
		\multicolumn{4}{l}{mean over training set} &  12.0894 & \\
		\multicolumn{4}{l}{previous day}  & 7.2937 & \\
		\multicolumn{4}{l}{EWMA with $\lambda = 0.94$}  & 7.8790 & \\
		\midrule
		\midrule
	\end{tabular}
\end{table}	
\renewcommand{\arraystretch}{1}
\normalsize
\newpage

\subsubsection*{Mean-variance trade-off in portfolio optimization}
\textcolor{black}{For additional economic evaluation of the forecasts, we construct portfolios based on each prediction model, which are mean-variance efficient. For a risk-averse investor we assume a quadratic utility function. Then, the problem to maximize the utility is reduced to finding the asset weights $\boldsymbol{w}$, which minimize the portfolio volatility $\sigma_{\text{p}}$ based on a fixed target expected return $\mu_{\text{p}}$ \citep{markowitz1952portfolio}. The optimal portfolio is obtained by solving the quadratic problem}

\textcolor{black}{\begin{align*}
	\min_{\boldsymbol{w}_{t+1}}\boldsymbol{w}_{t+1}'\hat{\boldsymbol{\Sigma}}_{t+1}\boldsymbol{w}_{t+1} \quad \text{s.t.\ } \  \boldsymbol{w}_{t+1}'\mathbb{E}\left[\boldsymbol{r}_{t+1}|\mathcal{F}_{t}\right] = \mu_{\text{p}} \quad \text{and} \quad \boldsymbol{w}_{t+1}'\boldsymbol{1}_d = 1,
	\end{align*}
	where $\boldsymbol{w}_{t+1}$ is the $d \times 1$ vector of portfolio weights chosen at day $t$ for $t+1$, $\boldsymbol{1}_d$ is a $d \times 1$ vector of ones, $\mu_{\text{p}}$ is the daily target expected return and $\hat{\boldsymbol{\Sigma}}_{t+1}$ is the conditional (with respect to the information set) covariance forecast at day $t$ for $t+1$. The latter corresponds to the realized covariance forecasts $\hat{\boldsymbol{Y}}_{t+1}$.}

\begin{figure}[t!]
	\centering
	\includegraphics[width=.75\linewidth]{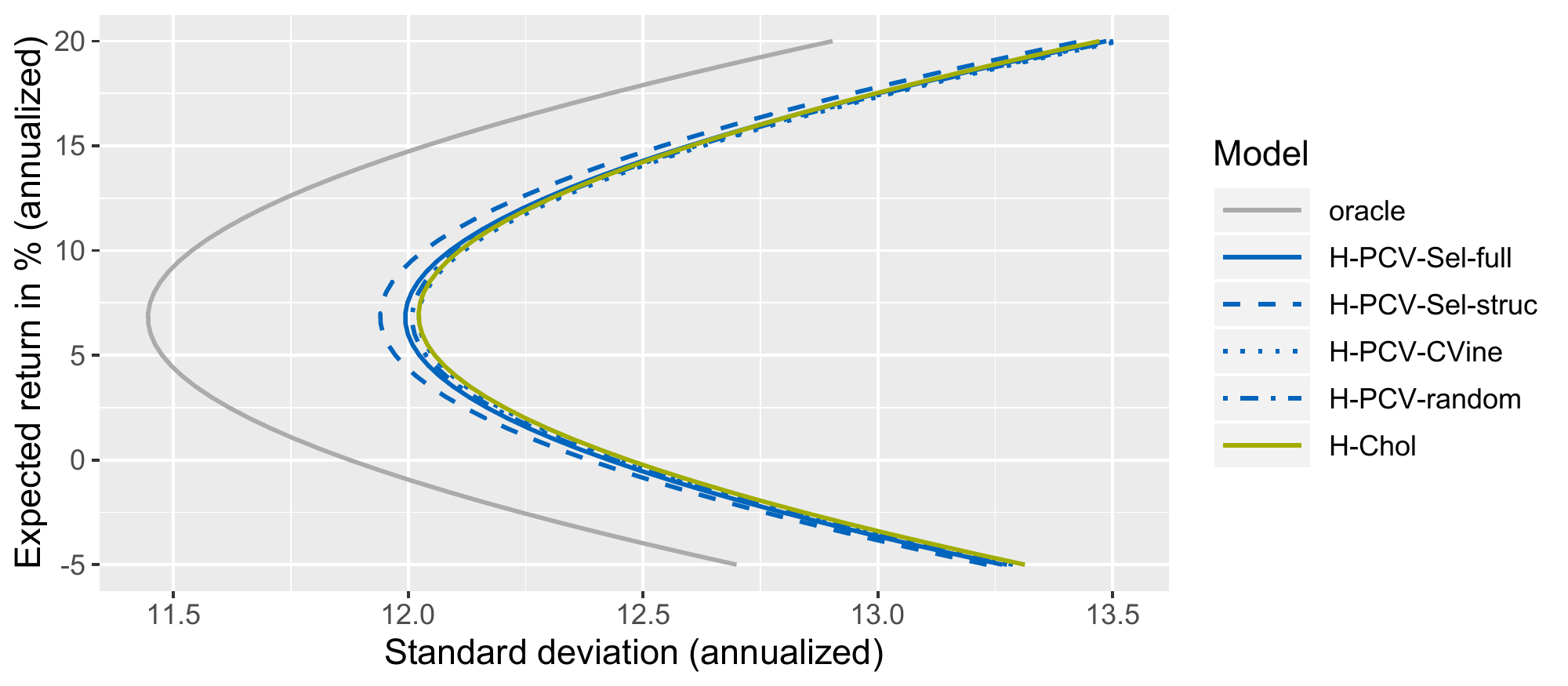}
	\includegraphics[width=.75\linewidth]{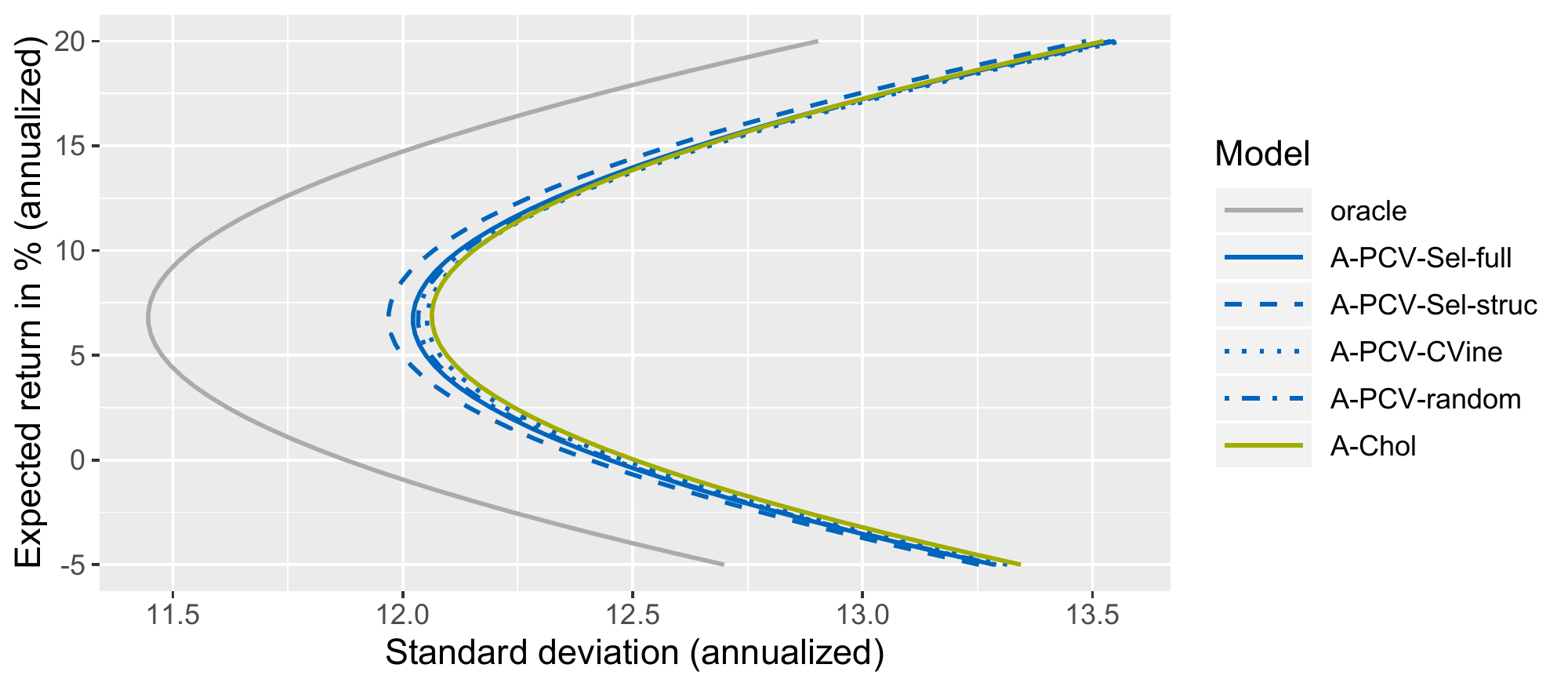}
	\caption{Efficient frontier for each HAR and ARFIMA based prediction model plotting the expected return versus its corresponding risk in terms of standard deviation. The curves are obtained as averages over the out-of-sample horizon (1368 days).}
	\label{fig:efffrontiers_HAR}
\end{figure}

\textcolor{black}{For each prediction model, we solve the above optimization problem for a daily target return $\mu_{\text{p}}$ for all 1368 days in the out-of-sample horizon. Based on the optimal portfolio weights $\boldsymbol{w}_t$ for day $t$ ($t=1,\ldots,1368$) the expected risk in terms of standard deviation, $\sqrt{\boldsymbol{w}_t'\hat{\boldsymbol{Y}}_t\boldsymbol{w}_t}$, corresponding to the target expected return $\mu_{\text{p}}$ can be calculated. Taking the averages over the forecasting horizon and repeating the procedure for a grid of target returns, results in an average efficient frontier for each prediction model. To obtain an average oracle efficient frontier, the true realized covariance matrices for each day $t$ are used. % Thus, the mean-variance trade-off achievable by the portfolio optimization strategy suggested by the different prediction models can be compared.}
	\autoref{fig:efffrontiers_HAR} shows the efficient frontiers for the considered HAR and ARFIMA
	based prediction models. %The results for ARFIMA based prediction models are nearly identical and are shown in \autoref{fig:efffrontiers_ARFIMA} in \autoref{Sec:AddOnEmpiricalStudy}. 
	All partial correlation vine data transformation based models show a clear improvement in terms of the expected mean-variance trade-off compared to the two Cholesky decomposition based prediction models.} 

\noindent

%\begin{figure}[h!]
%	\centering
%	\includegraphics[width=.75\linewidth]{plots/new_effFrontiers_ARFIMA}
%	\caption{Efficient frontier for each ARFIMA based prediction model plotting the expected return versus its corresponding risk in terms of standard deviation. The curves are obtained as averages over the out-of-sample horizon (1368 days).}
%	\label{fig:efffrontiers_ARFIMA}
%\end{figure}\noindent

\textcolor{black}{To validate this observation in an out-of-sample setting we calculate for each prediction model based on the corresponding optimal portfolio weights $\boldsymbol{w}_{t+1}$ estimated at day $t$ for $t+1$ ($t=0,\ldots,1367$) the ex-post realized portfolio return $r_{\text{p},{t+1}} = \boldsymbol{w}_{t+1}'\boldsymbol{r}_{t+1}$ and the ex-post realized portfolio volatility $\sigma_{\text{p},{t+1}} = \sqrt{\boldsymbol{w}_{t+1}'\boldsymbol{Y}_{t+1}\boldsymbol{w}_{t+1}}$. Here, $\boldsymbol{r}_{t+1}$ and $\boldsymbol{Y}_{t+1}$ are the true returns and the true covariance matrix, respectively, realized at day $t+1$. Given a small enough grid of target returns, we are able to obtain for each prediction model the series of ex-post portfolio standard deviation $\sigma_{\text{p},t+1}$, $t=0,\ldots,1367$, corresponding to a certain average ex-post realized return. For an average annualized ex-post realized portfolio return of approximately 7.5\%, 10\%, 12\% and 15\%, \autoref{table:EffFrontierMCS} shows the average annualized ex-post realized portfolio standard deviation for each prediction model. The set of models, which includes the model with the lowest standard deviation at a confidence level of 10\% based on the MCS approach of \cite{hansen2011model}, is highlighted in gray. The model with the lowest loss (deviation from zero) is highlighted in bold. In general, HAR based prediction models perform better than their ARFIMA based counterparts. In the ex-post analysis, all ARFIMA based partial correlation vine data transformation based models have the highest average standard deviation. This confirms the often seen phenomenon that models with the lowest statistical loss do not necessarily show superior results in economical applications \citep{laurent2013loss}. The HAR based model with R-vine structure selected according to \autoref{Sec:R-VineStructureSelection} and with reduced structured dependence among the model components is the best model at a confidence level of 10\% for all considered annualized ex-post realized portfolio returns. Comparing the average ex-post realized standard deviations of the HAR based prediction models, further demonstrates the strength of the proposed methodology irrespective of the R-vine structure used for data transformation.} 
\newpage
\renewcommand{\arraystretch}{1.0}
\begin{table}[t!]
	\centering
	\small
	\caption{Annualized average ex-post standard deviation corresponding to four levels of annualized ex-post realized return. The set of models, which includes the one with the smallest standard deviation at a confidence level of 10\%, are highlighted in gray. The last model to be eliminated is highlighted bold.}
	\label{table:EffFrontierMCS}
	%	\footnotesize
	\begin{tabular}{r c c c c } 
		\midrule
		\multicolumn{1}{c}{Model} & \multicolumn{4}{c}{Realized return in $\%$ (annualized)}\\
		& 7.5 & 10 & 12.5 & 15 \\
		\midrule
		\midrule
		A-PCV-Sel-full & 12.5217 & 12.9545 & 13.5492 & 14.2832 \\
		A-PCV-Sel-struc & 12.5055 & 12.9254 & 13.5086 & 14.2317 \\
		A-PCV-CVine & 12.4981 & 12.9269 & 13.5252 & 14.2722 \\
		A-PCV-random & 12.4949 & 12.9151 & 13.5027 & 14.2366 \\
		A-Chol & \cellcolor{lightgray}12.4616 & 12.8641 & 13.4352 & 14.1510 \\
		H-PCV-Sel-full & 12.4754 & 12.8748 & 13.4359 & 14.1363 \\
		H-PCV-Sel-struc & \cellcolor{lightgray}\textbf{12.4588} & \cellcolor{lightgray}\textbf{12.8447} & \cellcolor{lightgray}\textbf{13.3937} & \cellcolor{lightgray}\textbf{14.0796} \\
		H-PCV-CVine & \cellcolor{lightgray}12.4595 & 12.8606 & 13.4293 & 14.1429 \\
		H-PCV-random & 12.4680 & 12.8680 & 13.4349 & 14.1413 \\
		H-Chol & 12.4718 & 12.8729 & 13.4396 & 14.1475 \\
		\midrule
		\midrule
	\end{tabular}
\end{table}	
\renewcommand{\arraystretch}{1}

\section{Discussion}
In this paper, we introduce a novel approach to model and forecast time-series of realized covariance matrices. Realized variances and realized correlation matrices are jointly modeled. We address the challenge of generating symmetric and positive semi-definite correlation matrix forecasts by introducing partial correlation vines as a \textcolor{black}{tool to transform the series of realized correlation matrices.} Along with a real data example we explore in detail the benefits of the proposed methodology as compared to Cholesky decomposition based competitor models. Given the large number of R-vine structures for data transformation, we propose an R-vine structure selection method, which exclusively relies on historical information of the underlying data. This procedure allows the R-vine structure to dynamically change over time and therewith to adapt to market changes. \textcolor{black}{The selection method is motivated by the practical interpretation of the model components, which are proxies for the conditional variances and conditional correlations corresponding to a daily log-return series. High average correlation strengths are captured in lower tree levels of the R-vine structure leaving higher order realized partial correlation time-series, for which parsimonious univariate time-series modeling is sufficient. For the latter also dependence is negligible allowing for dimension reduction in the multivariate time-series model. The forecasting performance both in terms of statistical precision and in an economic evaluation, where ex-post realizations of mean-variance efficient portfolios are investigated, shows very good and in several settings even statistically significant superior prediction capability compared to the Cholesky decomposition based benchmark models. Given the excellent prediction power of the latter often demonstrated in literature, these findings provide strong evidence for the use of the partial correlation vine data transformation approach in practice.}    

\section*{Acknowledgments}
%\textcolor{black}{The authors thank the Co-Editor, the Associate Editor and the two reviewers for their valuable comments and good suggestions to further improve the manuscript and its presentation.}

Numerical calculations were performed on a Linux cluster supported by DFG grant INST 95/919-1 FUGG.

Funding: This work was supported by the Deutsche Forschungsgemeinschaft [DFG CZ 86/4, OK 103/5].

%\newpage
% references
\bibliographystyle{apalike}
\bibliography{References}
\appendix
\newpage

\section{Skewed generalized error distribution}\label{Sec:AppSGED}

The skewed generalized error distribution is specified by the location parameter $\mu$, the scale parameter $\sigma$, the shape parameter $\nu$ and the skewness parameter $\xi$. Its density function is given by
\begin{align*}
f\left(\varepsilon|\mu, \sigma, \nu, \xi \right) = \frac{C}{\sigma}\exp\left(-\frac{|\varepsilon-\mu+\delta\sigma|^{\nu}}{\left[1-\text{sign}\left(\varepsilon-\mu+\delta\sigma\right)\xi\right]^{\nu}\theta^{\nu}\sigma^{\nu}}\right)
\end{align*}
with
\begin{align*}
C &  = \frac{\nu}{2\theta}\Gamma\left(\frac{1}{\nu}\right)^{-1},\\
\theta & = \Gamma\left(\frac{1}{\nu}\right)^{1/2}\Gamma\left(\frac{3}{\nu}\right)^ {-1/2}S\left(\xi\right)^{-1},\\
\delta & = 2\xi A S\left(\xi\right)^{-1},\\
S\left(\xi\right) & = \sqrt{1 + 3 \xi^{2} - 4 A^{2} \xi^{2}},\\
A & = \Gamma\left(\frac{2}{\nu}\right) \Gamma\left(\frac{1}{\nu}\right)^{-1/2}\Gamma\left(\frac{3}{\nu}\right)^{-1/2}.
\end{align*} 
For the parameter specification $\nu = 2$ and $\xi = 0$ the normal distribution is obtained.

\section{R-vine copula models}\label{Sec:DepModelingVines}
R-vine distributions are also referred to as pair-copula constructions, since they assign to each of the $d\left(d-1\right)/2$ edges of a $d$-dimensional R-vine structure a bivariate unconditional copula (in tree $\mathcal{T}_1$) or a bivariate conditional copula (in trees $\mathcal{T}_2$ to $\mathcal{T}_{d-1}$). We consider the copula data $\left(U_1,\ldots,U_d\right)$ corresponding to the random vector $\left(X_1,\ldots,X_d\right)$ with marginal distribution functions $F_j$ ($j=1,\ldots,d$), i.e.\ $U_j = F_j\left(X_j\right)$. Since in this case the marginals of the underlying data are uniform, we speak of an R-vine copula. Following \cite{czado2010pair}, the $d$-dimensional R-vine copula density based on the R-vine structure $\mathcal{V}_d$ with edge set $E\left(\mathcal{V}_d\right) = E_1 \cup \cdots \cup E_{d-1}$  can be written as
\begin{align}
\scd&\left(u_1,\ldots,u_d\right)= \prod^{d-1}_{\ell = 1}\prod_{e \in E_\ell} \scd_{a_e,b_e;D_e}\lbrace \sC_{a_e|D_e}\left(u_{a_e}|\boldsymbol{u}_{D_e}\right),\sC_{b_e|D_e}\left(u_{b_e}|\boldsymbol{u}_{D_e}\right); \boldsymbol{u}_{D_e}\rbrace,\label{eq:RVineDensity}
\end{align}
where 
\begin{itemize}
	\item $\scd_{a_e,b_e;D_e}\left(\cdot,\cdot;\boldsymbol{u}_{D_e}\right)$ denotes the copula density corresponding to the conditional distribution of $\left(U_{a_e},U_{b_e}\right)'$ given $\boldsymbol{U}_{D_e} = \boldsymbol{u}_{D_e}$ with $\boldsymbol{U}_{D_e}$ the vector containing all variables corresponding to the conditioning set $D_e$. The corresponding copula will be denoted by  $\sC_{a_e,b_e;D_e}\left(\cdot,\cdot;\boldsymbol{u}_{D_e}\right)$. 
	\item $\sC_{a_e|D_e}\left(\cdot|\boldsymbol{u}_{D_e}\right)$ denotes the conditional distribution of $U_{a_e}$ given $\boldsymbol{U}_{D_e} = \boldsymbol{u}_{D_e}$.	 
\end{itemize}
Given the large number of valid R-vine structures and given that the pair-copulas corresponding to each edge of the underlying R-vine structure can be chosen and combined arbitrarily R-vine copulas clearly constitute a highly flexible class of dependence models.  

We assume that in \eqref{eq:RVineDensity} the conditional pair-copulas $\scd_{a_e, b_e;D_e}$ in trees $\mathcal{T}_\ell$ ($\ell = 2, \ldots, d-1$) do not depend on the conditioning vector $\boldsymbol{u}_{D_e}$. Their arguments  $\sC_{a_e|D_e}\left(u_{a_e}|\boldsymbol{u}_{D_e}\right)$ and $\sC_{b_e|D_e}\left(u_{b_e}|\boldsymbol{u}_{D_e}\right)$ indeed do depend on $\boldsymbol{u}_{D_e}$. For details on this simplifying assumption, see e.g.\ \cite{haff2010simplified} and \cite{stoeber2013simplified}.

\cite{Joe1997} provides the important result for pair-copula constructions that the conditional distributions $\sC_{a_e|D_e}\left(\cdot|\boldsymbol{u}_{D_e}\right)$ and $\sC_{b_e|D_e}\left(\cdot|\boldsymbol{u}_{D_e}\right)$, subsequently abbreviated as $\sC_{a|D}\left(\cdot|\boldsymbol{u}_{D}\right)$ and $\sC_{b|D}\left(\cdot|\boldsymbol{u}_{D}\right)$, can be evaluated using only the pair-copulas specified in lower tree levels of the underlying R-vine structure. Define for $i \in \lbrace a, b \rbrace$ the set $D_{+i} \coloneqq D \cup \lbrace i \rbrace$. Then,  
\begin{align*}
\sC_{a|D_{+b}}&\left(u_{a}|\boldsymbol{u}_{D_{+b}}\right) = h_{a|b;D}\lbrace \sC_{a|D}\left( u_a|\boldsymbol{u}_{D}\right)\big\vert \sC_{b|D}\left(u_b|\boldsymbol{u}_{D}\right)\rbrace
\end{align*}
and
\begin{align*}
\sC_{b|D_{+a}}&\left(u_{b}|\boldsymbol{u}_{D_{+a}}\right) = h_{b|a;D}\lbrace \sC_{b|D}\left( u_b|\boldsymbol{u}_{D}\right)\big\vert \sC_{a|D}\left(u_a|\boldsymbol{u}_{D}\right)\rbrace, 
\end{align*}
where
\begin{align*}
h_{a|b;D}\lbrace \sC_{a|D}\left( u_a|\boldsymbol{u}_{D}\right)\big\vert \sC_{b|D}\left(u_b|\boldsymbol{u}_{D}\right)\rbrace \coloneqq \frac{\partial}{\partial u} \sC_{a,b;D}\lbrace \sC_{a|D}\left(u_a|\boldsymbol{u}_{D}\right), u\rbrace \bigg\vert_{u = \sC_{b|D}\left(u_b|\boldsymbol{u}_{D}\right)}.
\end{align*}
and
\begin{align*}
h_{b|a;D}\lbrace \sC_{b|D}\left( u_b|\boldsymbol{u}_{D}\right)\big\vert \sC_{a|D}\left(u_a|\boldsymbol{u}_{D}\right)\rbrace \coloneqq \frac{\partial}{\partial u} \sC_{a,b;D}\lbrace u, \sC_{b|D}\left(u_b|\boldsymbol{u}_{D}\right)\rbrace \bigg\vert_{u = \sC_{a|D}\left(u_a|\boldsymbol{u}_{D}\right)}.
\end{align*}
are the h-functions corresponding to the pair-copula $\sC_{a,b;D}$. Clearly, the arguments of the h-functions can again be expressed in terms of h-functions such that a recursive representation of $\sC_{a|D}\left(u_{a}|\boldsymbol{u}_{D}\right)$ and $\sC_{b|D}\left(u_{b}|\boldsymbol{u}_{D}\right)$ in terms of lower tree pair-copulas is obtained.

R-vine copulas have been extensively studied in the recent years including the development of comprehensive statistical software available in the \texttt{R}-package \texttt{VineCopula}  \citep{schepsmeier2014vinecopula}.

\section{Additional results for the empirical study}\label{Sec:AddOnEmpiricalStudy}

\renewcommand{\arraystretch}{1.0}
\begin{table}[h!]
	\centering
	\footnotesize
	\caption{RMSE with respect to the complete out-of-sample forecasting horizon (1632 days) for the model components in the Cholesky decomposition based model. The set of superior models according to the MCS approach at a confidence level of 10\% is highlighted in gray. The lowest RMSE is highlighted in bold.}
	\label{table:timeSeriesPerformanceCholesky}
	%	\footnotesize
	\begin{tabular}{c c c c c c c c} 
		\midrule
		& mean & HAR & HN & HSGED & ARFIMA & AN & ASGED \\
		\midrule
		\midrule
		{\tiny AXP,AXP} & 0.5215 & 0.2355 & 0.2358 & 0.2359 & \cellcolor{lightgray}\textbf{0.2334} & \cellcolor{lightgray}0.2336 & \cellcolor{lightgray}0.2340\\
		{\tiny AXP,C} & 0.6789 & \cellcolor{lightgray}0.3730 & \cellcolor{lightgray}0.3706 & \cellcolor{lightgray}0.3769 & \cellcolor{lightgray}0.3698 & \cellcolor{lightgray}\textbf{0.3684} & \cellcolor{lightgray}0.3750\\
		{\tiny C,C} & 0.4579 & \cellcolor{lightgray}0.2069 & \cellcolor{lightgray}0.2076 & 0.2094 & \cellcolor{lightgray}\textbf{0.2063} & \cellcolor{lightgray}0.2070 & \cellcolor{lightgray}0.2081\\
		{\tiny AXP,GE} & 0.4142 & \cellcolor{lightgray}0.2648 & \cellcolor{lightgray}0.2646 & 0.2682 & \cellcolor{lightgray}0.2618 & \cellcolor{lightgray}\textbf{0.2616} & \cellcolor{lightgray}0.2659\\				
		{\tiny C,GE} & 0.2431 & 0.1725 & 0.1726 & 0.1735 & \cellcolor{lightgray}\textbf{0.1714} & \cellcolor{lightgray}0.1718 & \cellcolor{lightgray}0.1727\\
		{\tiny GE,GE} & 0.3676 & 0.2111 & \cellcolor{lightgray}0.2108 & 0.2112 & \cellcolor{lightgray}0.2102 & \cellcolor{lightgray}0.2099 & \cellcolor{lightgray}\textbf{0.2096}\\		
		{\tiny AXP,HD} & 0.4624 & 0.2976 & 0.2998 & 0.3023 & \cellcolor{lightgray}\textbf{0.2944} & 0.2972 & 0.2997\\
		{\tiny C,HD} & 0.2732 & \cellcolor{lightgray}0.2160 & \cellcolor{lightgray}0.2176 & \cellcolor{lightgray}0.2165 & \cellcolor{lightgray}0.2155 & \cellcolor{lightgray}0.2163 & \cellcolor{lightgray}\textbf{0.2153}\\		
		{\tiny GE,HD} & 0.2352 & \cellcolor{lightgray}0.1906 & \cellcolor{lightgray}0.1913 & 0.1925 & \cellcolor{lightgray}\textbf{0.1900} & \cellcolor{lightgray}0.1904 & \cellcolor{lightgray}0.1917\\		
		{\tiny HD,HD} & 0.3516 & \cellcolor{lightgray}0.2165 & \cellcolor{lightgray}0.2165 & \cellcolor{lightgray}0.2169 & \cellcolor{lightgray}0.2158 & \cellcolor{lightgray}0.2157 & \cellcolor{lightgray}\textbf{0.2156}\\
		{\tiny AXP,IBM} & 0.3102 & \cellcolor{lightgray}0.2190 & \cellcolor{lightgray}0.2180 & \cellcolor{lightgray}0.2202 & \cellcolor{lightgray}0.2174 & \cellcolor{lightgray}\textbf{0.2166} & \cellcolor{lightgray}0.2189\\
		{\tiny C,IBM} & 0.1929 & \cellcolor{lightgray}0.1529 & \cellcolor{lightgray}0.1539 & \cellcolor{lightgray}0.1534 & \cellcolor{lightgray}\textbf{0.1527} & \cellcolor{lightgray}0.1533 & \cellcolor{lightgray}0.1532\\		
		{\tiny GE,IBM} & 0.1768 & \cellcolor{lightgray}0.1401 & \cellcolor{lightgray}0.1404 & \cellcolor{lightgray}0.1409 & \cellcolor{lightgray}\textbf{0.1399} & \cellcolor{lightgray}0.1401 & \cellcolor{lightgray}0.1404\\		
		{\tiny HD,IBM} & 0.1321 & 0.1260 & 0.1257 & 0.1254 & \cellcolor{lightgray}0.1248 & \cellcolor{lightgray}0.1246 & \cellcolor{lightgray}\textbf{0.1246}\\		
		{\tiny IBM,IBM} & 0.3390 & \cellcolor{lightgray}0.2021 & \cellcolor{lightgray}\textbf{0.2020} & \cellcolor{lightgray}0.2021 & \cellcolor{lightgray}\textbf{0.2020} & \cellcolor{lightgray}0.2022 & \cellcolor{lightgray}0.2026\\ 		
		{\tiny AXP,JPM} & 0.6646 & \cellcolor{lightgray}0.3918 & \cellcolor{lightgray}0.3895 & \cellcolor{lightgray}0.3946 & \cellcolor{lightgray}0.3894 & \cellcolor{lightgray}\textbf{0.3879} & \cellcolor{lightgray}0.3931\\ 		
		{\tiny C,JPM} & 0.4075&  \cellcolor{lightgray}0.2777 & \cellcolor{lightgray}0.2804 & \cellcolor{lightgray}0.2803 & \cellcolor{lightgray}\textbf{0.2767} & \cellcolor{lightgray}0.2789 & \cellcolor{lightgray}0.2784\\
		{\tiny GE,JPM} & 0.1981 & 0.1758 & 0.1757 & 0.1756 & \cellcolor{lightgray}0.1745 & \cellcolor{lightgray}\textbf{0.1743} & \cellcolor{lightgray}0.1744\\		
		{\tiny HD,JPM} & 0.1619 & \cellcolor{lightgray}0.1552 & \cellcolor{lightgray}0.1554 & \cellcolor{lightgray}0.1556 & \cellcolor{lightgray}\textbf{0.1550} & \cellcolor{lightgray}0.1551 & \cellcolor{lightgray}0.1556\\
		{\tiny IBM,JPM} & 0.1559 & \cellcolor{lightgray}0.1498 & \cellcolor{lightgray}0.1497 & \cellcolor{lightgray}0.1500 & \cellcolor{lightgray}\textbf{0.1496} & \cellcolor{lightgray}0.1498 & \cellcolor{lightgray}0.1500\\		
		{\tiny JPM,JPM} & 0.4461 & \cellcolor{lightgray}0.2114 & \cellcolor{lightgray}0.2112 & \cellcolor{lightgray}0.2123 & \cellcolor{lightgray}\textbf{0.2107} & \cellcolor{lightgray}\textbf{0.2107} & \cellcolor{lightgray}0.2116\\
		\midrule
		\midrule
	\end{tabular}
\end{table}	
\renewcommand{\arraystretch}{1}

\begin{figure}[h!]
	\centering
	\includegraphics[width=.75\linewidth]{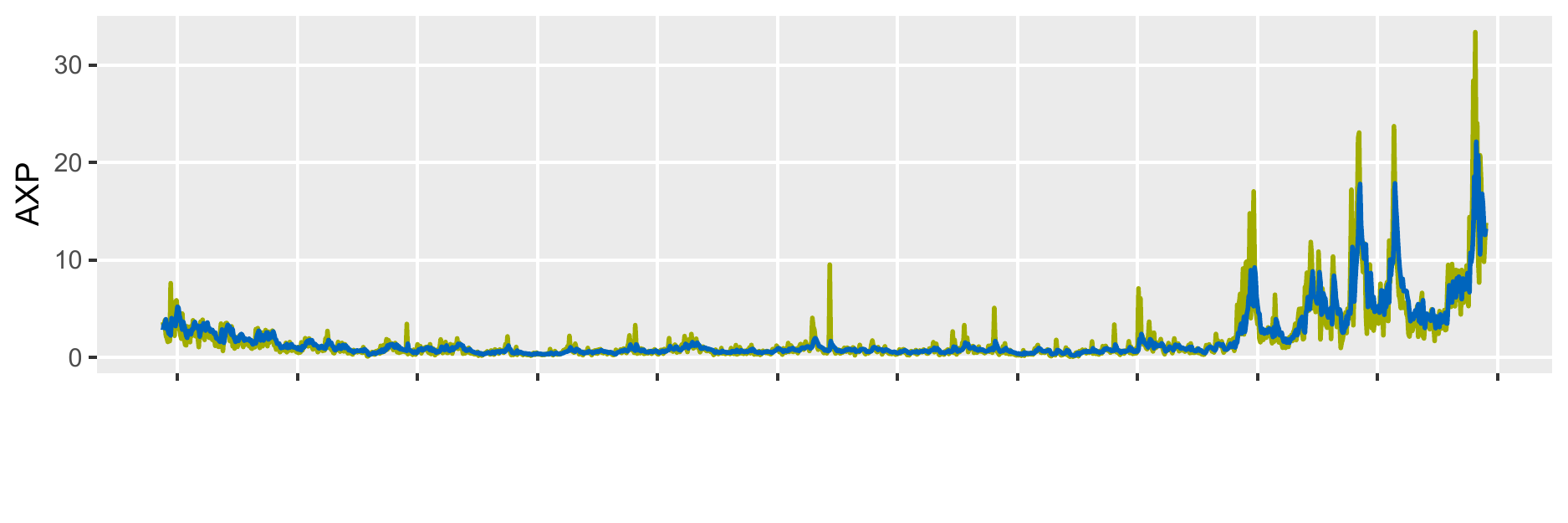}\vspace*{-1.15cm}\\
	\includegraphics[width=.75\linewidth]{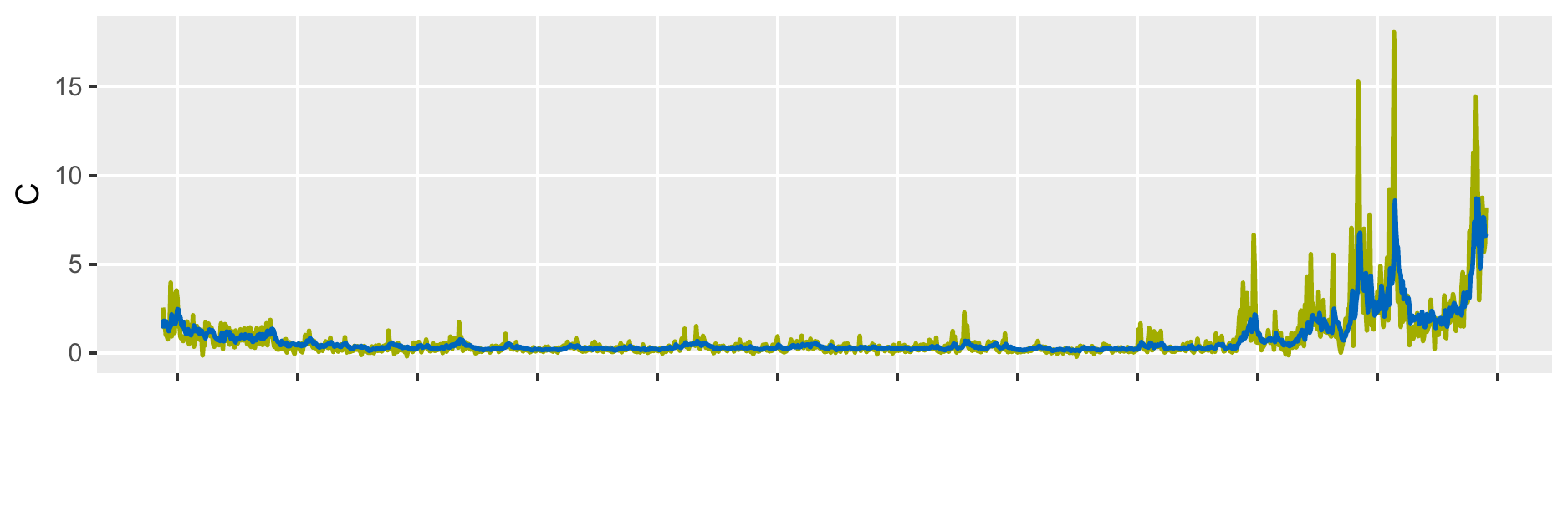}\vspace*{-1.15cm}\\	\includegraphics[width=.75\linewidth]{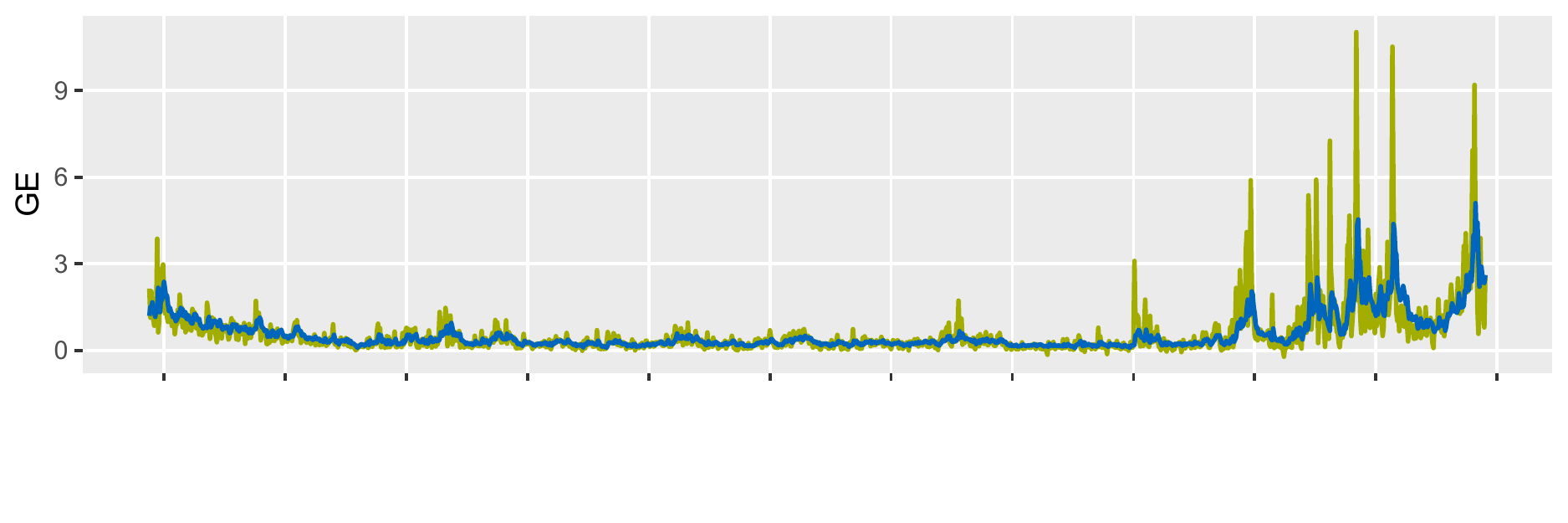}\vspace*{-1.15cm}\\
	\includegraphics[width=.75\linewidth]{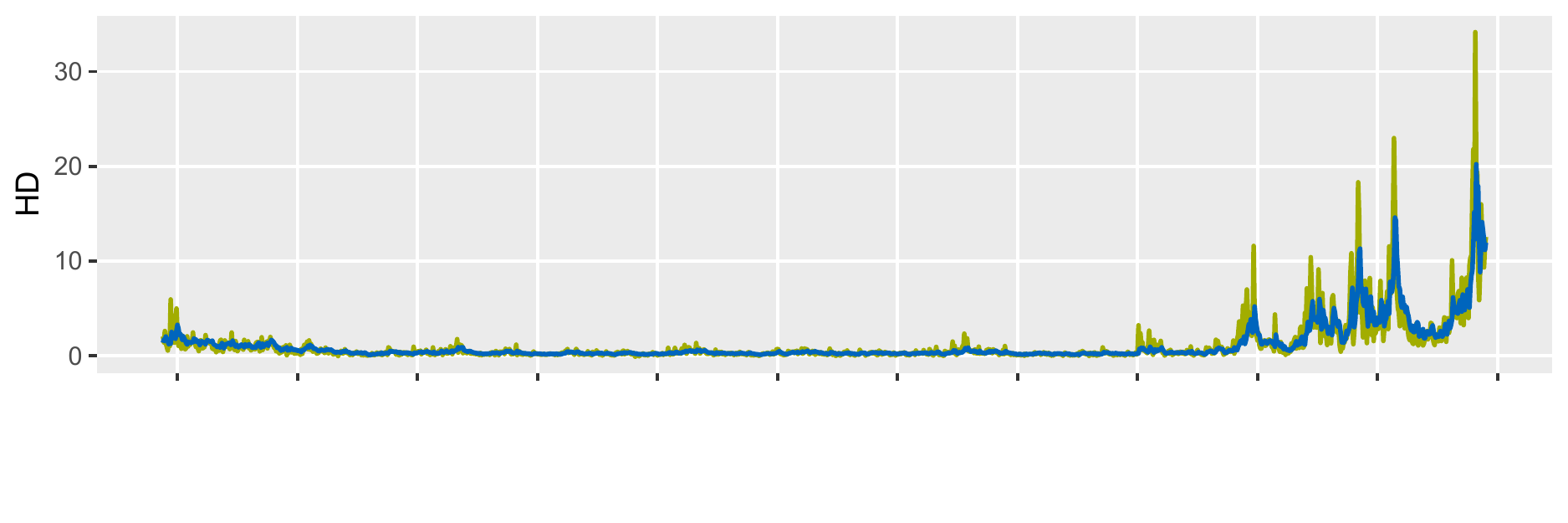}\vspace*{-1.15cm}\\
	\includegraphics[width=.75\linewidth]{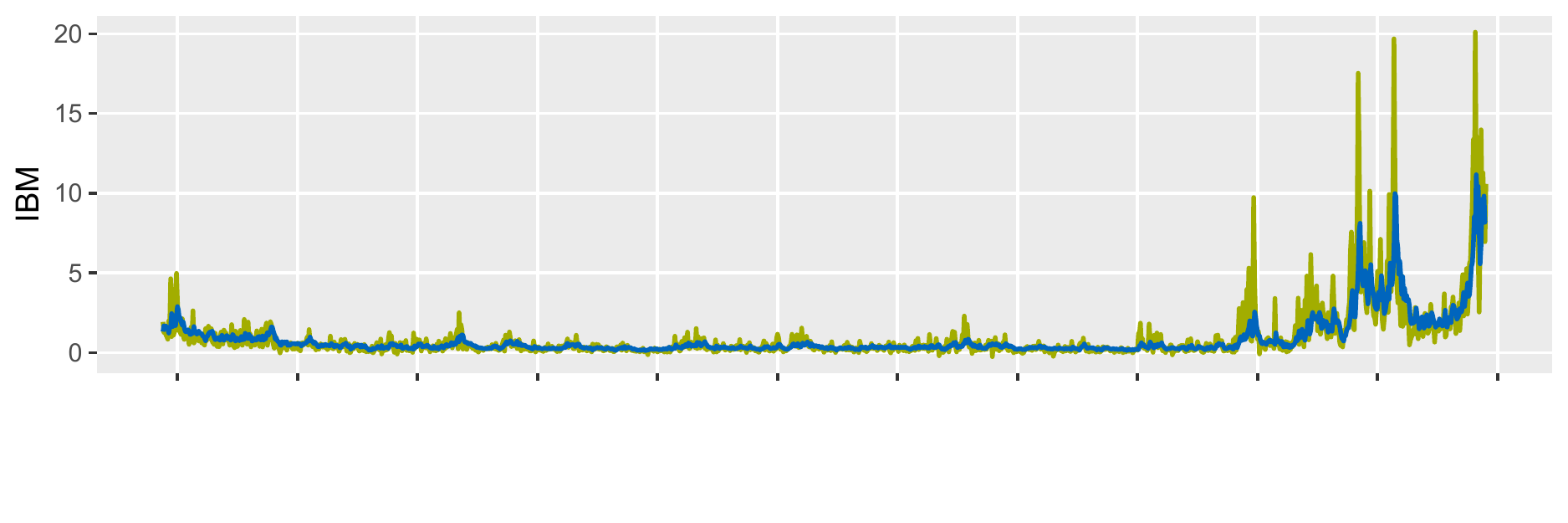}\vspace*{-1.15cm}\\
	\includegraphics[width=.75\linewidth]{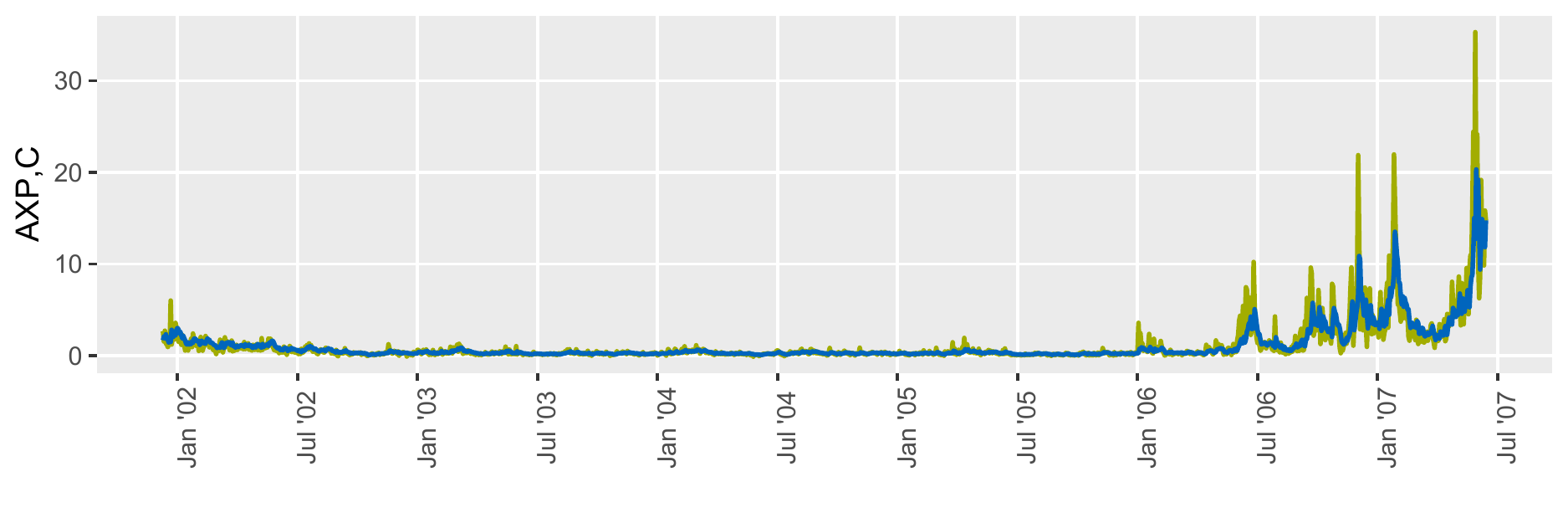}
	\renewcommand{\arraystretch}{0.75}
	\begin{tabular}{p{.05cm}lp{.05cm}l}
		\cellcolor{TUMgreen} & {\scriptsize historical data}
		& \cellcolor{TUMblue} & {\scriptsize one-day-ahead forecasts}\\
		\vspace{-3cm}
	\end{tabular}	
	\renewcommand{\arraystretch}{1}
	\caption{(part 1/3) Daily realized variance time-series and daily realized covariance time-series together with the time-series of the corresponding daily forecasts based on the partial correlation vine data transformation approach.}
	%\label{fig:tsplot_app1}
\end{figure}

\setcounter{figure}{13}

\begin{figure}[h!]
	\centering
	\includegraphics[width=.75\linewidth]{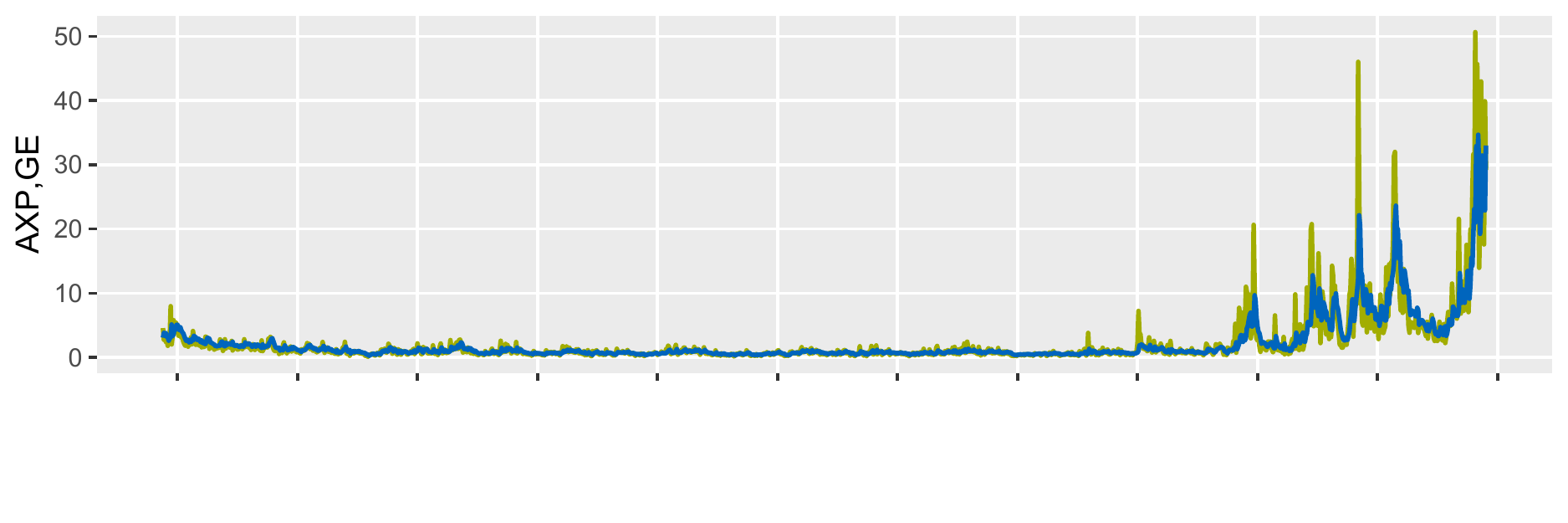}\vspace*{-1.15cm}\\
	\includegraphics[width=.75\linewidth]{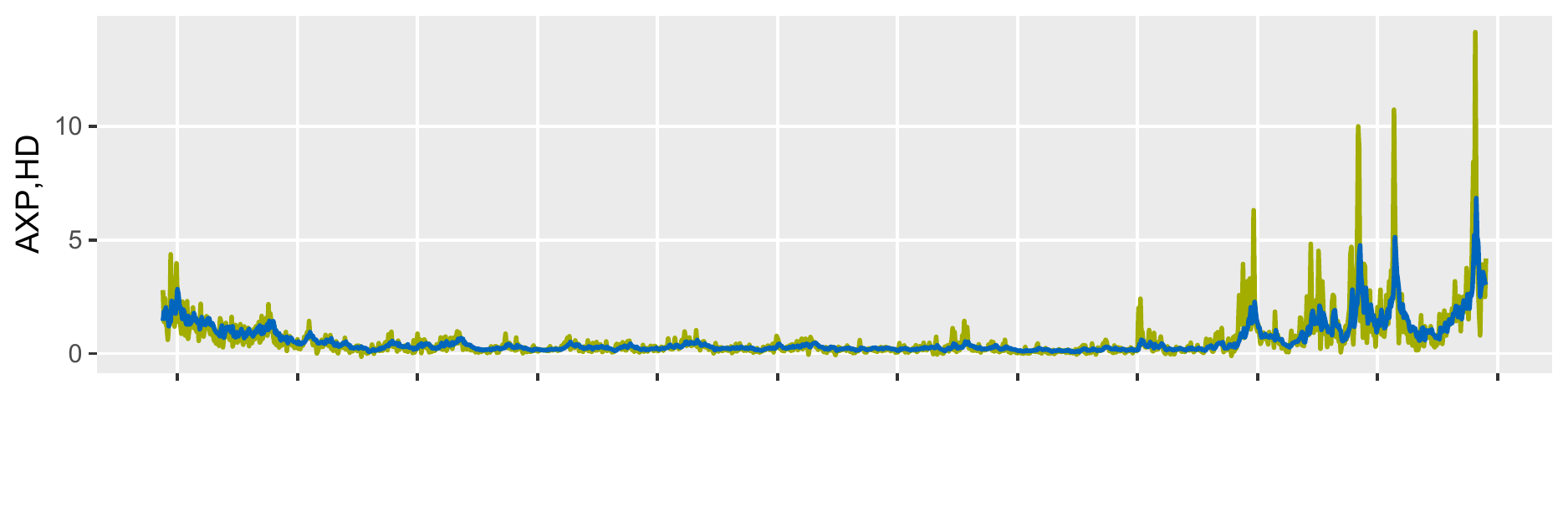}\vspace*{-1.15cm}\\	\includegraphics[width=.75\linewidth]{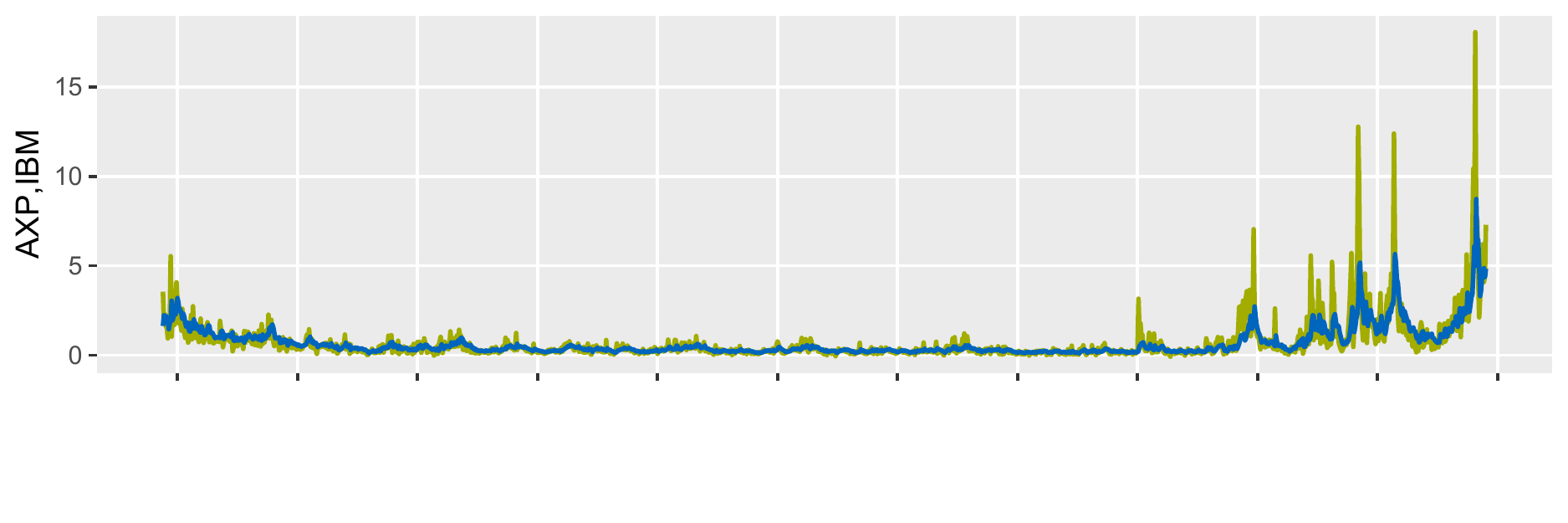}\vspace*{-1.15cm}\\
	\includegraphics[width=.75\linewidth]{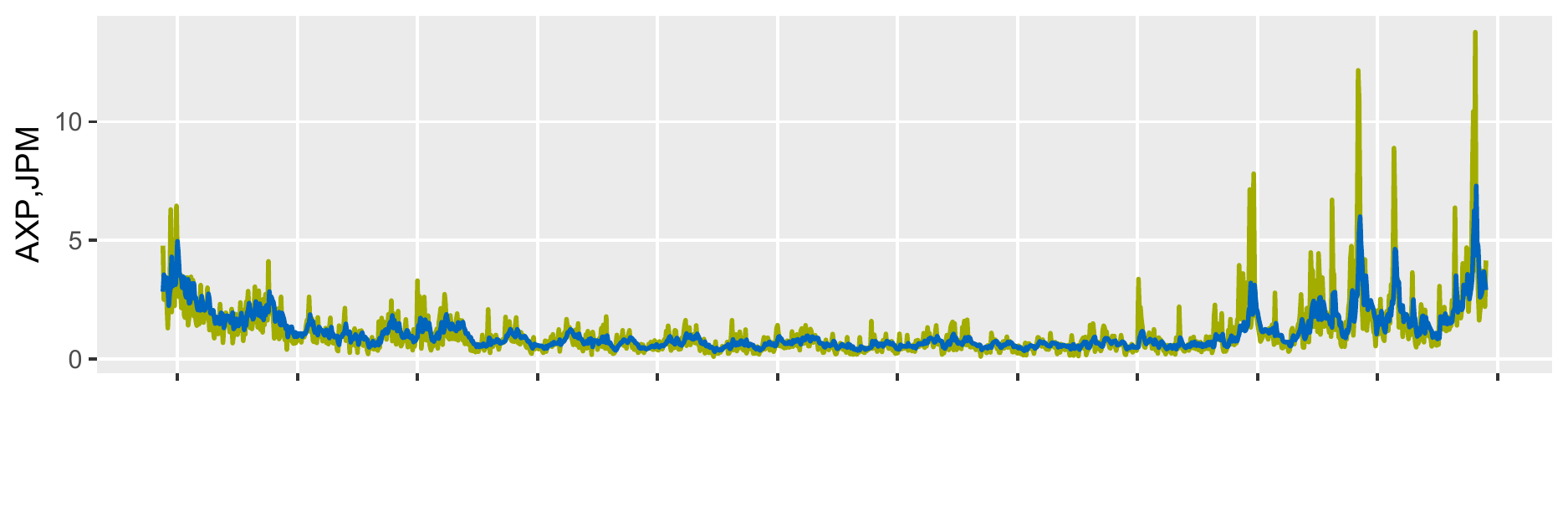}\vspace*{-1.15cm}\\
	\includegraphics[width=.75\linewidth]{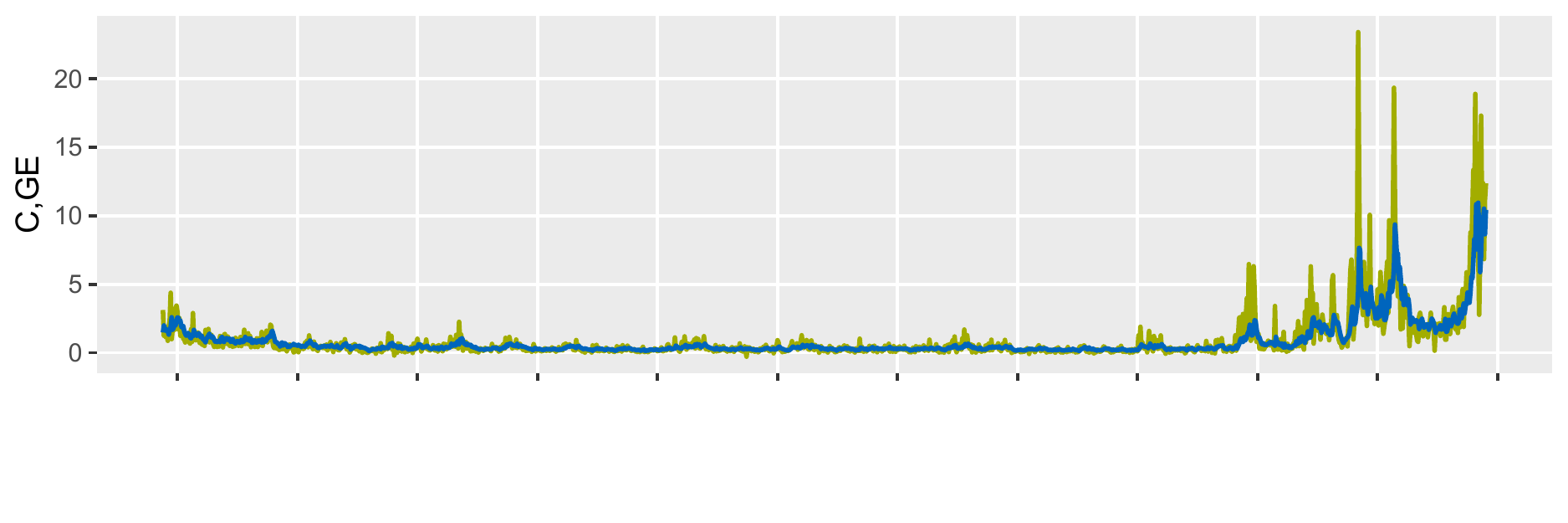}\vspace*{-1.15cm}\\
	\includegraphics[width=.75\linewidth]{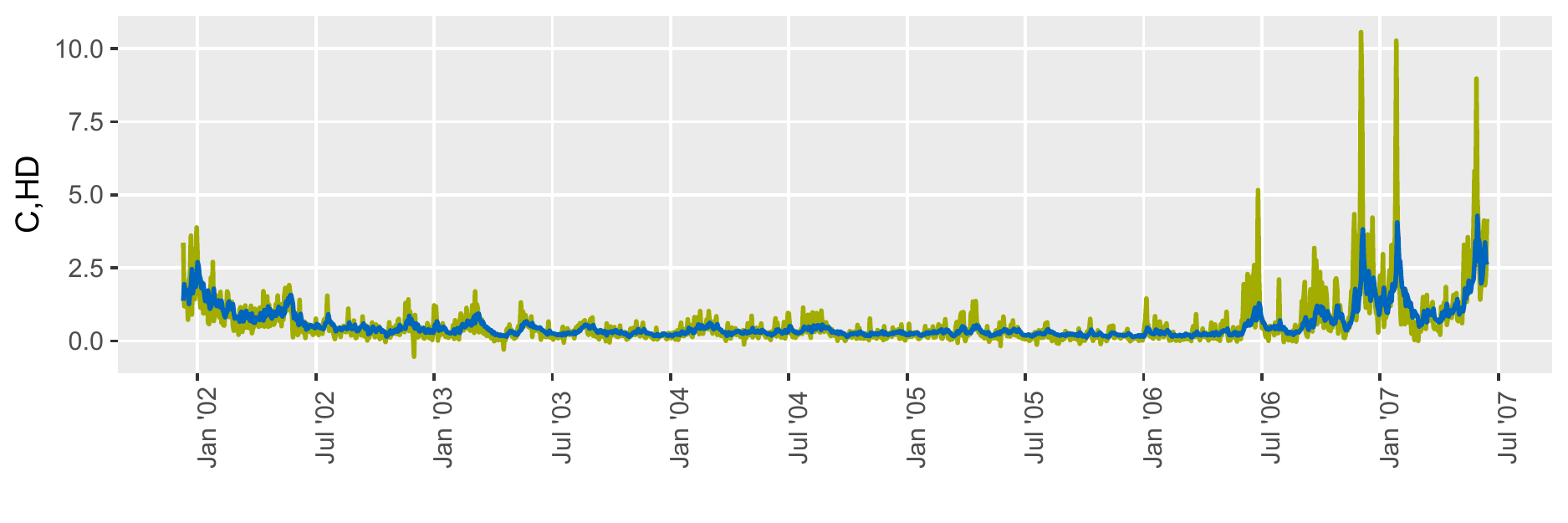}
	\renewcommand{\arraystretch}{0.75}
	\begin{tabular}{p{.05cm}lp{.05cm}l}
		\cellcolor{TUMgreen} & {\scriptsize historical data}
		& \cellcolor{TUMblue} & {\scriptsize one-day-ahead forecasts}\\
		\vspace{-3cm}
	\end{tabular}	
	\renewcommand{\arraystretch}{1}
	\caption{(part 2/3) Daily realized variance time-series and daily realized covariance time-series together with the time-series of the corresponding daily forecasts based on the partial correlation vine data transformation approach.}
	%\label{fig:tsplot_app2}
\end{figure}

\setcounter{figure}{13}

\begin{figure}[h!]
	\centering
	\includegraphics[width=.75\linewidth]{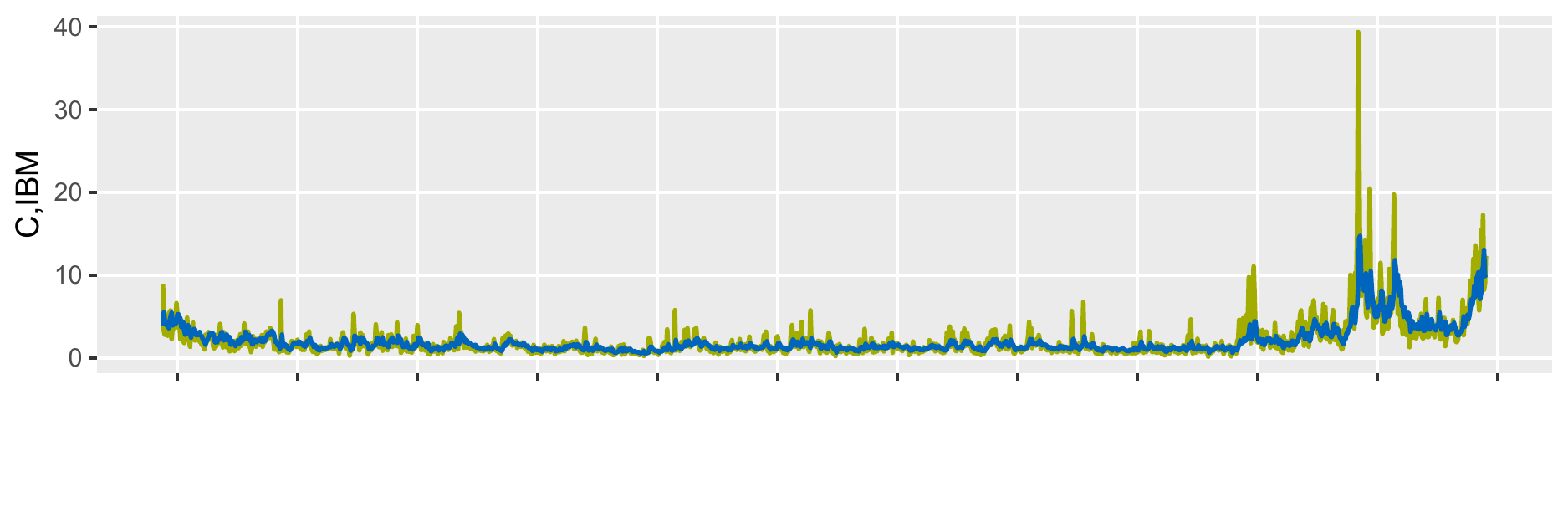}\vspace*{-1.15cm}\\
	\includegraphics[width=.75\linewidth]{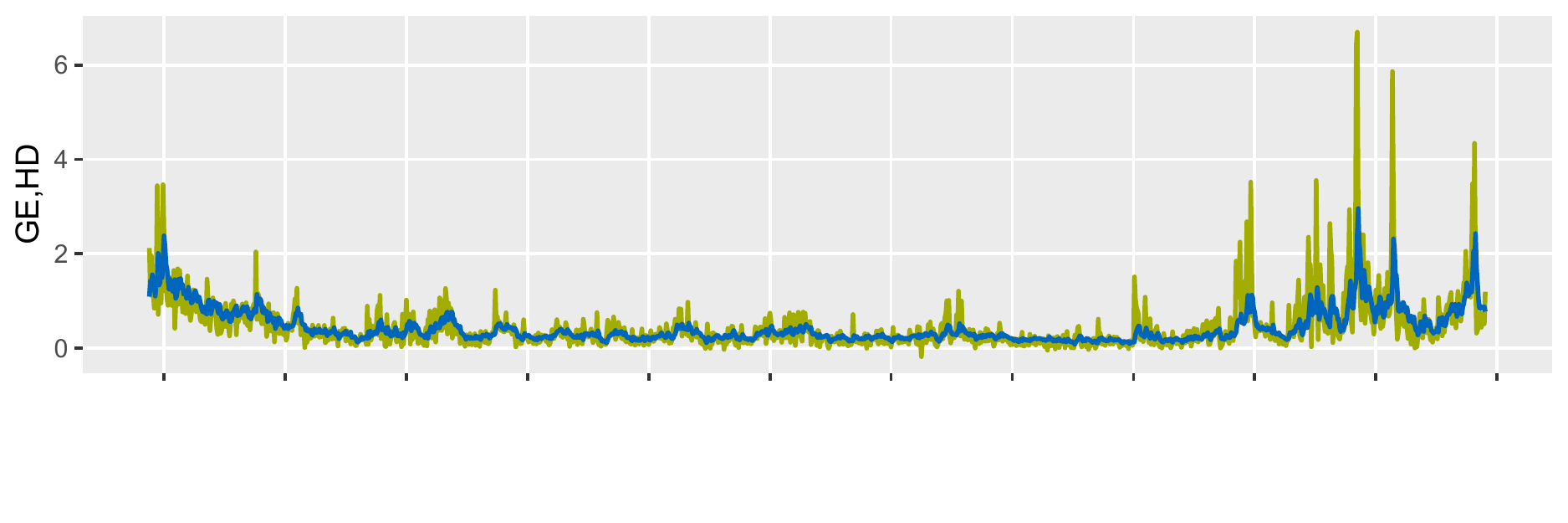}\vspace*{-1.15cm}\\
	\includegraphics[width=.75\linewidth]{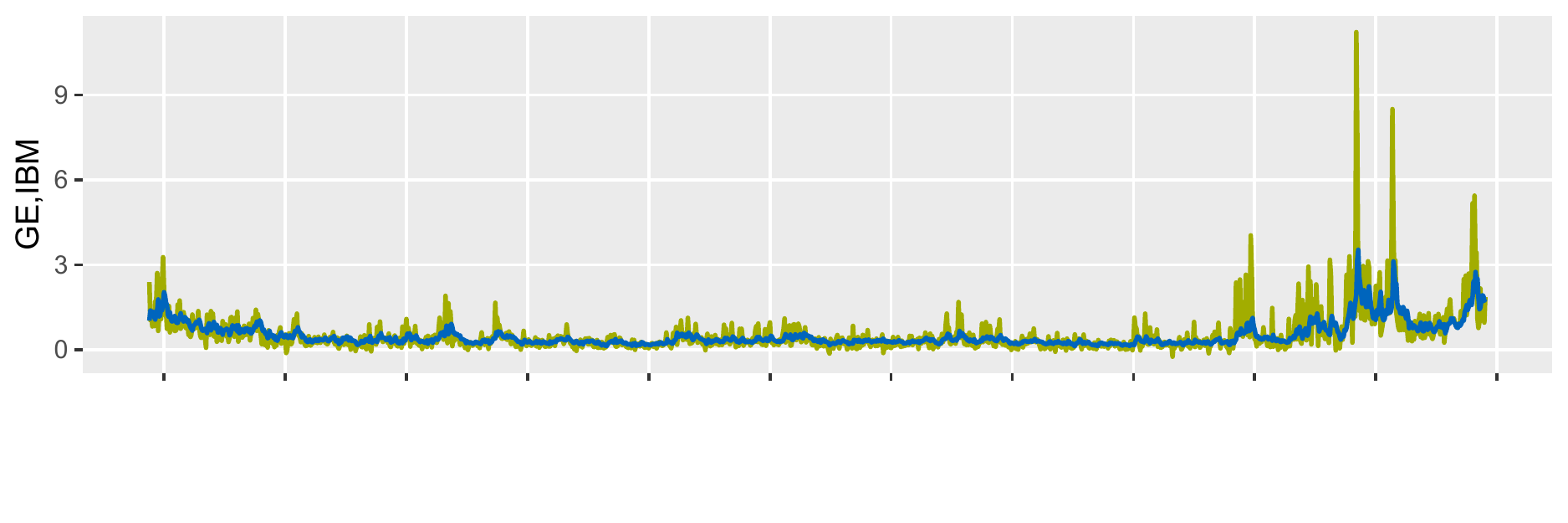}\vspace*{-1.15cm}\\
	\includegraphics[width=.75\linewidth]{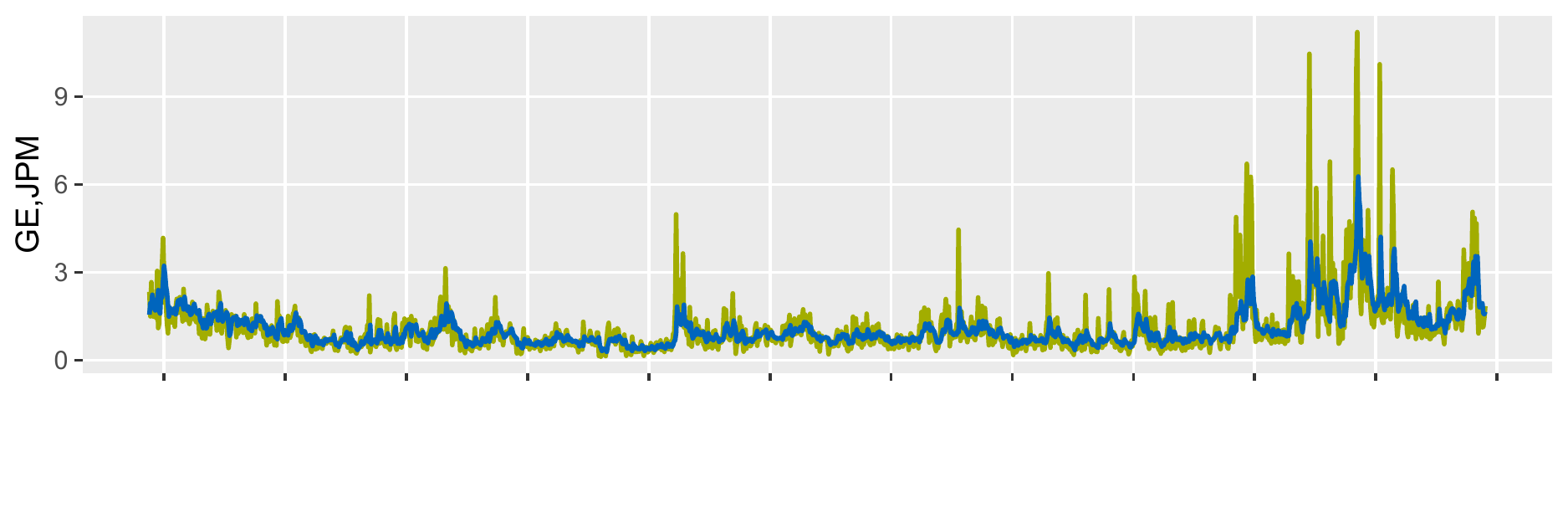}\vspace*{-1.15cm}\\
	\includegraphics[width=.75\linewidth]{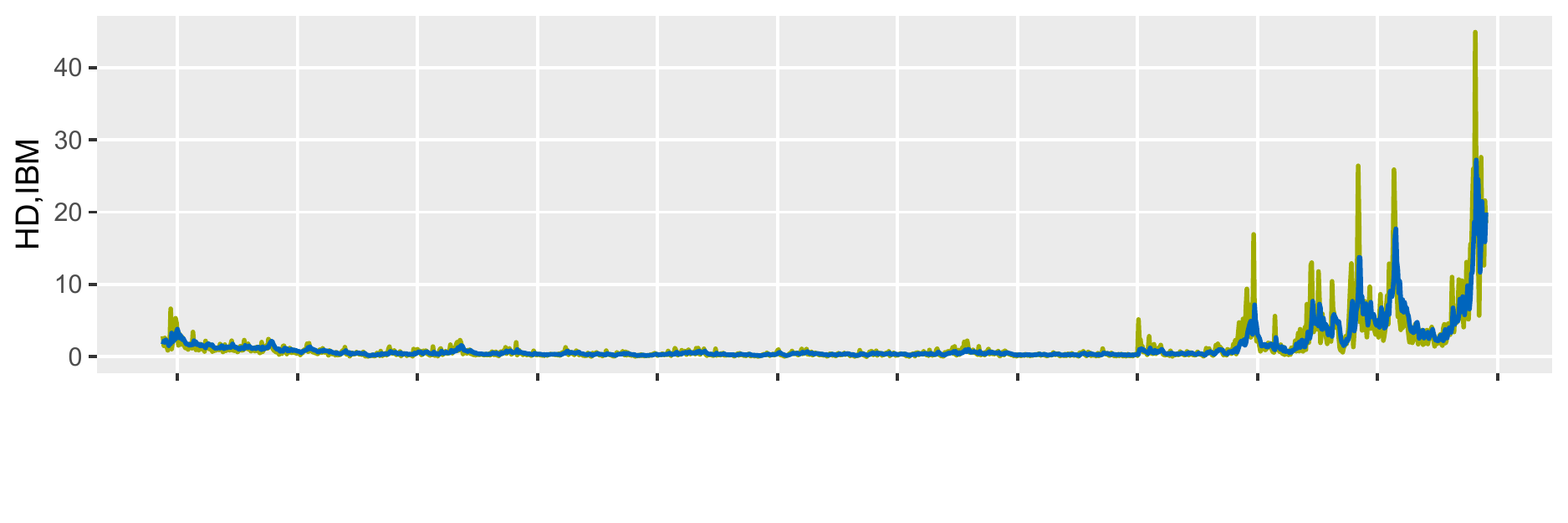}\vspace*{-1.15cm}\\
	\includegraphics[width=.75\linewidth]{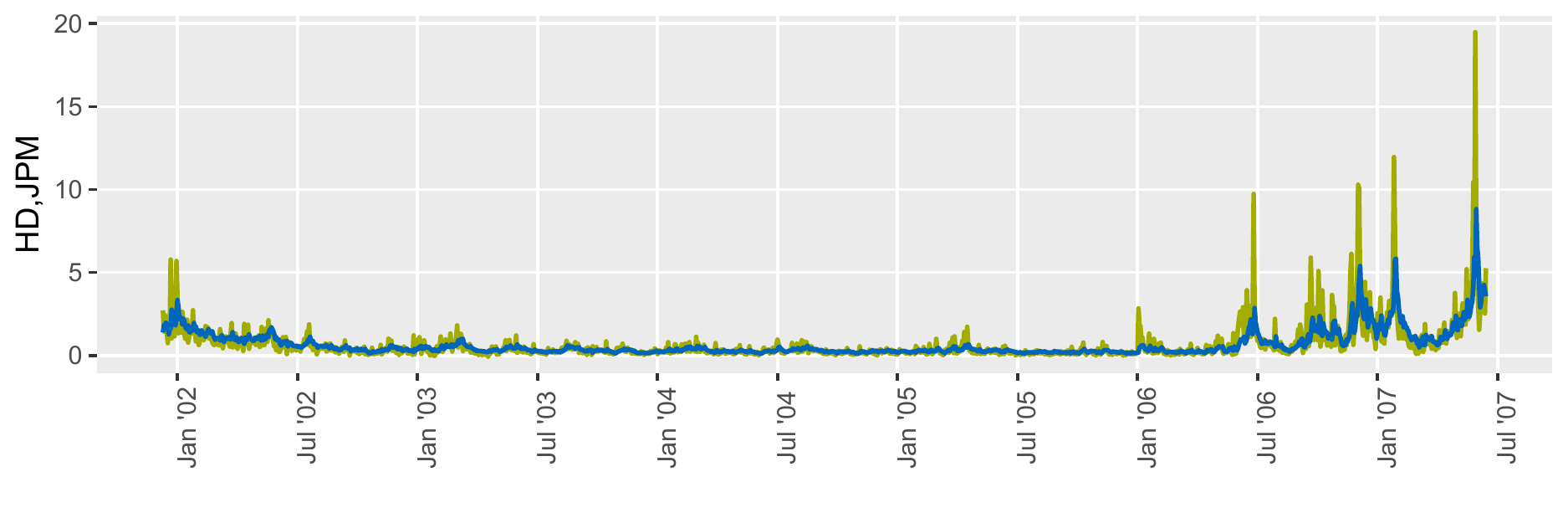}
	\renewcommand{\arraystretch}{0.75}
	\begin{tabular}{p{.05cm}lp{.05cm}l}
		\cellcolor{TUMgreen} & {\scriptsize historical data}
		& \cellcolor{TUMblue} & {\scriptsize one-day-ahead forecasts}\\
		\vspace{-3cm}
	\end{tabular}	
	\renewcommand{\arraystretch}{1}
	\caption{(part 3/3) Daily realized variance time-series and daily realized covariance time-series together with the time-series of the corresponding daily forecasts based on the partial correlation vine data transformation approach.}
	\label{fig:tsplot_app3}
\end{figure}

\end{document}